\documentclass[useAMS,usenatbib]{mn2e}
\usepackage{graphicx}
\usepackage{threeparttable}
\usepackage{subfig}
\title[3 mm line survey of two lines of sight of GC molecular clouds]{3 mm spectral line survey of two lines of sight toward two typical 
cloud complexes in the Galactic Center}
\author[Armijos-Abenda\~no et al. (2013)]{J. Armijos-Abenda\~no$^1$\thanks{E-mail:armijosaj@cab.inta-csic.es}, 
J. Mart\'in-Pintado$^1$, M. A. Requena-Torres$^2$, S. Mart\'in$^3$
\newauthor and A. Rodr\'iguez-Franco$^{4}$\\
$^{1}$Centro de Astrobiolog\'ia (INTA-CSIC), Ctra a Ajalvir, km 4, 28850, Torrej\'on de Ardoz, Madrid, Spain\\
$^{2}$Max-Planck Institut f\"ur Radioastronomie, Auf dem H\"ugel 69, D-53121 Bonn, Germany\\
$^{3}$European Southern Observatory, Alonso de C\'ordova 3107, Vitacura, casilla 19001, Santiago 19, Chile\\
$^{4}$Facultad de \'Optica y Optometr\'ia, Departamento de Matem\'atica Aplicada (Biomatem\'atica), Universidad Complutense de Madrid, Avenida\\
de arcos de Jal\'on, 118, E-28037 Madrid, Spain}
\begin{document}

\date{Accepted 2013 March 15. Received 2013 March 30; in original form 1988 October 11}

\pagerange{\pageref{firstpage}--\pageref{lastpage}} \pubyear{2002}

\maketitle

\label{firstpage}

\begin{abstract}

We present the results of two Mopra 3-mm spectral line surveys of the Lines of Sight (\emph{LOS}) toward the Galactic Center (GC) molecular complexes Sgr B2 (\emph{LOS}+0.693) and 
Sgr A (\emph{LOS}$-$0.11). The spectra covered the frequency ranges of $\sim$77-93 GHz and $\sim$105-113 GHz.
We have detected 38 molecular species and 25 isotopologues. The isotopic ratios derived from column density ratios are consistent with the canonical 
values, indicating that chemical isotopic fractionation and/or selective photodissociation can be considered negligible ($<$10\%) for the GC physical 
conditions. The derived abundance and rotational temperatures are very similar for both \emph{LOSs}, indicating very similar chemical and excitation 
conditions for the molecular gas in the GC. The excitation conditions are also very similar to those found for the nucleus of the starburst galaxy 
NGC 253. We report for the first time the detection of HCO and HOC$^+$ emission in \emph{LOS}+0.693. Our comparison of the abundance ratios 
between CS, HCO, HOC$^+$ and HCO$^+$ found in the two \emph{LOSs} with those in typical Galactic photodissociation regions (PDRs) and starbursts 
galaxies does not show any clear trend to distinguish between UV and X-ray induced chemistries. We propose that the CS/HOC$^+$ ratio could be
used as a tracer of the PDR components in the molecular clouds in the nuclei of galaxies.
\end{abstract}


\section{Introduction}

In this paper we study the physical conditions and chemical complexity of the quiescent molecular gas along two Lines of Sight \emph{(LOS)}, one
toward the Sgr B2 complex (\emph{LOS}+0.693) and the other one toward the Sgr A complex (\emph{LOS}$-$0.11), both complexes located in the Galactic Center (GC).
Molecular cloud complexes inside the Central Molecular Zone (CMZ, \citet{Morris}) in the GC show very different characteristics that the clouds in the Galactic disc. GC clouds are 
characterized by high gas-kinetic temperatures of $\ga$100 K \citep{Hutteme,nemesi01} and cold dust temperatures (T$_{dust}$) of $\la$30 K \citep{nemesi04}.

\begin{figure*}
  \includegraphics[width=135mm,angle=0]{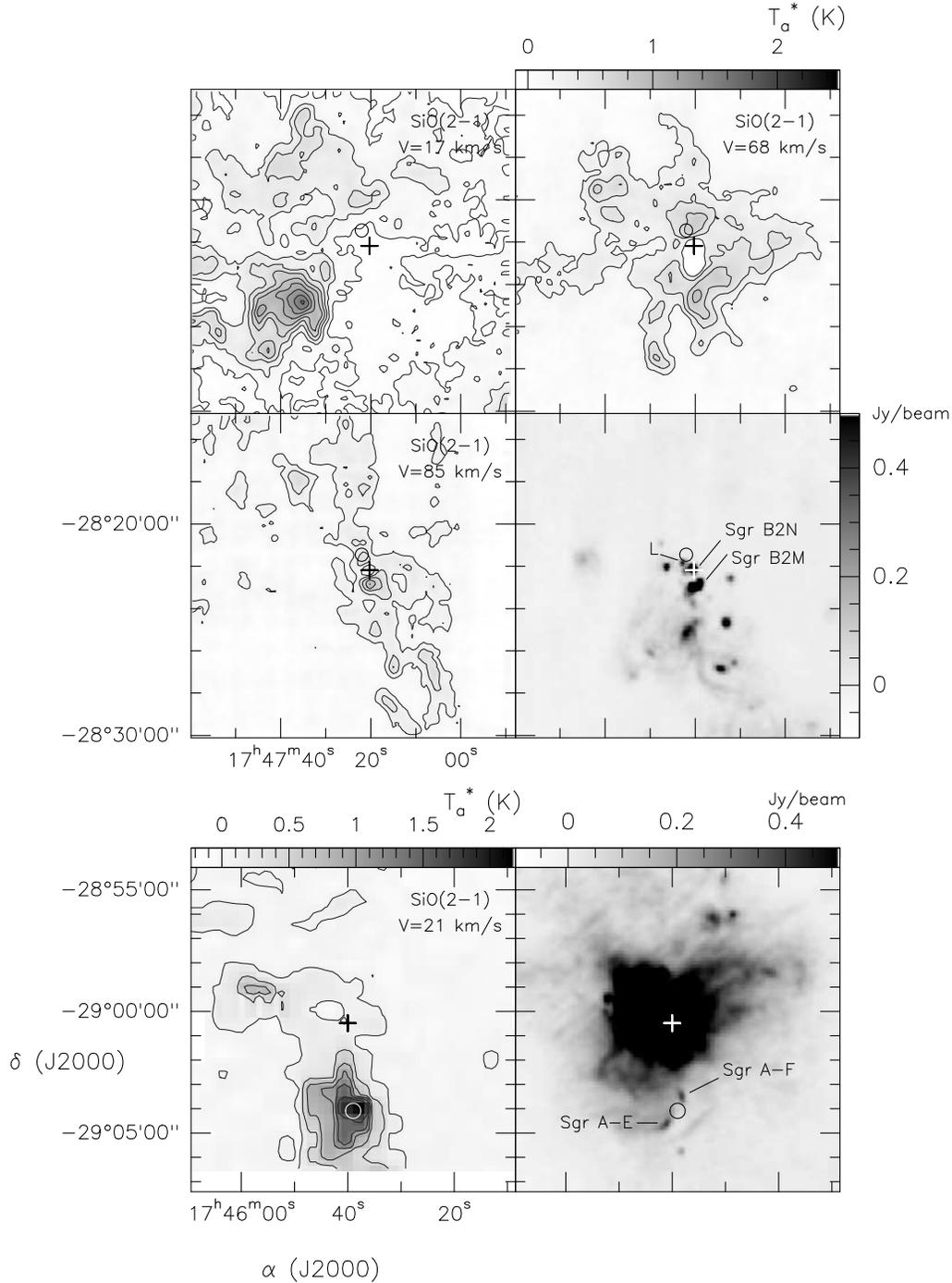}
  \caption{\textbf{(Upper and middle panels)} From left to right and top to bottom, SiO(2-1) large scale emission (Mart\'in-Pintado, priv. communication) maps for velocities of
   17, 68 and 85 km s$^{-1}$ and 20 cm radio continuum map \citep{Yusef04} of the Sgr B2 complex. \emph{LOS}+0.693 is shown as a circle with the size
   of the Mopra telescope beam (38$^{''}$ at 90 GHz) on the SiO(2-1) and 20 cm radio continuum maps. The crosses indicate the
   position of the massive hot core Sgr B2N on the four maps. The position of the HII region L is indicated on the 20 cm radio continuum map. \textbf{(Bottom panels)} From left
   to right, SiO(2-1) large scale emission (Amo-Baladr\'on, priv. communication) map for the velocity of 21 km s$^{-1}$ and 20 cm radio continuum
   map \citep{Yusef04} of the Sgr A complex. \emph{LOS}$-$0.11 is shown as a circle with the size of the Mopra telescope beam on the SiO(2-1)
   and 20 cm radio continuum maps. The crosses on the two maps show the position of the compact radio source Sgr A$^*$ corresponding to the central supermassive
   black hole. The features Sgr A-E and Sgr A-F are indicated with lines on the 20 cm radio continuum map.}
\label{fig1}
\end{figure*}

The Sgr B2 complex contains one of the most outstanding massive star formation sites in the Galaxy and consists of several star-forming cores embedded in a
lower density envelope \citep{Gordon}.
Sgr B2 hosts many dozens of compact and hypercompact HII regions, e.g., \citet{Pree96,Pree98} concentrated in two regions, Sgr B2N and Sgr B2M.
\emph{LOS}+0.693 toward the Sgr B2 molecular complex studied in this paper, shown in Fig.~\ref{fig1} (upper and middle panels), is outside the HII
region L \citep{Mehringer} and the main massive star forming regions Sgr B2M and Sgr B2N \citep{Jesus90}.
Therefore, \emph{LOS}+0.693 does not appear to be affected strongly by UV radiation, but otherwise this \emph{LOS} may be
subjected to significant X-ray irradiation since the Sgr B2 cloud is a well established X-ray reflection 
nebula due to its strong Fe K$\alpha$ line emission at 6.4 keV \citep{Koyama}.

The Sgr A molecular complex contains two massive molecular clouds at 20 km s$^{-1}$ and 50 km s$^{-1}$ believed to be interacting with the Circumnuclear Disk (CND)
surrounding the central supermassive black hole Sgr A$^*$ and two supernova remnants, Sgr A East and G359$-$0.09 \citep{Coil, Herr2005,Ferriere}.
\emph{LOS}$-$0.11 toward the 20 km s$^{-1}$ molecular cloud, shown in Fig.~\ref{fig1} (bottom panels),  is located in projection
216$^{''}$ ($\sim$8.6 pc) south of the black hole Sgr A$^*$. \emph{LOS}$-$0.11, lies near the elongated non-thermal features Sgr A-F and Sgr A-E \citep{Yus87,Lu,Yusef05} 
indicated in Fig.~\ref{fig1}. Recent star formation (compact HII regions G$-$0.02-0.007) around Sgr A$^*$ is mainly concentrated toward the 50 km s$^{-1}$ molecular cloud \citep{Mills}.


Unlike the star forming regions of Sgr B2M and Sgr B2N, the selected lines of sight toward the cloud positions in the Sgr B2 and Sgr A complexes
studied in this paper do not show any signposts of massive star formation like H$_2$O masers, ultracompact HII regions, hot cores or
recombination line emission \citep{Hutteme,Jesus97}.

Several unbiased spectral line surveys have been carried out toward the Sgr B2 star forming cores, positions N \citep{Nummelin98,Nummelin00,Friedel,Belloche}, M \citep{sutton,Belloche}, S and
OH \citep{Friedel}. \citet{Nummelin98,Nummelin00} also carried out a spectral survey toward a quiescent region in Sgr B2. They detected 26 species and showed 
the large difference in chemical complexity and excitation between the quiescent and the star forming clouds.
Our line survey of two quiescent lines of sight were selected from the systematic study of the HNCO/CS ratios carried out by \citet{sergio2008}.
In this study both locations were among the objects which showed the highest ratios indicating that the chemistry and likely the heating of
both GC \emph{LOSs} are mainly
dominated by low velocity shocks. This is also supported by \citet{miguel08}, who also studied several GC sources, including our two
\emph{LOSs}, finding that the abundances of complex organic molecules like CH$_2$OHCHO, (CH$_3$)$_2$O, CH$_3$CHO, among
others, are larger in these GC \emph{LOSs} than those in hot cores of the Galactic disk, suggesting ejection of molecules 
from dust grains by low velocity C-type shocks. 
Previous surveys of the Sgr B2 quiescent clouds by \citet{Nummelin98,Nummelin00} at high frequencies did not have the 
sensitivity to detect weak molecular lines from these species.

In this paper, we present a 3 mm spectral line survey of the quiescent molecular gas along \emph{LOS}+0.693 and \emph{LOS}$-$0.11 toward 
the Sgr B2 and Sgr A complexes, respectively. Both complexes, inside the CMZ, are outstanding regions affected by high energy phenomena \citep{Koyama,terrier,Ponti} and large 
scale shocks \citep{Jesus01}.
Our sensitivity allowed us to detect, for the first time, the emission of the HCO and HOC$^+$ molecules, which are considered to be tracers
of UV radiation and X-ray chemistry in molecular clouds.
In Sec.~\ref{observations}, we present our observations and data reduction. In Sec.~\ref{results}, the details of the line identification (Sec.~\ref{Line_identification}),
line profiles (Sec.~\ref{Lineprofiles}) and analysis (Sec.~\ref{analysis}) are reported.
In Sec.~\ref{isotopic_ratios} we derive six isotopic ratios and discuss their implications.
In Sec.~\ref{diss_abundances}, we discuss the molecular abundance and excitation conditions found in the GC and compare them with those in galactic nuclei (Sec.~\ref{nucli_galaxies}), and 
the implication of our detection of HCO and HOC$^+$ in the
UV and X-ray induced chemistry in galactic nuclei (Sec.~\ref{PDR_XDRtracers}).
Finally, the conclusions are summarized in Sec.~\ref{conclusions}.

\section{Observations and data reduction}\label{observations}

The observations were carried out with the 22-m Mopra radio telescope\footnote{Mopra is operated by the Australia Telescope National 
Facility, CSIRO and the University of New South Wales.} on November of 2007. We used the the dual 3 mm Monolithic Microwave Integrated Circuit (MMIC) 
receiver connected to the 8 GHz Spectrometer, which provided a velocity resolution of $\sim$0.9 km s$^{-1}$ at 90 GHz. Spectra in two polarizations were observed simultaneously.
Two frequency ranges in the 3 mm window were covered, from $\sim$77 to 93 GHz and from $\sim$105 to 113 GHz. The beam size of the 
telescope was 38$''$ at 90 GHz and 30$''$ at 115 GHz. We used position switching as observing mode with 
the emission free reference positions selected from the CS maps obtained by \citet{Bally}. The nominal positions used for the observation of \emph{LOS}+0.693 and \emph{LOS}$-$0.11 were 
$\alpha_{J2000}=17^{\rmn{h}} 47^{\rmn{m}} 22\fs0$, $\delta_{J2000}=-28\degr 21\arcmin 27\farcs0$ and $\alpha_{J2000}=17^{\rmn{h}} 45^{\rmn{m}} 39\fs0$, 
$\delta_{J2000}=-29\degr 04\arcmin 05\farcs0$, respectively. Fig.~\ref{fig1} shows the two observed positions superimposed on the SiO(2-1) (Mart\'in-Pintado \& Amo-Baladr\'on, priv.
communication) and the 20 cm radio continuum maps \citep{Yusef04} of the Sgr B2 and Sgr A complexes. The reference positions were $\alpha_{J2000}=17^{\rmn{h}} 46^{\rmn{m}} 23\fs0$, 
$\delta_{J2000}=-28\degr 16\arcmin 37\farcs3$ and $\alpha_{J2000}=17^{\rmn{h}} 46^{\rmn{m}} 00\fs1$, $\delta_{J2000}=-29\degr 16\arcmin 47\farcs2$ 
for \emph{LOS}+0.693 and \emph{LOS}$-$0.11, respectively. 

The raw data were reduced by using the ATNF Spectral Analysis Package (ASAP) at Mopra telescope to create the fits files for further 
processing with the MASSAIJ package\footnote{This package have been developed at the Centro de Astrobiolog\'ia. More information about this package in 
http://damir.iem.csic.es/mediawiki-1.12.0/index.php/Portada.}, where baseline subtraction 
were applied. In several regions of the spectra there were baseline ripples, which were partially 
corrected by editing the data in the Fourier transformed plane. The two polarizations of the spectra were averaged to improve the signal 
to noise ratio. The spectra were then smoothed to a velocity resolution of $\sim$3.5 km s$^{-1}$, appropriate for the linewidths of $\sim$20 km s$^{-1}$
observed toward molecular clouds in the GC. Our line intensities in T$_{\rm A}^*$ units are affected by 20\%-30\% 
uncertainties in the calibration procedure based on a noise diode and an ambient temperature load.

\begin{figure*}
\includegraphics[width=140mm]{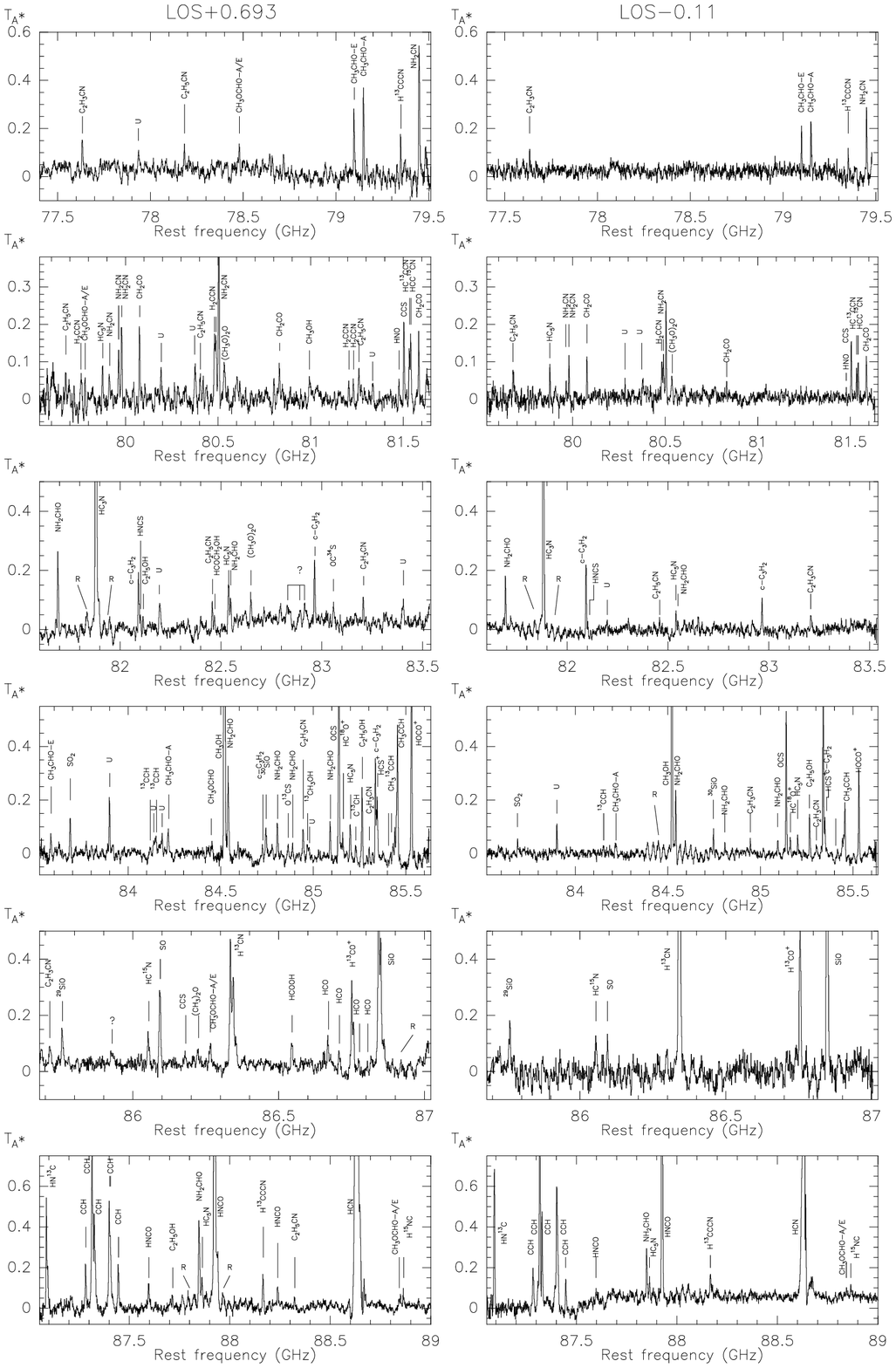}
\caption{3 mm spectra of \emph{LOS}+0.693 (left panels) and \emph{LOS}$-$0.11(right panels). Each panel shows a spectra with a 
$\sim$2 GHz coverage. Detected transitions of known molecules are indicated, while the unidentified lines are labeled as ``U''. Some regions of the 
spectra affected by ripples are indicated with the letter R.}
\label{fig21}
\end{figure*}
\begin{figure*}
\hspace{1.8cm}
\includegraphics[width=148mm]{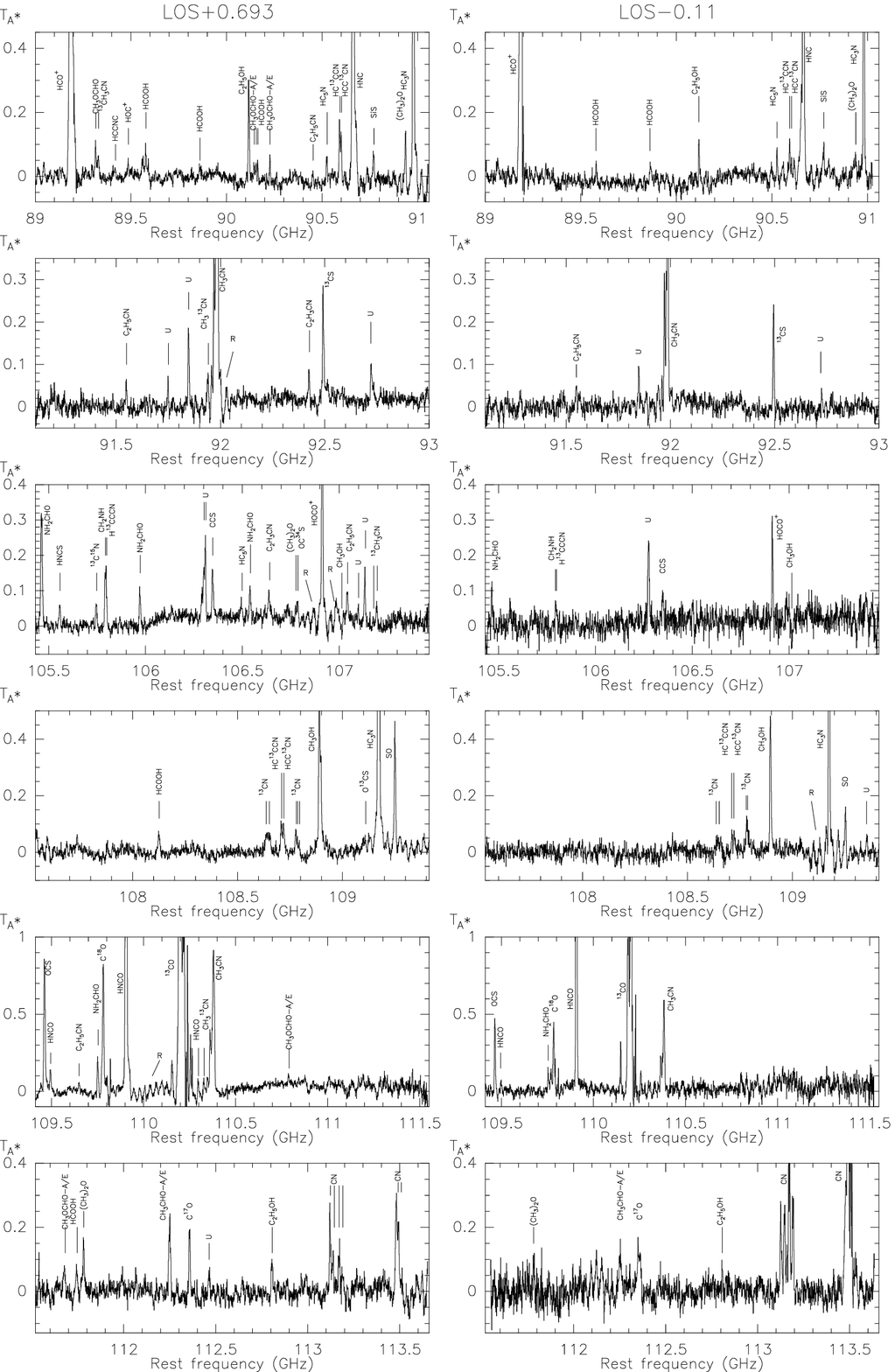}
\contcaption{}
\end{figure*}

\section{Results}\label{results}

\subsection{Line identification}\label{Line_identification}

The 3 mm spectra observed toward both \emph{LOSs} are shown in Fig.~\ref{fig21}, where the identified molecules and the unidentified 
lines are indicated. A representative sample of line profiles for several molecular transitions  and their isotopologues for \emph{LOS}+0.693 and 
\emph{LOS}$-$0.11 are shown in Fig.~\ref{fig3} and~\ref{fig4}, respectively. 
The molecular identification and analysis were carried out using the frequencies and the spectroscopic information from the JPL \citep{Pickett} and 
CDMS \citep*{Muller01,Muller05} catalogs contained in the MASSAIJ package. For \emph{LOS}+0.693, we found 38 molecular species and 25 isotopologues, as well as 18 unidentified lines. 
In contrast, for \emph{LOS}$-$0.11, we only found 34 molecular species and 18 isotopologues, as well as 8 unidentified lines. 

In our survey, we have detected, for the first time, the HCO and HOC$^+$ emission toward the quiescent gas of \emph{LOS}+0.693 in the GC outside the hot cores of Sgr B2N and Sgr B2M. 
We have detected four HCO hyperfine lines (F=2-1, 1-0, 1-1 and 0-1) and one HOC$^+$(1-0) line toward \emph{LOS}+0.693 (see Fig.~\ref{fig21} and~\ref{fig3}). HCO, HOC$^+$, HC$_2$NC and HCOCH$_2$OH 
are not detected toward \emph{LOS}$-$0.11 (see Fig.~\ref{fig21} and~\ref{fig4}). HCO and HOC$^+$ molecules have been proposed to be tracers of PDRs \citep{Apponi,Javi09b,sergio2009b} and XDRs \citep{usero}. 
Tables \ref{table1} and \ref{table2} show the line parameters derived by Gaussian fitting to almost all observed molecular lines for both \emph{LOSs}.  
The tables also include, for both \emph{LOSs}, upper limits to the emission of several molecular species, like HOC$^+$ and HCO, which are relevant for 
the discussion. These upper limits (4 for \emph{LOS}+0.693 and 16 for \emph{LOS}$-$0.11) correspond to 3$\sigma$ of the peak and 
velocity-integrated intensities (see Tables \ref{table1} and \ref{table2}).


\subsection{Line profiles}\label{Lineprofiles}


Most of line profiles from each cloud trace just their kinematic structure, i.e. three velocity components ($\sim$17, 68 and 
85 km s$^{-1}$) in \emph{LOS}+0.693 and one velocity component ($\sim$20 km s$^{-1}$) in \emph{LOS}$-$0.11. Additional
double-peaked line profiles of certain strong molecular species like HCO$^+$, HNC and HCN are likely a result of optical
depth effects, as this double-peaked structure is not present in the line profiles of their optically-thin isotopologues.
In this paper, we will concentrate on the emission from the 68 and 85 km s$^{-1}$ clouds along \emph{LOS}+0.693 and the
$\sim$20 km s$^{-1}$ cloud along \emph{LOS}$-$0.11.

To compare molecular abundances derived from optically-thick lines (e.g. CN, HNC, HCN, HCO$^+$ and CH$_3$OH) with those showing simpler profiles we have calculated, in Tables  \ref{table1} and \ref{table2}, their 
velocity-integrated intensities over the selected velocity ranges given in column 4 of these tables. For \emph{LOS}+0.693 we have considered two velocity ranges of 40$-$80 km s$^{-1}$ and 80$-$110 km s$^{-1}$ corresponding 
to the emission of the $\sim$68 km s$^{-1}$ and $\sim$85 km s$^{-1}$ clouds, respectively.  For \emph{LOS}$-$0.11, we have derived velocity-integrated intensities only for HCO$^+$(1-0) and HCN(1-0) lines and their corresponding isotopologues. 

Fig.~\ref{fig3} and~\ref{fig4} present the line profiles of all detected methanol transitions for \emph{LOS}+0.693 and \emph{LOS}$-$0.11, respectively.  
Interestingly the CH$_3$OH(3$_{1,3}$-4$_{0,4}$) line is the only methanol line observed 
in absorption in our survey, while the CH$_3$OH(0$_{0,0}$-1$_{-1,1}$) line appears to be optically thick in both \emph{LOSs}. Furthermore, the CH$_3$OH(5$_{-1,5}$-4$_{0,4}$) line at 84.5 GHz is intense 
and it has been reported to be masing in some sources \citep{Zuck,Batrla}.

\subsection{Analysis}\label{analysis}

Assuming optically thin emission and excitation in Local Thermodynamic Equilibrium (LTE), the relation between the LTE total molecular column density $N$ and the measured integrated line intensity, $W=\int T^{*}_A dv$ in K km s$^{-1}$, of 
a transition is given by:

\begin{equation}
 \frac{N_u}{g_u}=\frac{N}{Q_{rot}} e^{-E_u/kT_{rot}}=\frac{1.67\times10^{14}W}{\nu \mu^2 S}
\label{ecua1}
\end{equation}

where $N_u$ is the molecular column density in the upper energy level, $g_u$ and $E_u$/$k$ are the degeneracy and the energy (in K) of the upper level of the transition 
with a frequency $\nu$ in GHz, a dipole moment $\mu$ in Debye and a line strength $S$. T$_{rot}$ is 
the LTE rotational temperature in K, $Q_{rot}$ is the rotational partition function.
Taking the natural logarithm in (\ref{ecua1}) we obtain that the main equation of the rotational diagrams (RDs) is given by:

\begin{equation}
 \ln \frac{1.67\times10^{14}W}{\nu \mu^2 S}=\ln \frac{N}{Q_{rot}}- \frac{E_u}{k T_{rot}}.
 \label{ecua2}
\end{equation}

\begin{figure}
\includegraphics[width=84mm]{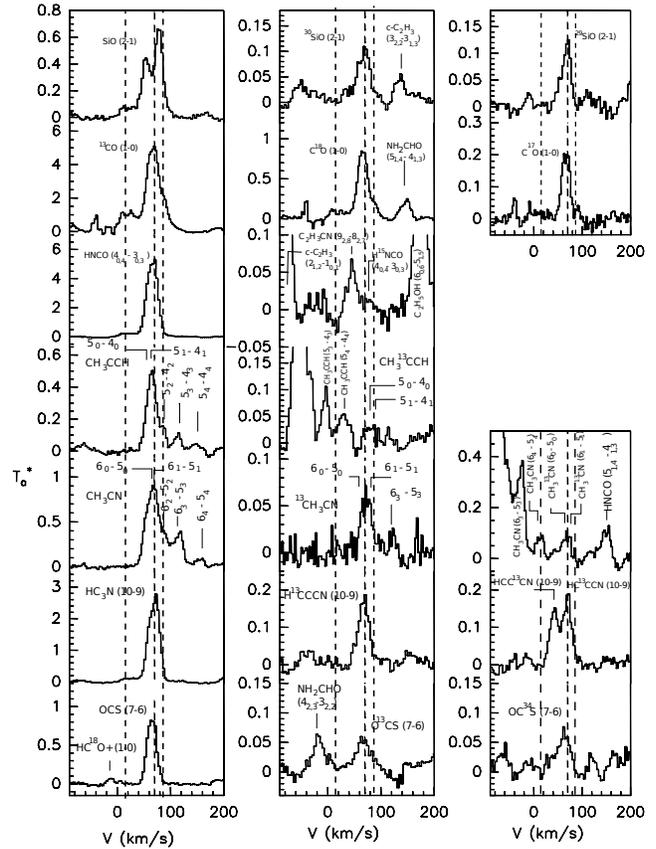}
\caption{Sample of some molecular lines and their isotopologues observed toward \emph{LOS}+0.693. We used the column densities derived from these transitions and 
others to estimate the carbon, nitrogen, oxygen, sulfur and silicon isotopic ratios. The vertical dashed lines show the LSR velocities of 17, 68 and 85 
km s$^{-1}$.}
\label{fig3}
\end{figure}

\begin{figure}
\includegraphics[width=84mm]{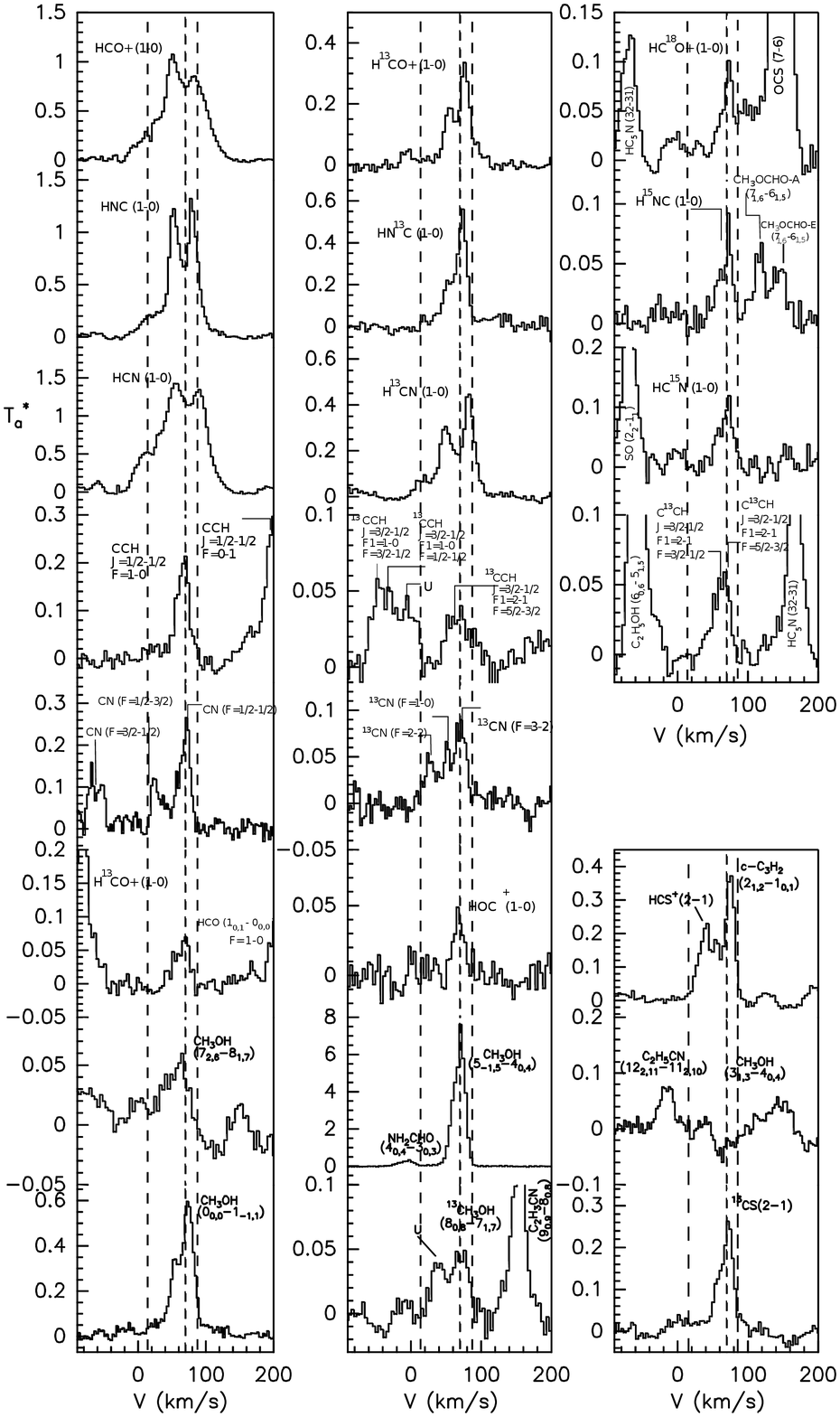}
\contcaption{}
\end{figure}

\begin{figure}
\includegraphics[width=84mm]{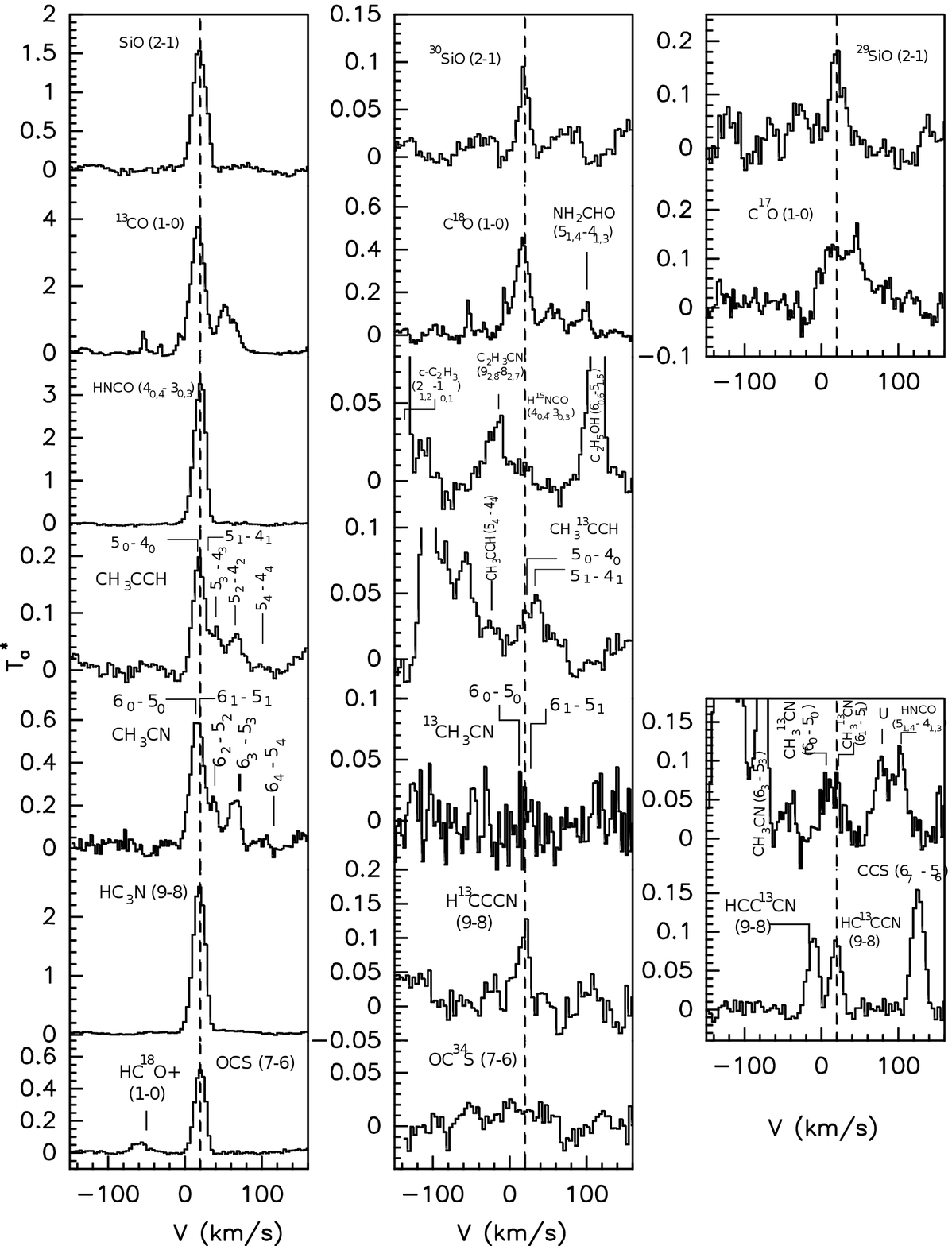}
\caption{Sample of some molecular lines and theirs isotopologues observed toward \emph{LOS}$-$0.11. We used the column densities derived from these transitions and 
others to estimate the carbon, nitrogen, oxygen, sulfur and silicon isotopic ratios. The vertical dashed line shows the LSR velocity of 20 km s$^{-1}$.}
\label{fig4}
\end{figure}

\begin{figure}
\includegraphics[width=84mm]{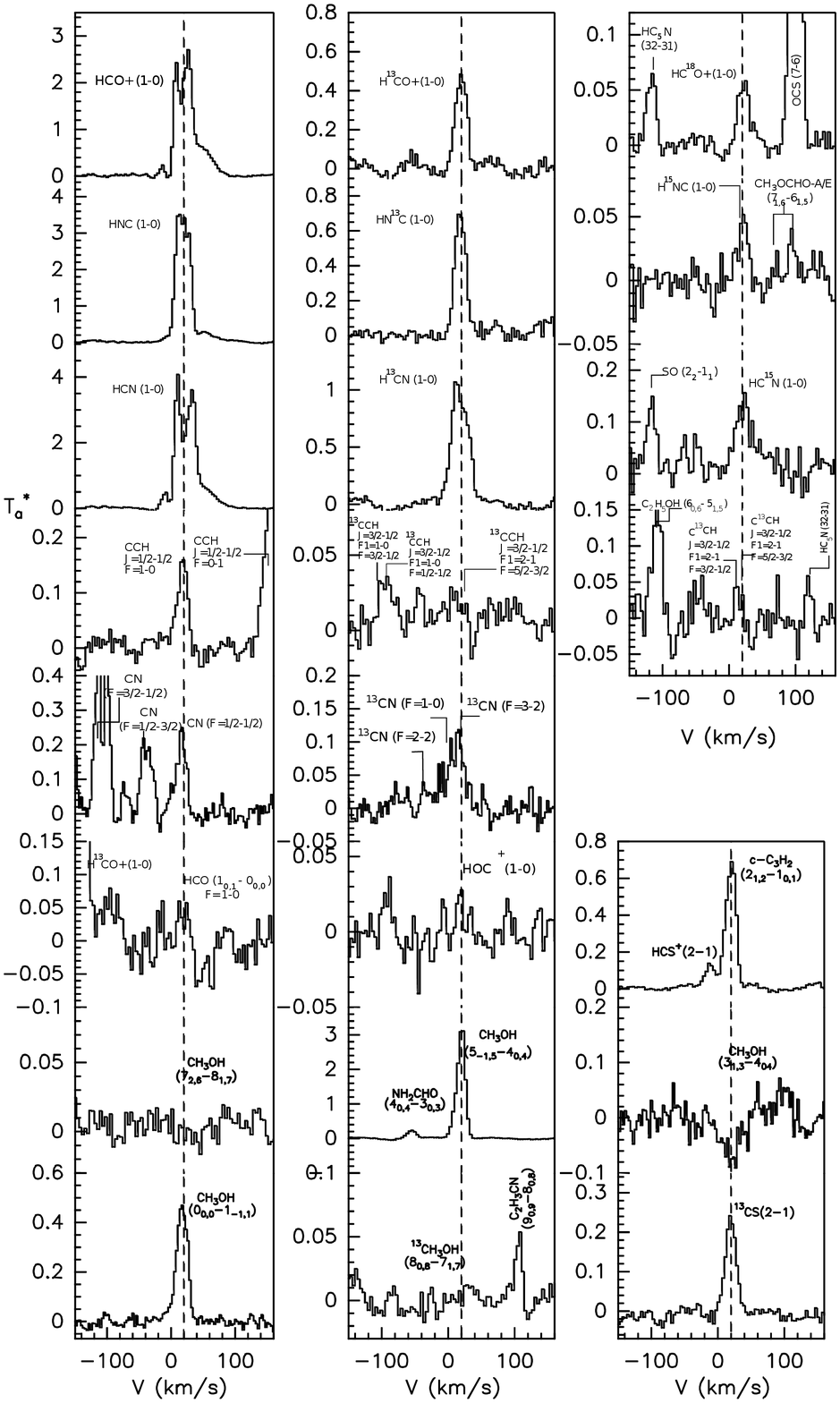}
\contcaption{}
\end{figure}

By observing more than one transition from the same molecule, we can estimate the excitation temperature of the rotational levels, T$_{rot}$,
and the LTE total column densities (by using the $Q_{rot}$ for T$_{rot}$) from a linear regression fit to equation (\ref{ecua2}).
Since the typical uncertainties of assuming the Rayleigh-Jeans approximation are less than the calibration uncertainties, they are ignored.

\subsubsection{Rotational temperatures and molecular column densities }\label{column_densities}

We have used the MASSAIJ package with the spectroscopic information ($A_{ul}$, $g_u$, $E_u$, $\mu^2S$, $Q_{rot}$, etc) from the JPL and CDMS catalogs to generate the RDs for all molecules with more than one observed transition. 
Fig.~\ref{fig5} shows RDs (for both GC sources) for selected molecules, and the derived T$_{rot}$ from RDs are listed in Table \ref{table3}. 
All RDs are fitted with a single T$_{rot}$. The uncertainties in the T$_{rot}$ are derived from the 1$\sigma$ uncertainties in the integrated line intensities and also take into
account the uncertainty due to the linear regression fit in the RDs. In our survey, the derived T$_{rot}$ ranges from 5 K to 73 K (see Table \ref{table3}),
however, most of the derived T$_{rot}$ are rather low compared with the estimated mean kinetic temperature of $\approx$100 K found in the GC \citep{Gustem,Hutteme}, indicating
subthermal excitation due to relatively low H$_2$ densities (see Sec. \ref{densities}).

Molecular lines contaminated by the emission from unidentified molecular species (see notes in Tables \ref{table1} and \ref{table2}) were not included in the RDs. 
When required, we have also properly taken into account the hyperfine  structure  (see notes in Tables \ref{table1} and \ref{table2}) to estimate the total column densities given in Table \ref{table3}. 
We have used equation (\ref{ecua1}) to estimate the total column densities for molecules with only one detected transition by assuming a T$_{rot}$ of 10 K, the average value of the low T$_{rot}$ derived from other molecules (see Table \ref{table3}). We have also used a T$_{rot}$ of 10 K to derive the column densities of molecular species with several observed 
transitions but with insufficient dynamical range in $E_u$ ($\la$2 K) to derive a reliable T$_{rot}$.

To avoid the uncertainties introduced by optical depth effects in the estimated column densities of molecules like HCN, HNC, HCO$^+$, in Table \ref{table3}, when possible, we have derived 
them from the optically thin lines (see note b in Table \ref{table3}) of their rarer isotopologues, assuming the typical GC isotopic ratios (\citet{wilson94}, hereafter 
W\&R94). The LTE approximation used in our analysis provides beam-averaged abundances which are relatively similar to those obtained by using non-LTE statistical equilibrium methods.
The difference in the estimated HC$_3$N column densities using the LTE and non-LTE analysis is less than a factor of $\sim$2 for both
GC sources (see Sec.~\ref{densities}). 

To derive the CH$_3$OH column density for \emph{LOS}+0.693 we have used the 7$_{2,6}$-8$_{1,7}$ transition and the T$_{rot}$ of 14 K derived by 
\citet{miguel08}. This transition is selected since it shows thermal emission, is likely optically thin, and is observed in 
emission toward \emph{LOS}+0.693 (see Sec. \ref{Lineprofiles}).
For \emph{LOS}$-$0.11 this methanol transition is not detected, so we have derived only an upper limit to the methanol column density by assuming the T$_{rot}$ of 13 K derived by \citet{miguel08}.

\subsubsection{H$_2$ column densities and molecular abundances}\label{abundances}

To estimate the molecular fractional abundances relative to H$_2$ we need to estimate the H$_2$ column density (N$_{\rm H_2}$) toward both \emph{LOSs}. We have used
the C$^{18}$O lines to estimate the H$_2$ column densities by assuming a $^{16}$O/$^{18}$O isotopic ratio of 250 (W\&R94) and a relative abundance of CO to H$_2$ of 10$^{-4}$ \citep{Frerking}. We found N$_{\rm H_2}$ of 5.9 (0.2)$\times$10$^{22}$ cm$^{-2}$
and 2.4 (0.2)$\times$10$^{22}$ cm$^{-2}$ for \emph{LOS}+0.693 and \emph{LOS}$-$0.11, respectively.
The estimated relative abundances obtained by dividing the molecular column densities by their respective N$_{\rm H_2}$ are shown in 
Table \ref{table3} for the molecules and velocity components identified in our spectral line survey. Fig.~\ref{fig6} summarizes the derived fractional abundances from Table \ref{table3}
for all species except for $^{13}$C$^{15}$N (its 1-0 transition is contaminated by emission from an unknown molecular species), O$^{13}$C$^{34}$S (there is no lower limit for 
\emph{LOS}$-$0.11) and HCO$^+$, HCN, HNC and their $^{13}$C isotopologues, whose abundances are obtained
from their less abundant isotopologues.
 
It is remarkable, that two \emph{LOSs} separated by more than $\sim$120 pc in the GC show very similar 
abundances, within a factor of 2, of $\sim$80\% of the detected molecular species, including the most complex organic molecules like C$_2$H$_5$OH, C$_2$H$_5$CN and (CH$_3$)$_2$O (see Fig.~\ref{fig6}). 
This finding will be discussed in detail in section \ref{galactic_center}.

\subsubsection{Kinetic temperatures}\label{Kinetic_temperatures}

Symmetric rotors are usually used to estimate the kinetic temperature, T$_{kin}$, of the molecular clouds because the radiative transitions between levels of ladders with $\Delta$K$\neq$0 are forbidden \citep{turner}, and their excitation is dominated by collisions with H$_2$. 
In our survey we have detected transitions from CH$_3$CN, $^{13}$CH$_3$CN, CH$_3$$^{13}$CN, CH$_3$CCH and CH$_3$$^{13}$CCH. We have derived the T$_{rot}$ for both \emph{LOSs} from the RDs of the  6$_K$-5$_K$ (K=0, 1, 2, 3 and 4) transitions of 
CH$_3$CN and the 6$_K$-5$_K$ (K=0, 1, 2 and 3) transitions of CH$_3$CCH. In addition for \emph{LOS}+0.693 we have also used the 6$_K$-5$_K$ (K=0, 1 and 3) transitions of $^{13}$CH$_3$CN (the K=2 line is not used because it is blended with the K=0 and 1 lines). 

As expected, the symmetric rotors CH$_3$CN, CH$_3$CCH and $^{13}$CH$_3$CN show the highest T$_{rot}$ of $\approx$55-73 K for both GC sources (see Table \ref{table3}), indicating that the T$_{kin}$ must be larger than 73 K. 
For a T$_{kin}$ of 100 K, statistical equilibrium calculations reveal that T$_{rot}$ derived from CH$_3$CN transitions 
with $\Delta$K$\neq$0 (K$<$4) approach the T$_{kin}$ only at H$_2$ densities of $\sim$10$^6$ cm$^{-3}$. 
For H$_2$ densities of $\la$10$^5$ cm$^{-3}$ and a T$_{kin}$ of 100 K, these CH$_3$CN transitions show T$_{rot}$ of $\la$70 K \citep{Churchwell}.
Our results are consistent with the mean T$_{kin}$ of $\approx$100 K derived by \citet{Gustem}, \citet{Hutteme} and \citet{nemesi01} toward molecular clouds distributed over the central 500 pc of the GC.
Thus, for the low H$_2$ densities ($\sim$10$^4$ cm$^{-3}$, see Sec. \ref{densities}) found in both GC sources we will consider in the following discussions a T$_{kin}$ of $\approx$100 K. 



\begin{figure*}
\subfloat[CCS, T$_{rot}$=7.3 (0.5) K]{\includegraphics[width=63mm]{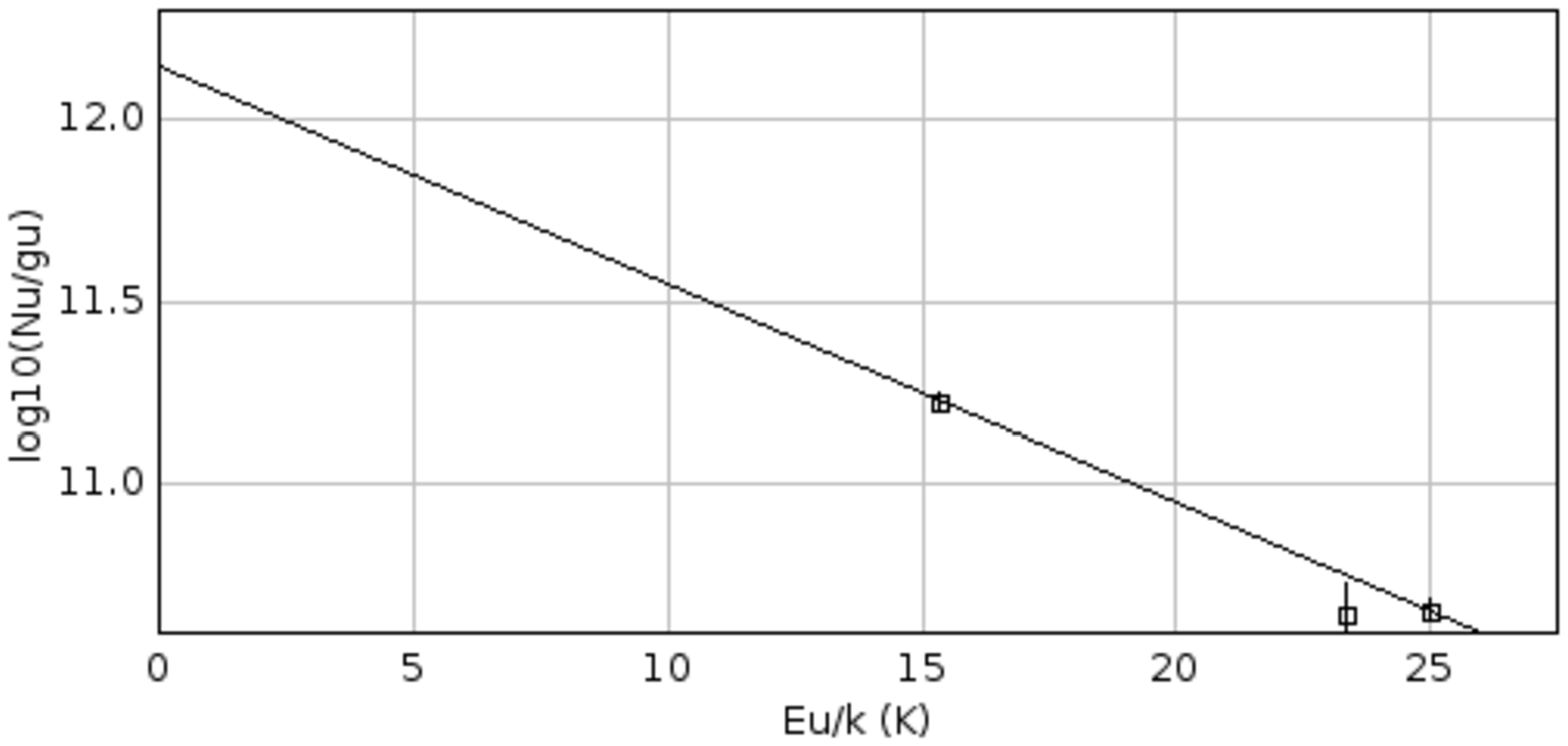}}\hfil
\subfloat[CCS, T$_{rot}$=8.6 (2.3) K]{\includegraphics[width=63mm]{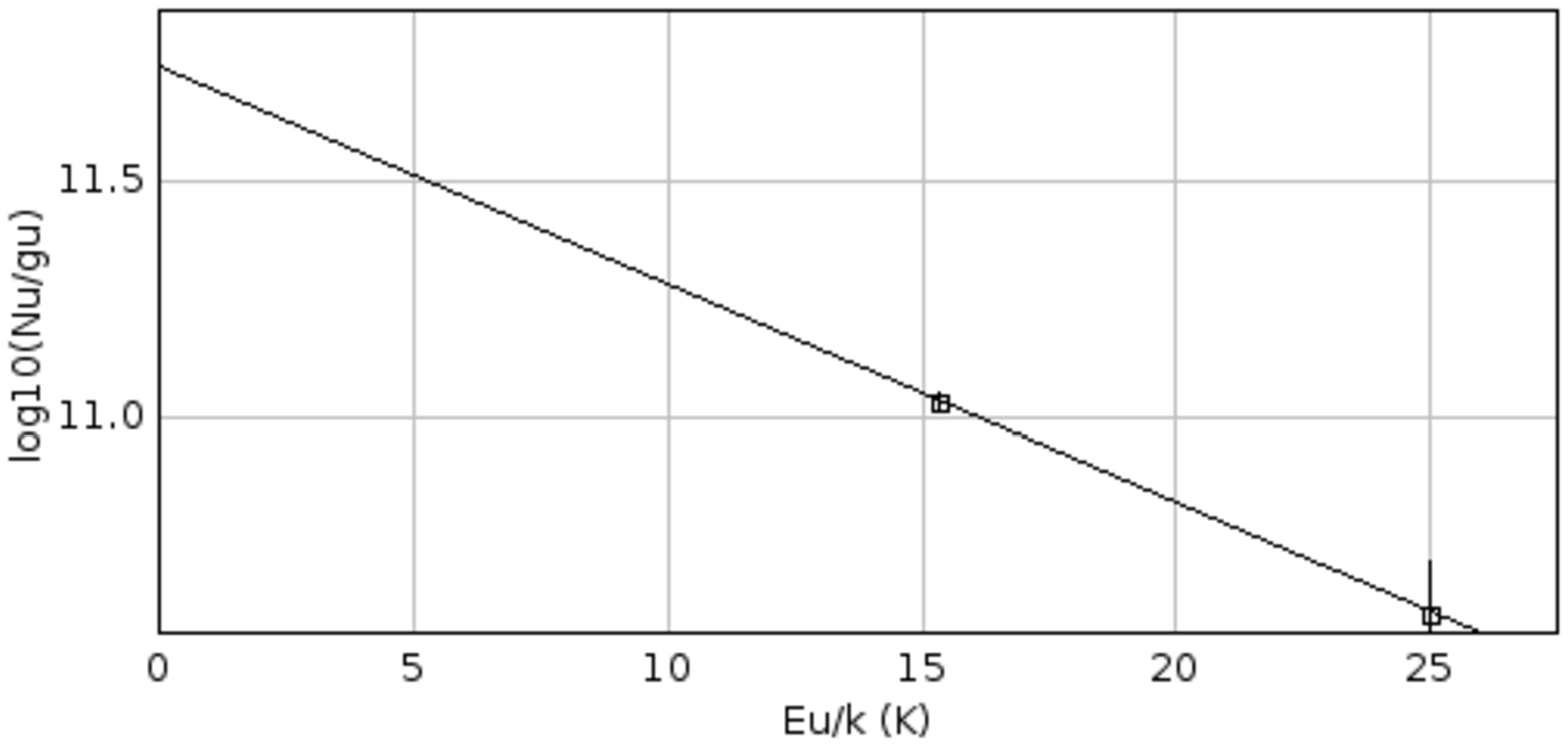}}

\subfloat[OCS, T$_{rot}$=17.6 (1.1) K]{\includegraphics[width=63mm]{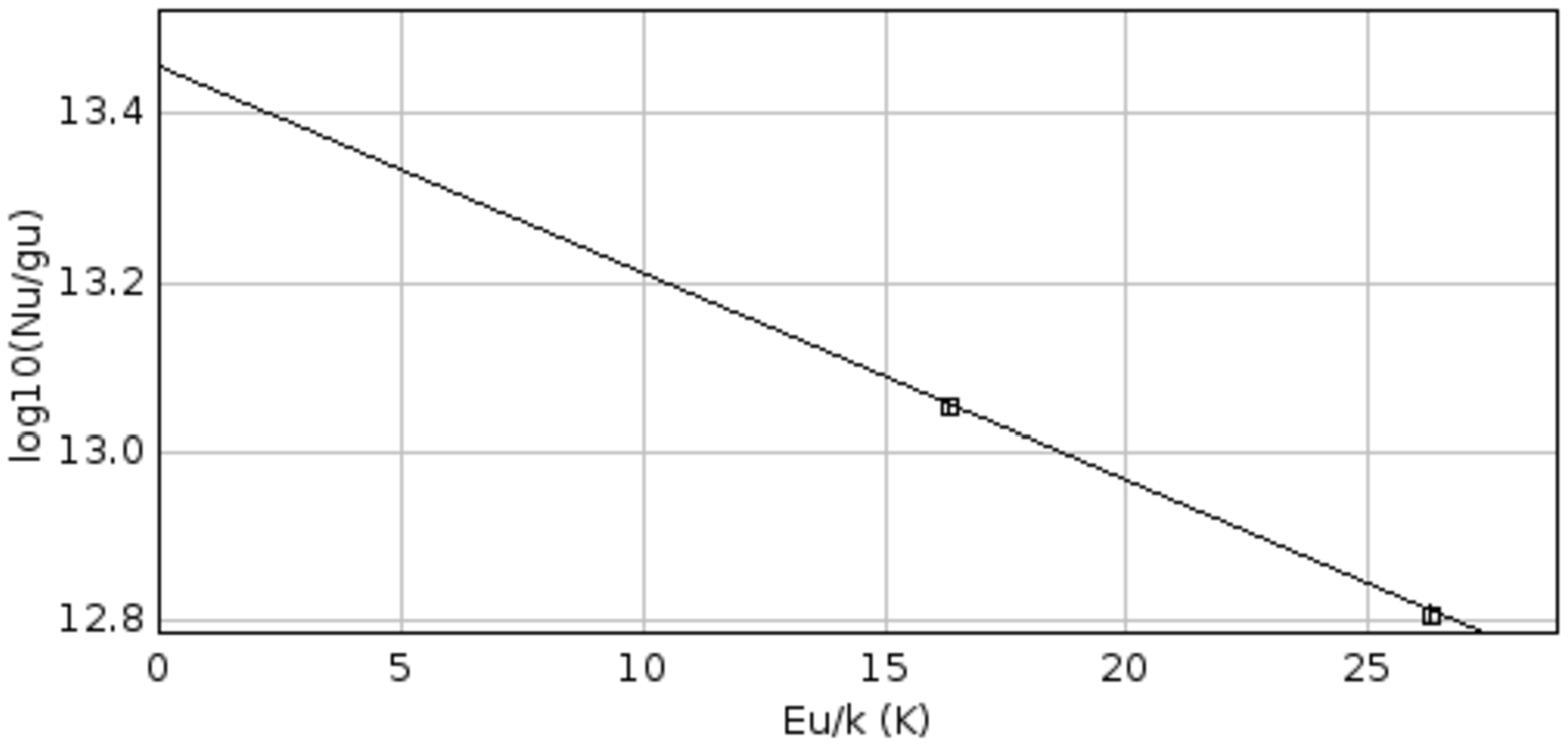}}\hfil
\subfloat[OCS, T$_{rot}$=15.3 (1.4) K]{\includegraphics[width=63mm]{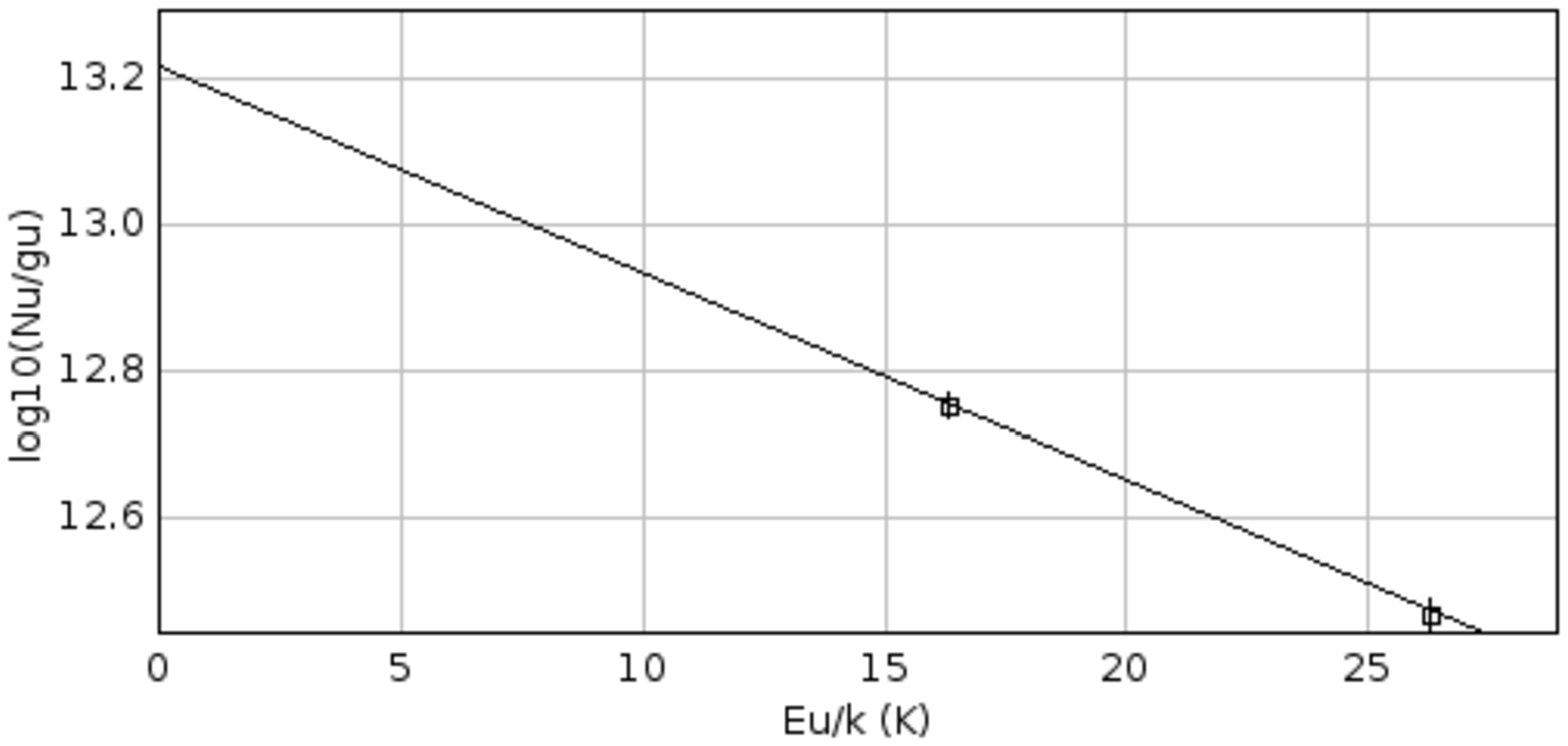}}

\subfloat[HNCO, T$_{rot}$=11.2 (0.2) K]{\includegraphics[width=63mm]{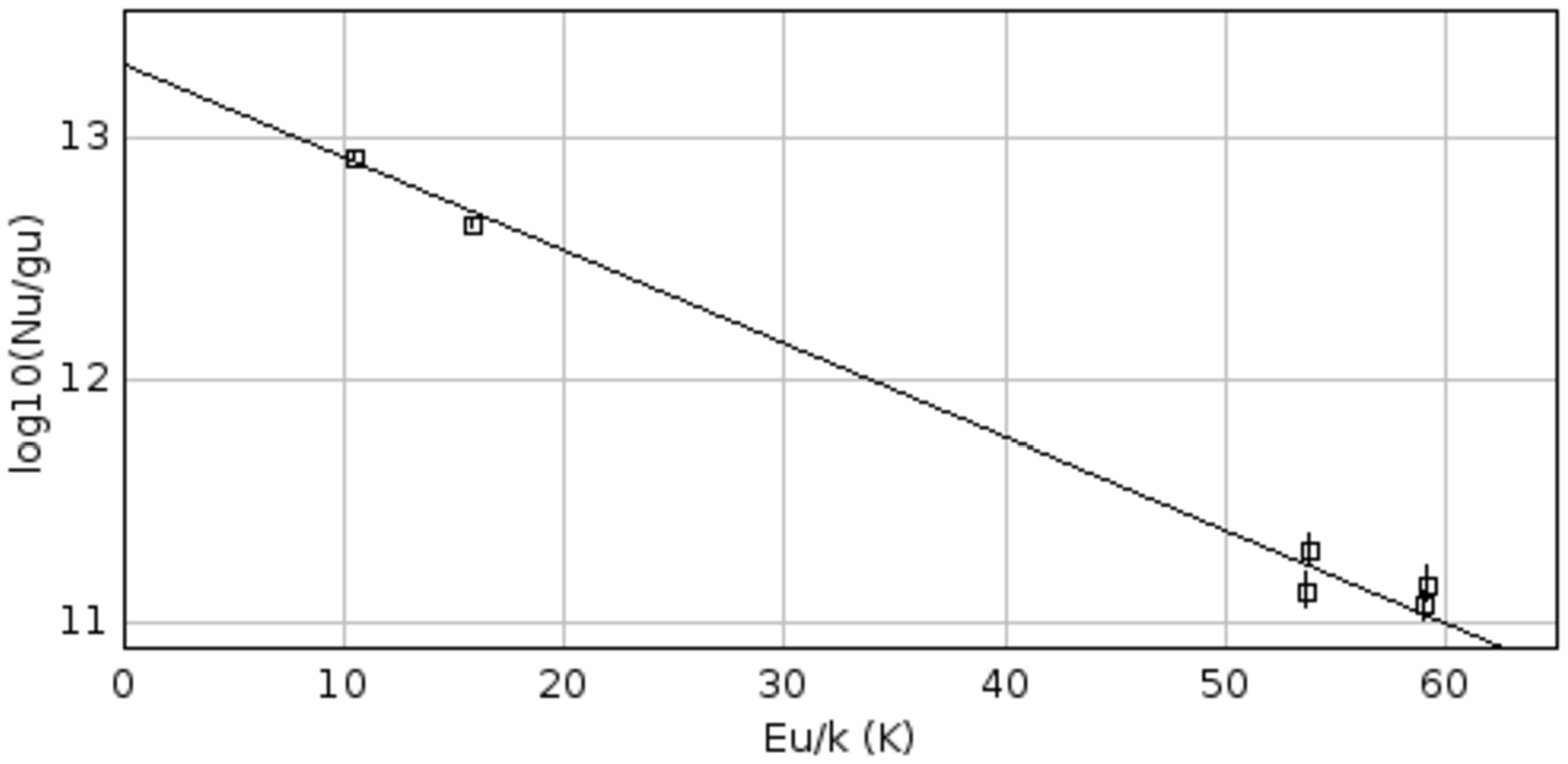}}\hfil
\subfloat[HNCO, T$_{rot}$=11.5 (0.5) K]{\includegraphics[width=63mm]{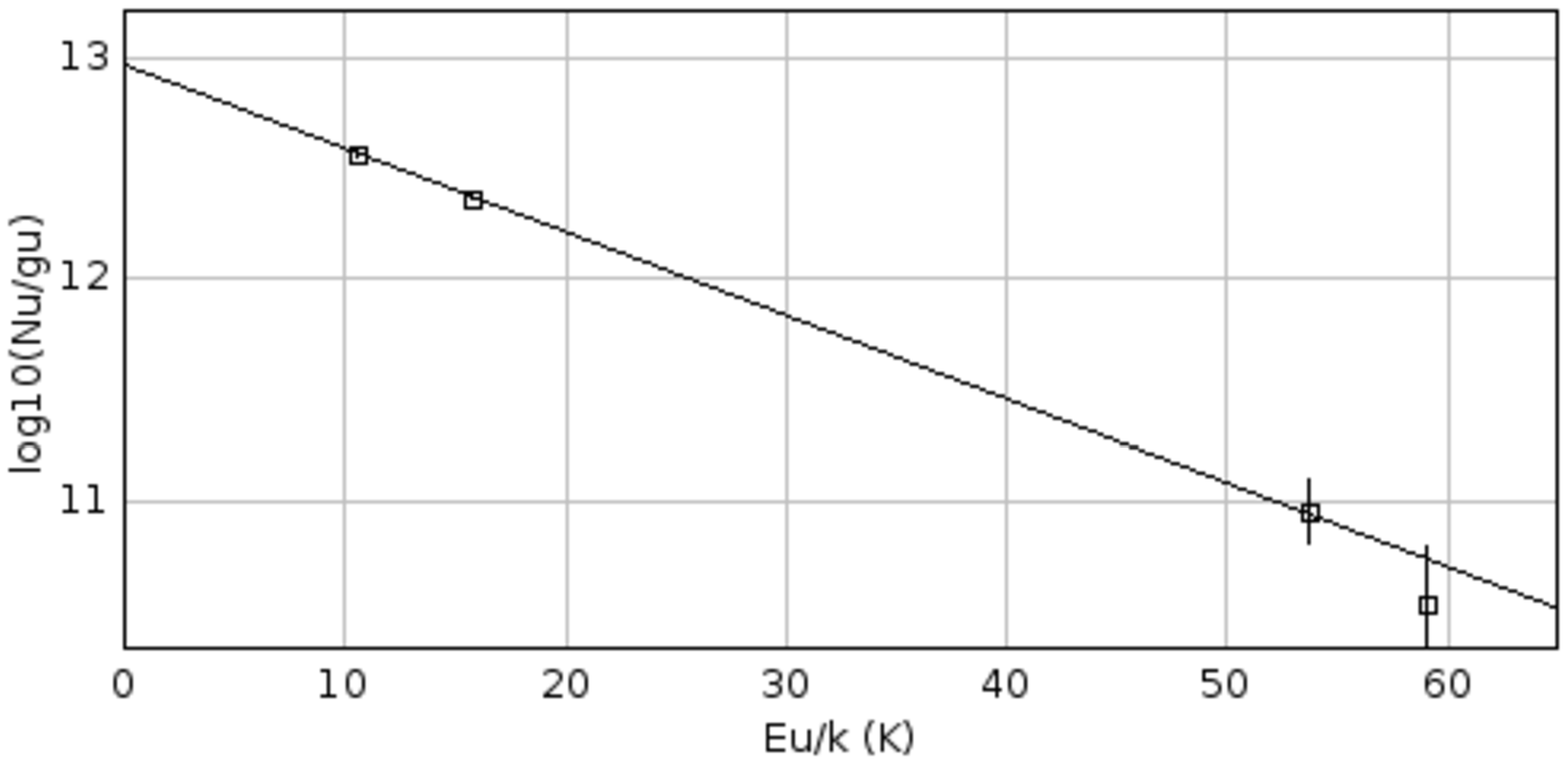}}

\subfloat[CH$_2$CO, T$_{rot}$=29.4 (5.1) K]{\includegraphics[width=63mm]{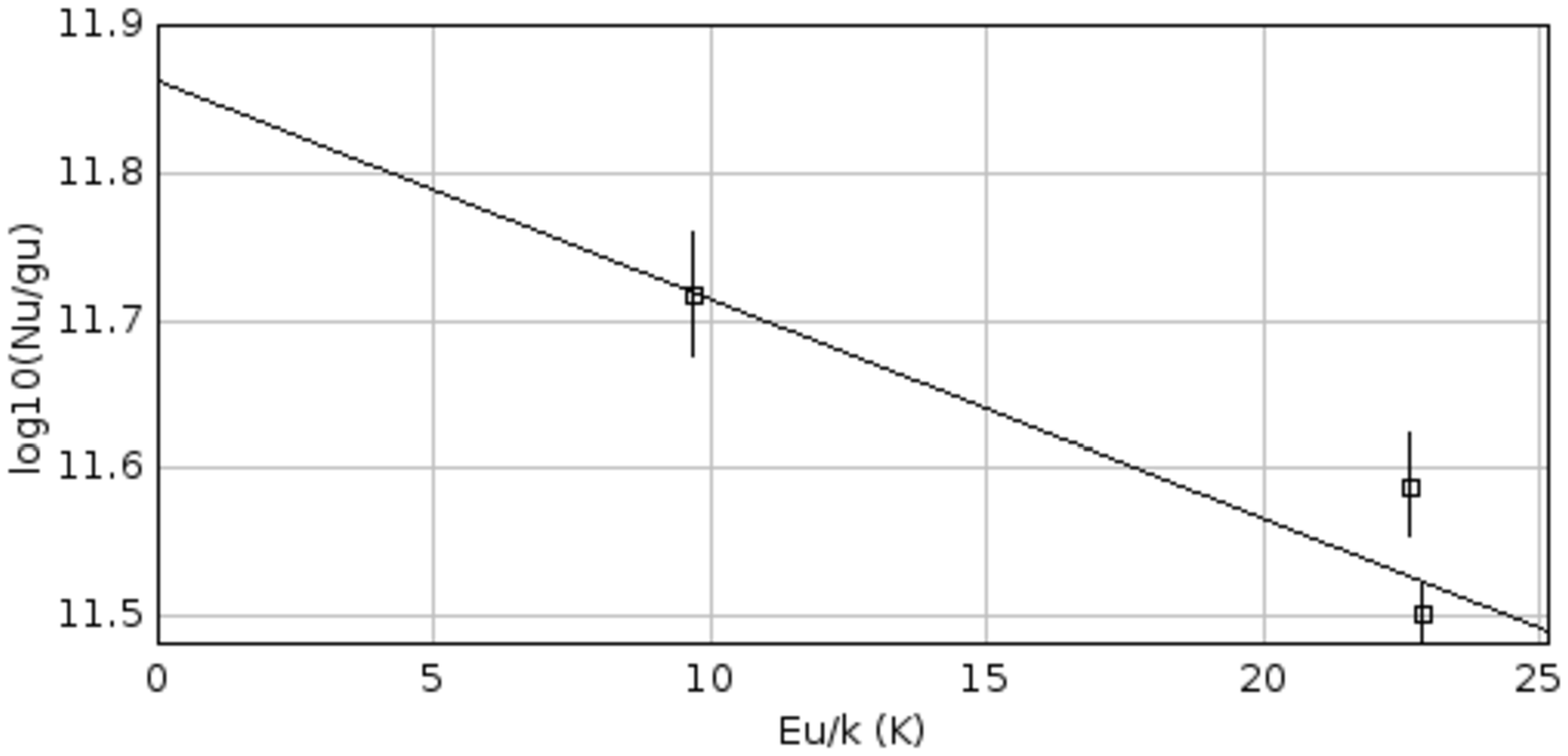}}\hfil
\subfloat[CH$_2$CO, T$_{rot}$=30.6 (14.8) K]{\includegraphics[width=63mm]{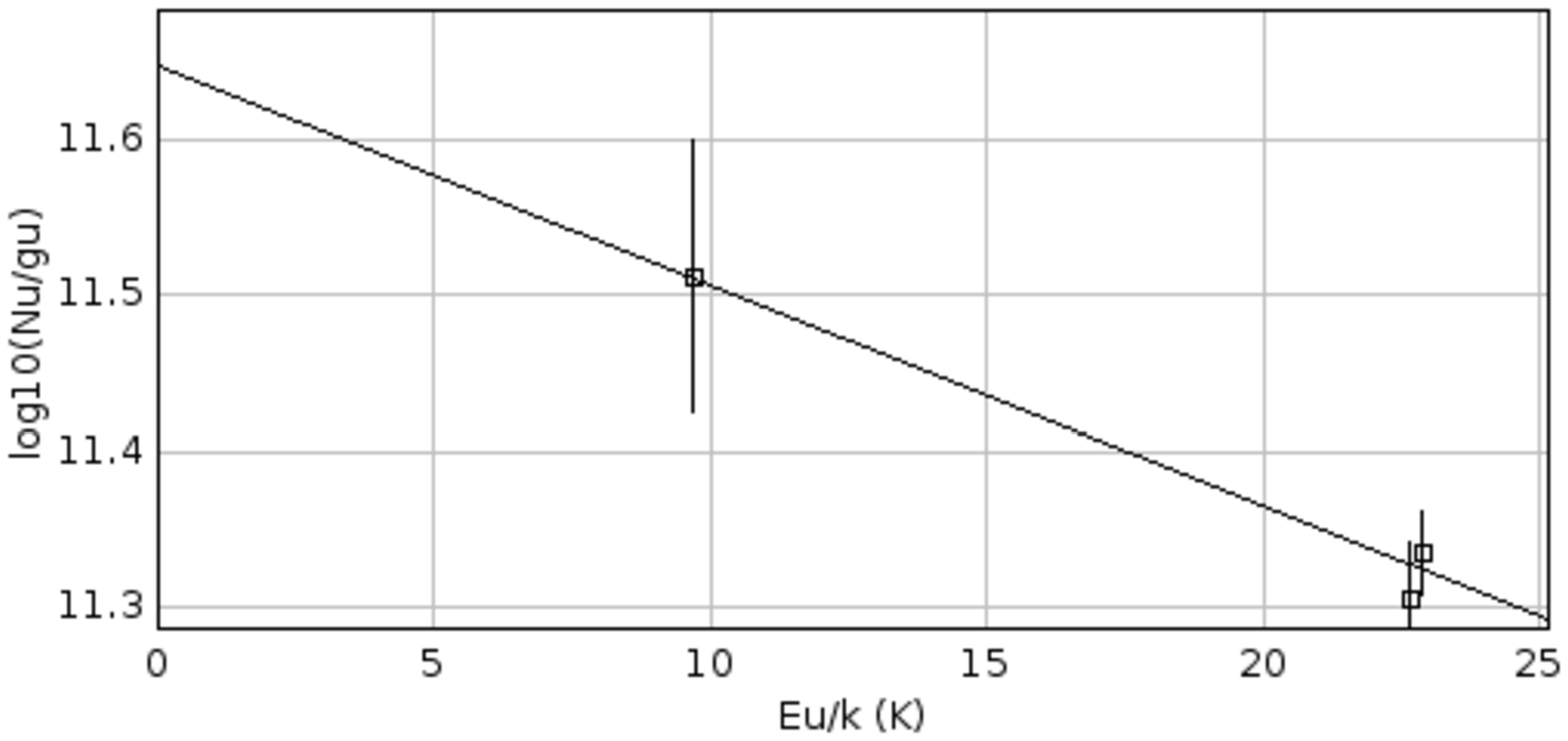}}

\subfloat[HC$_3$N, T$_{rot}$=15.4 (0.6) K]{\includegraphics[width=63mm]{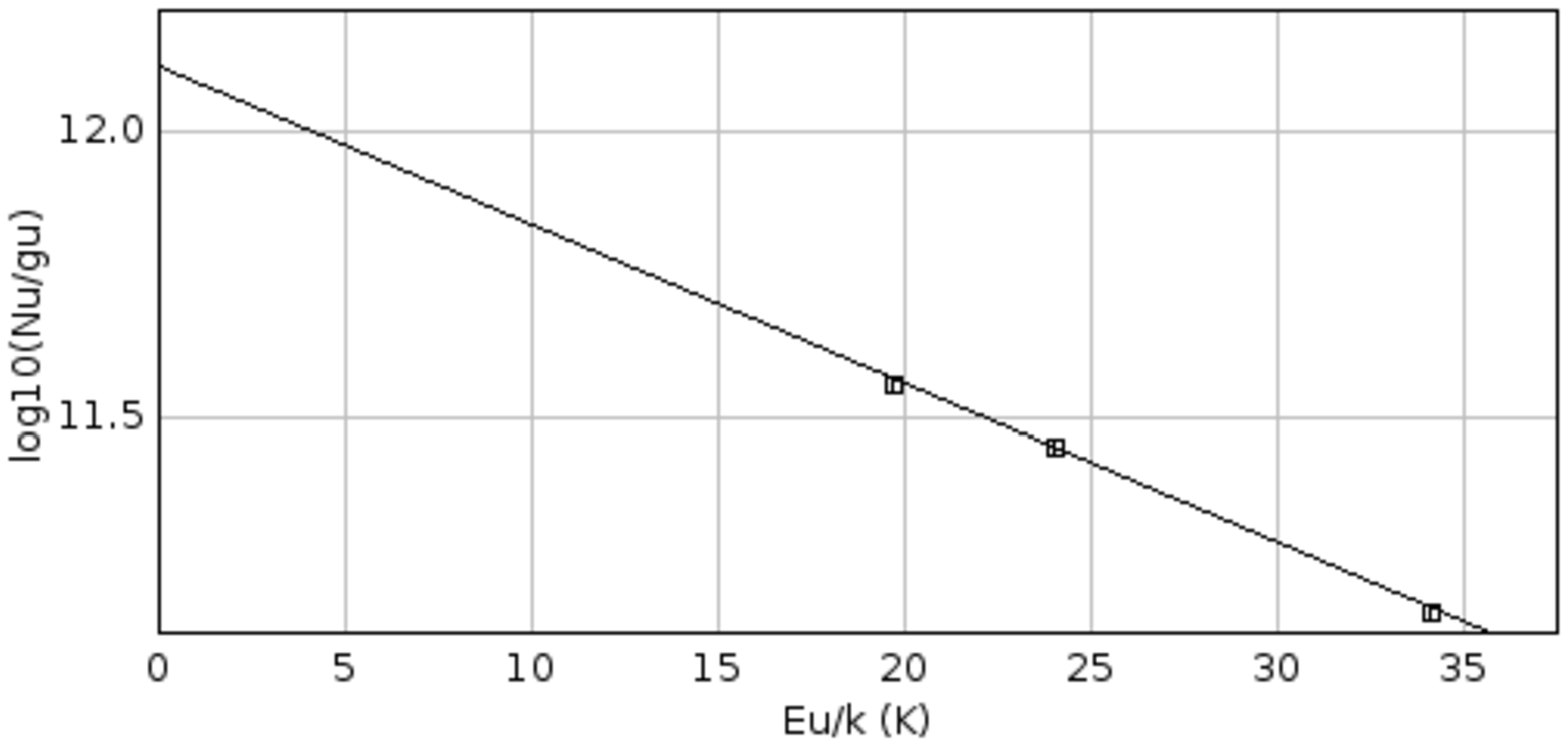}}\hfil
\subfloat[HC$_3$N, T$_{rot}$=11.2 (0.2) K]{\includegraphics[width=63mm]{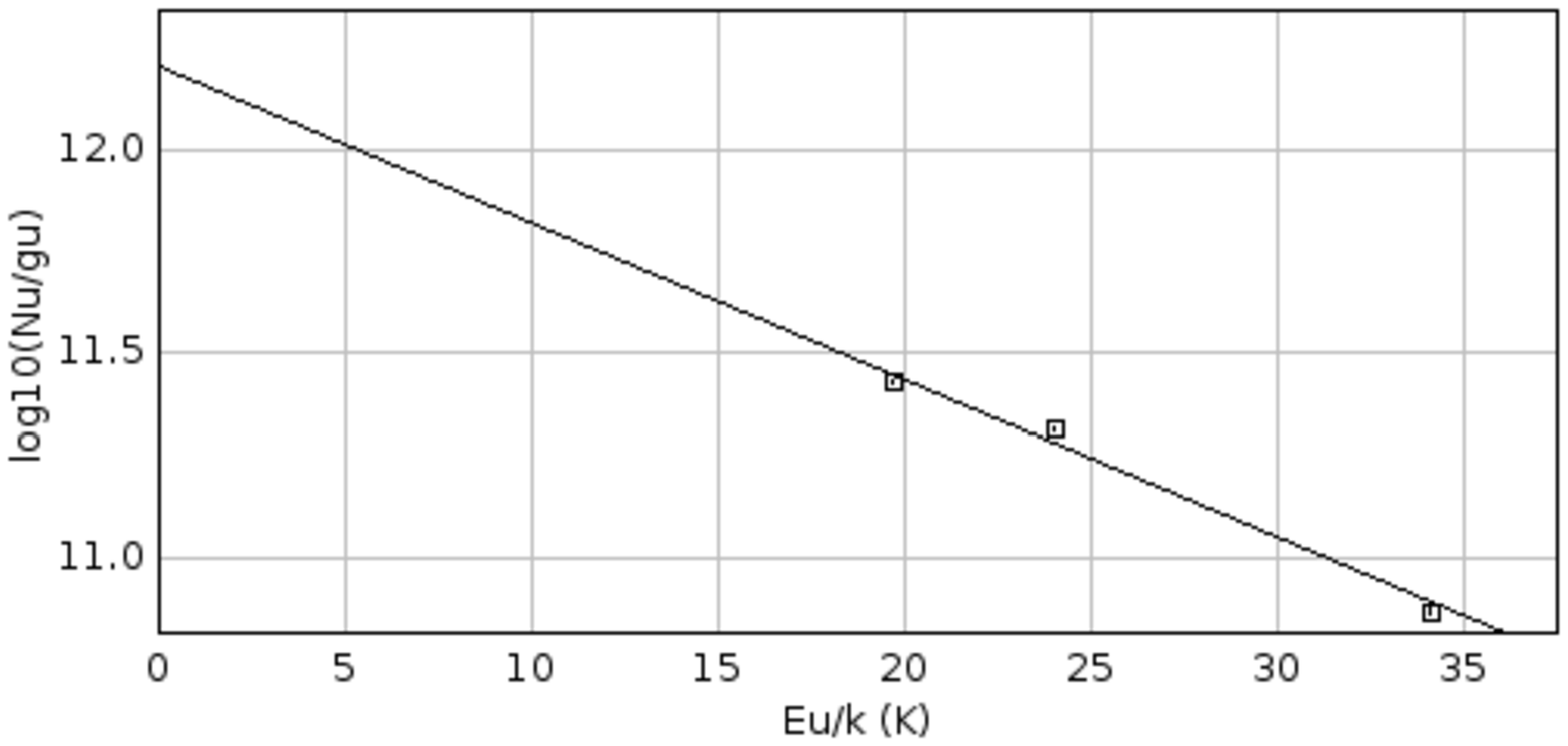}}

\subfloat[(CH$_3$)$_2$O, T$_{rot}$=20.3 (1.2) K]{\includegraphics[width=63mm]{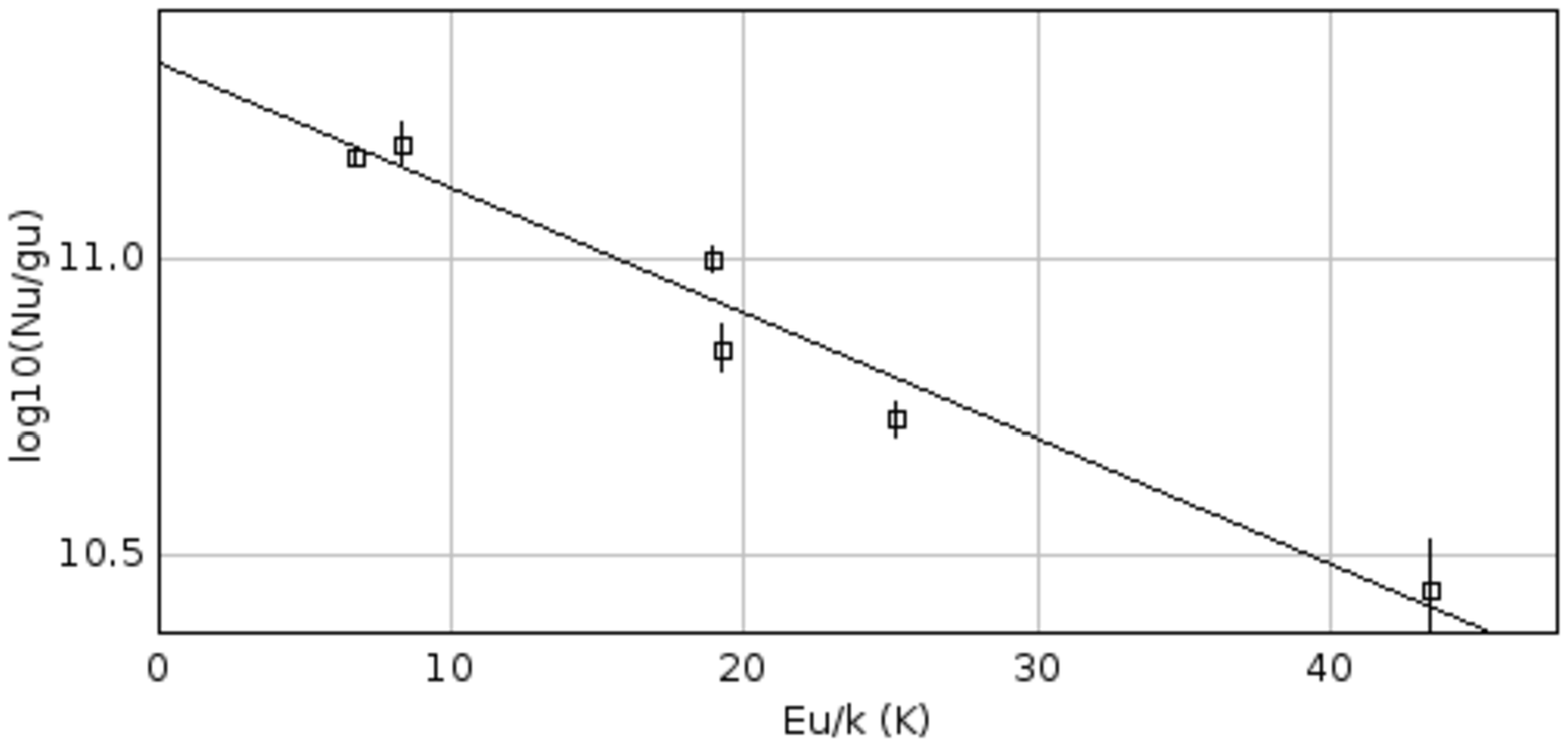}}\hfil
\subfloat[(CH$_3$)$_2$O, T$_{rot}$=15.8 (7.6) K]{\includegraphics[width=63mm]{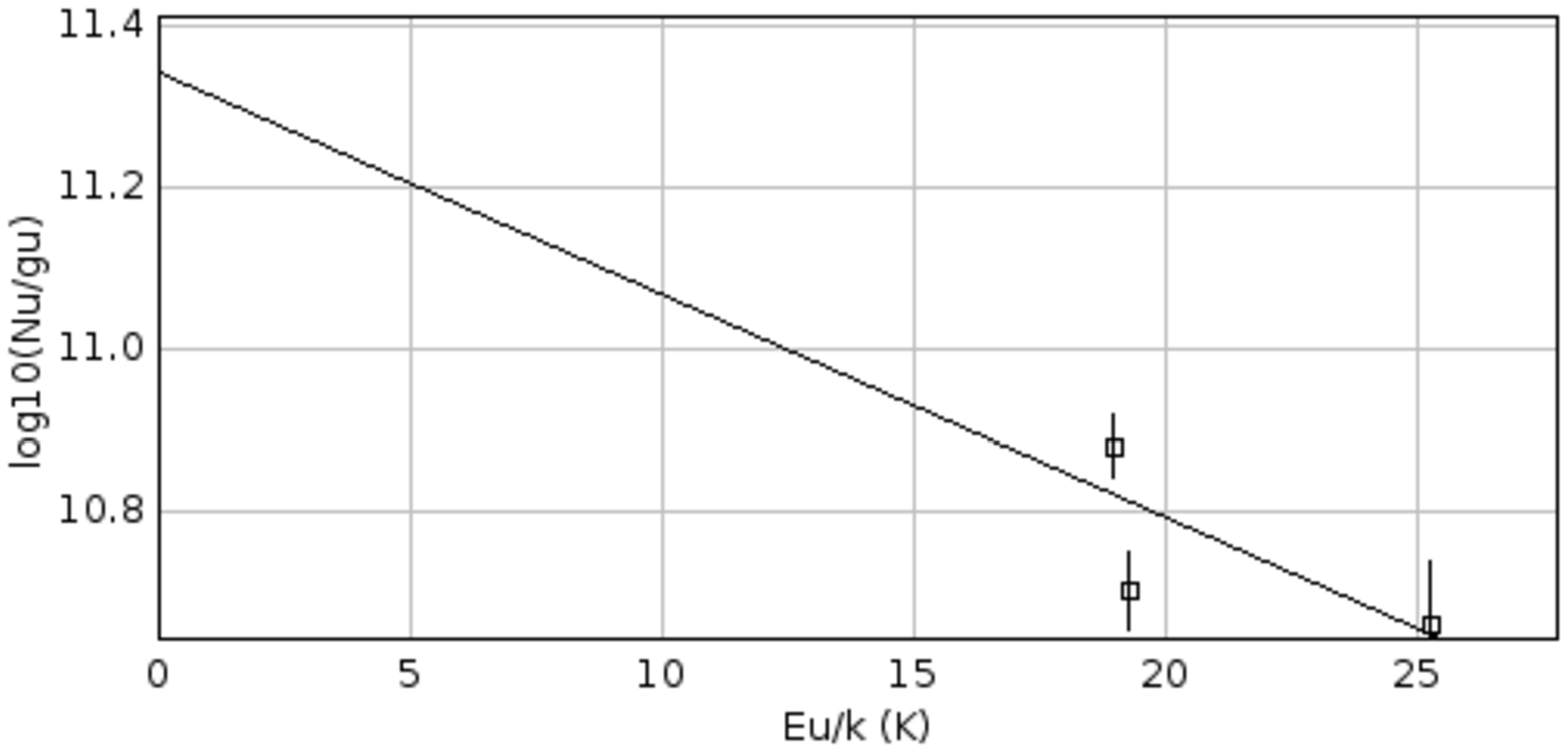}}
\caption{Rotational diagrams of some molecules for \emph{LOS}+0.693 (left panels) and \emph{LOS}$-$0.11 (right panels).}
\label{fig5}
\end{figure*}

\begin{figure*}

\subfloat[NH$_2$CN, T$_{rot}$=47.6 (3.1) K]{\includegraphics[width=63mm]{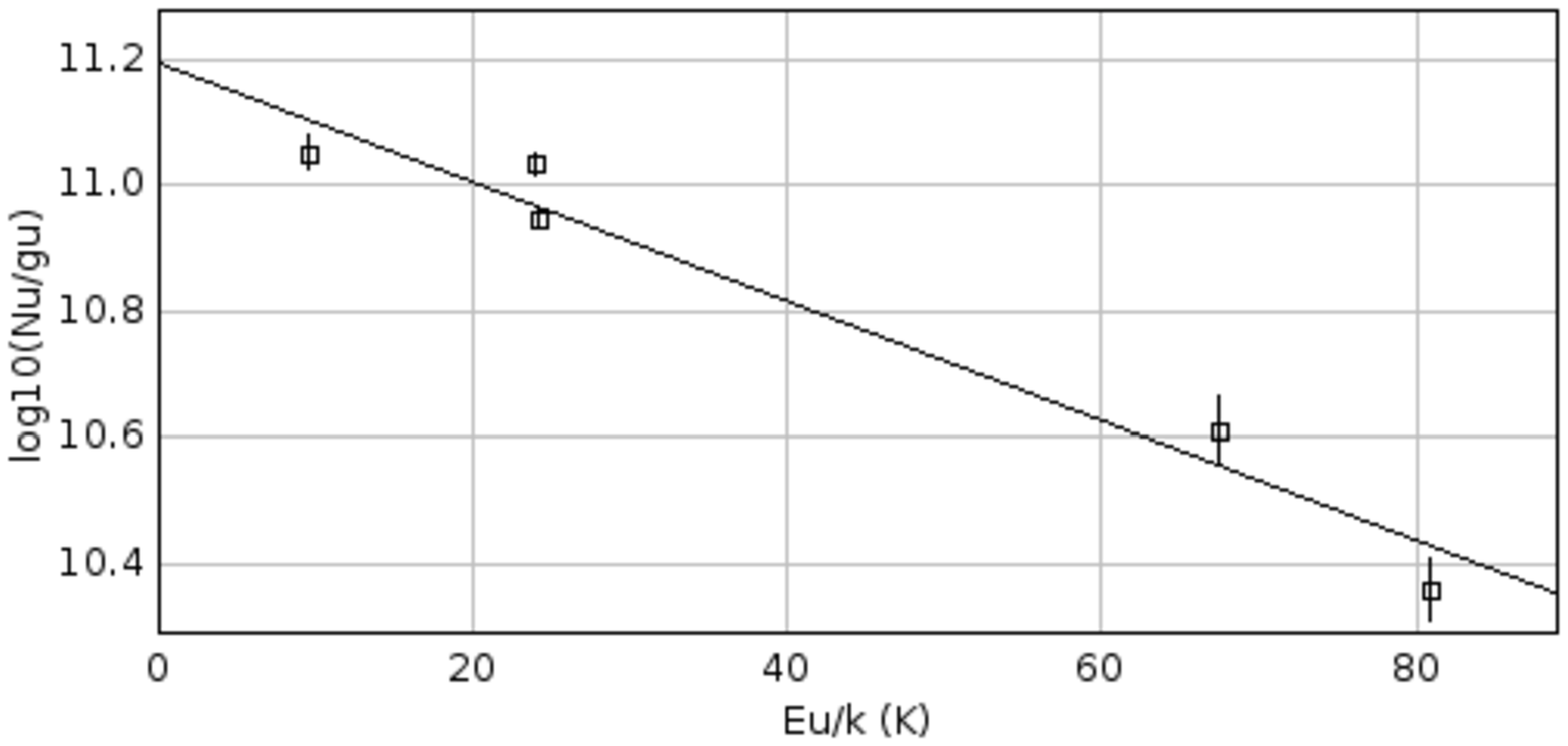}}\hfil
\subfloat[NH$_2$CN, T$_{rot}$=53.6 (10.3) K]{\includegraphics[width=63mm]{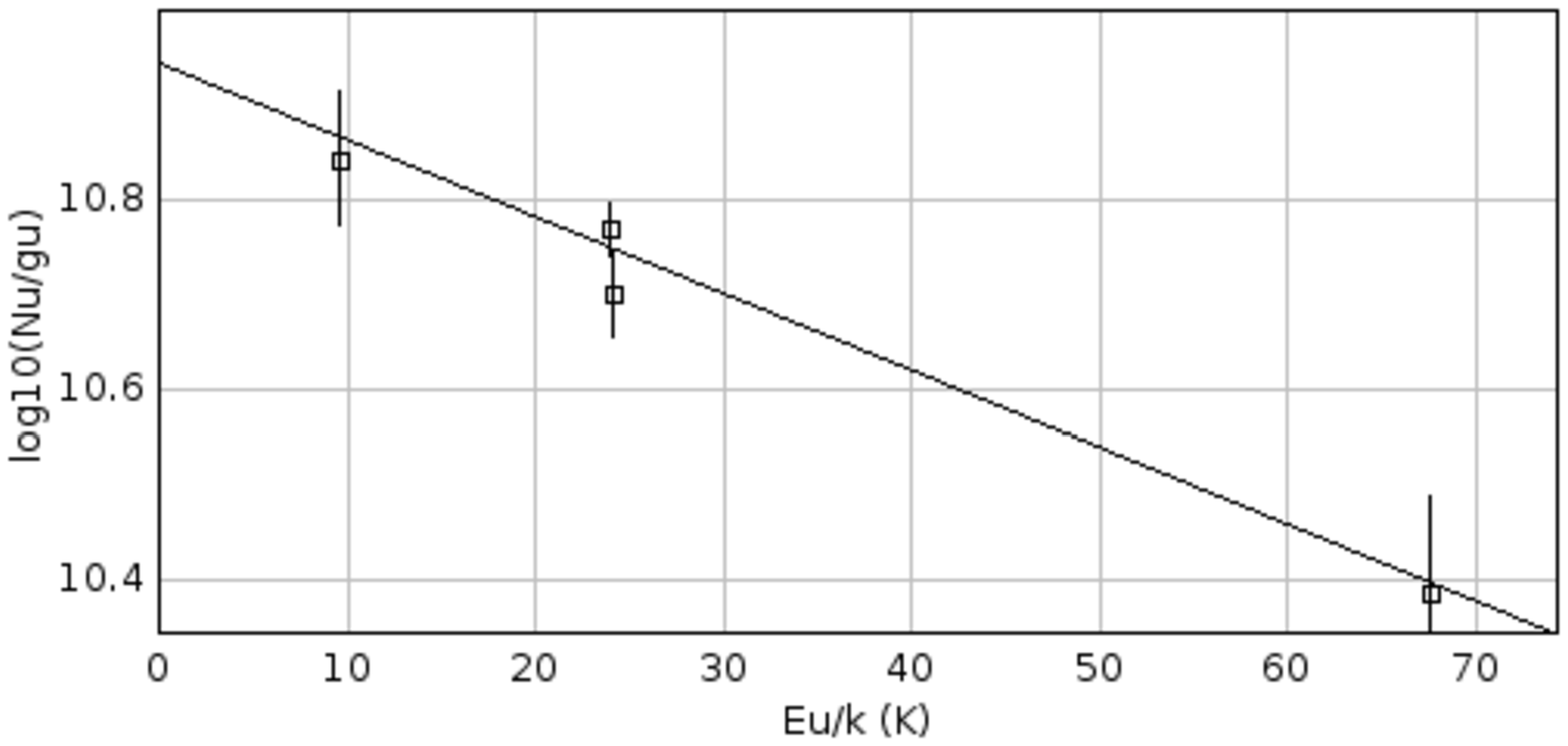}}

\subfloat[CH$_3$CN, T$_{rot}$=65.2 (6.5) K]{\includegraphics[width=63mm]{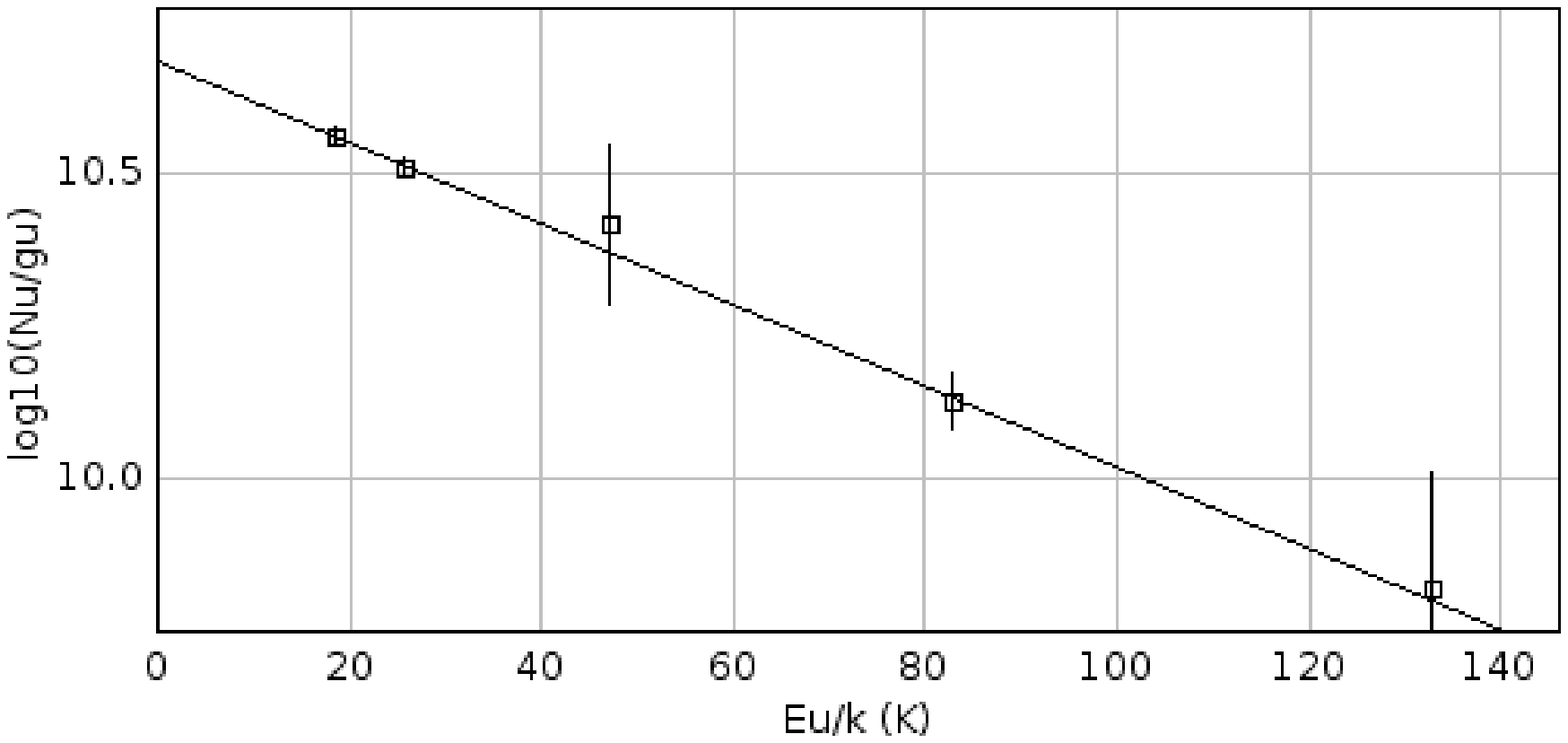}}\hfil
\subfloat[CH$_3$CN, T$_{rot}$=63.5 (8.7) K]{\includegraphics[width=63mm]{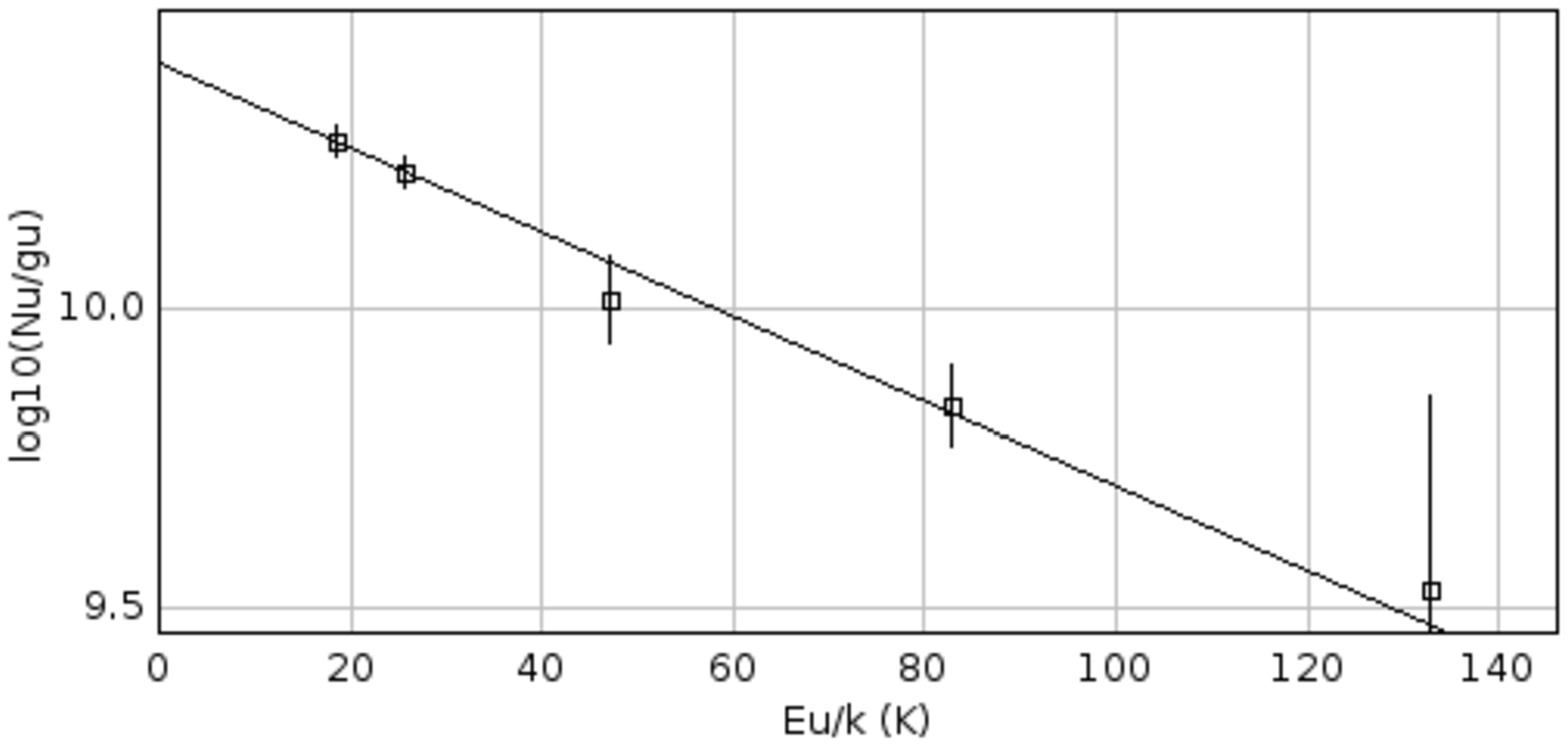}}

\subfloat[CH$_3$CCH, T$_{rot}$=61.1 (10.9) K]{\includegraphics[width=63mm]{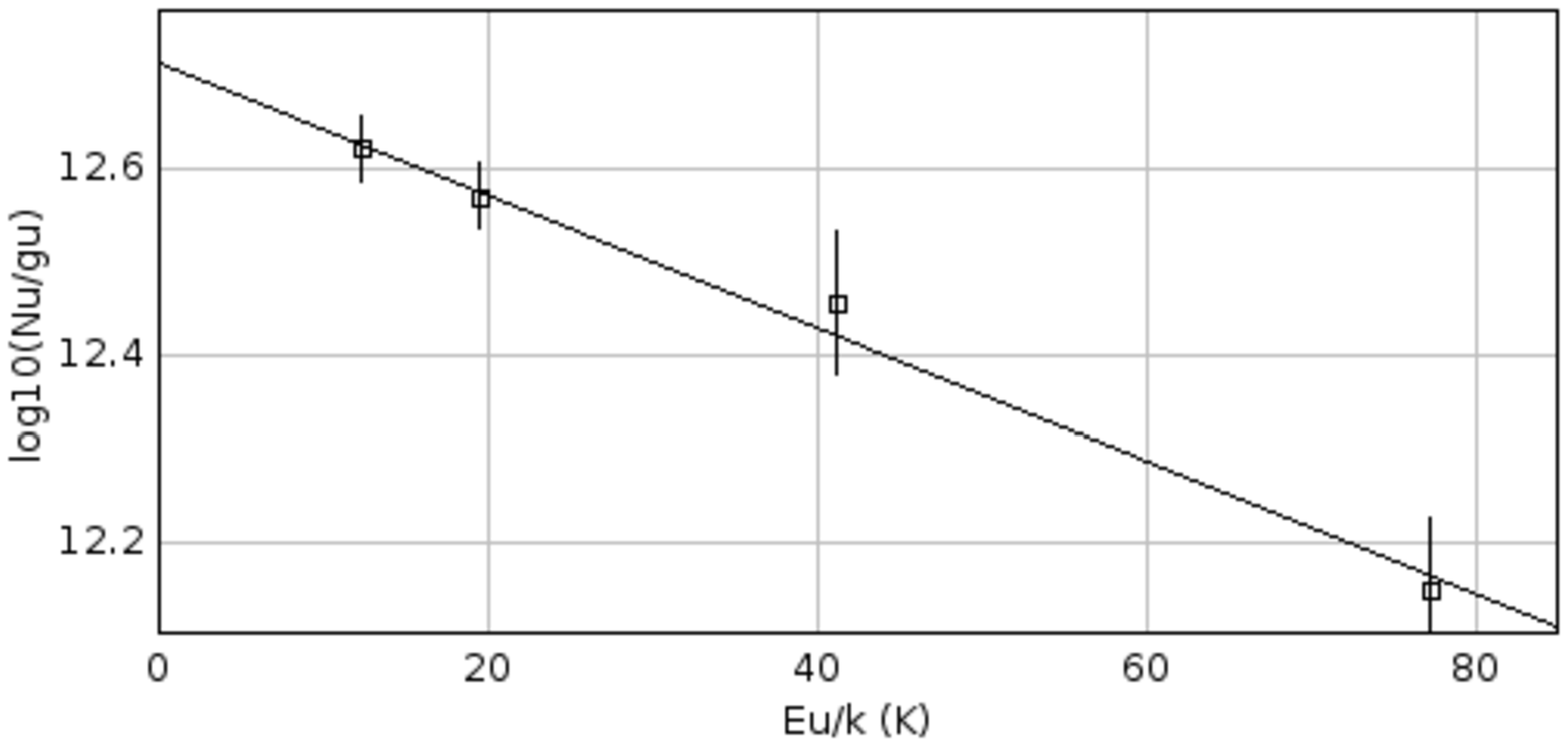}}\hfil
\subfloat[CH$_3$CCH, T$_{rot}$=54.8 (5.5) K]{\includegraphics[width=63mm]{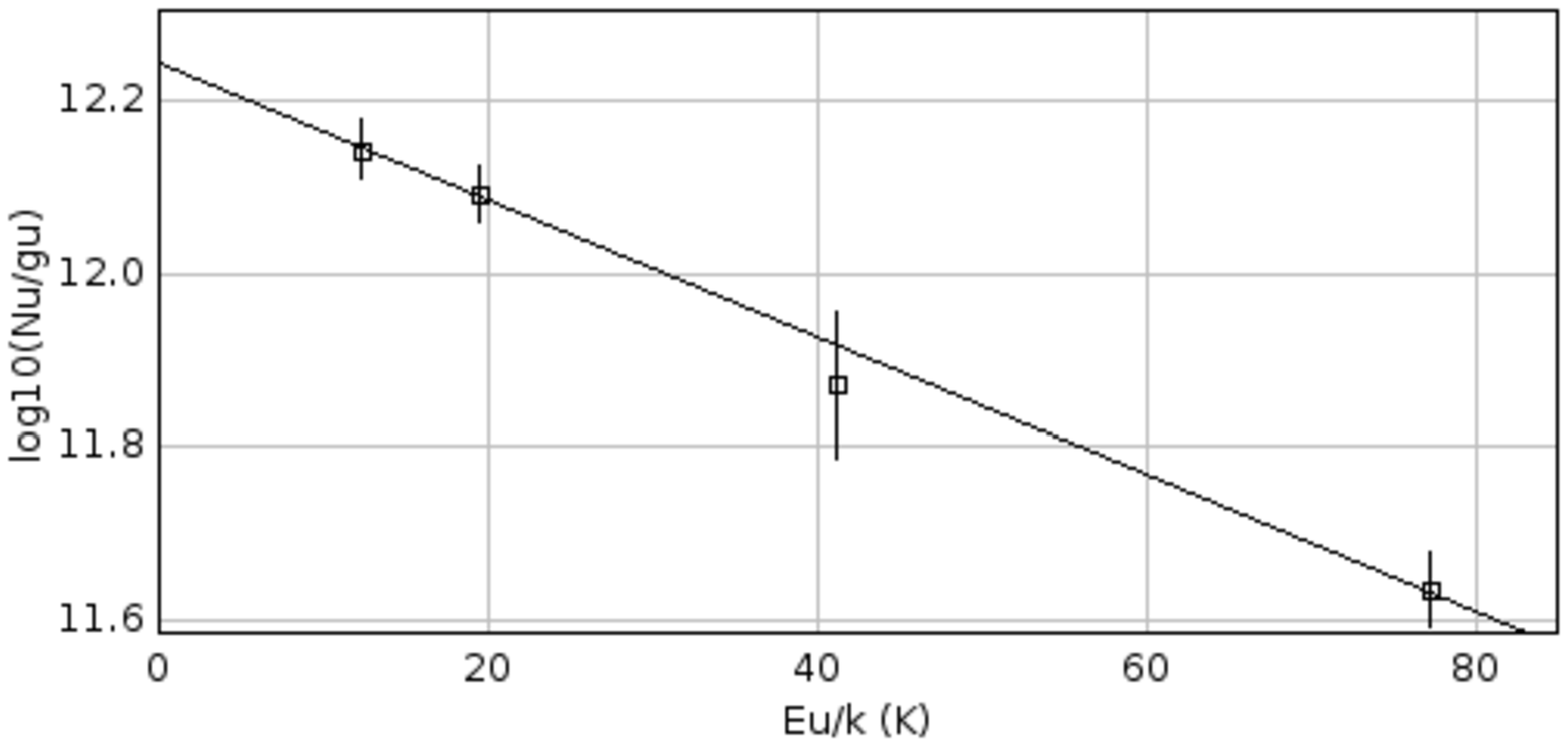}}

\subfloat[HC$_5$N, T$_{rot}$=32.1 (4.7) K]{\includegraphics[width=63mm]{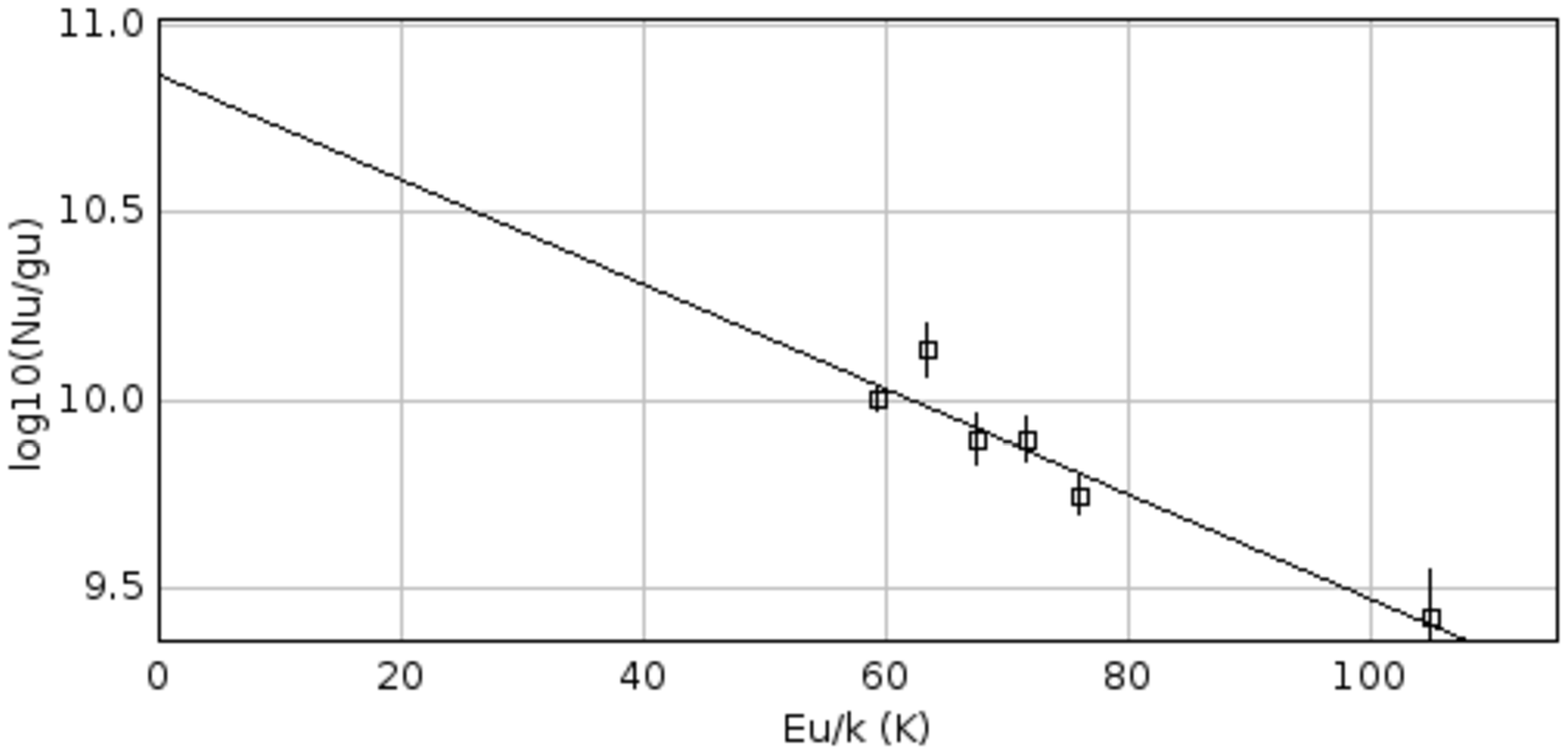}}\hfil
\subfloat[HC$_5$N, T$_{rot}$=44.5 (19.1) K]{\includegraphics[width=63mm]{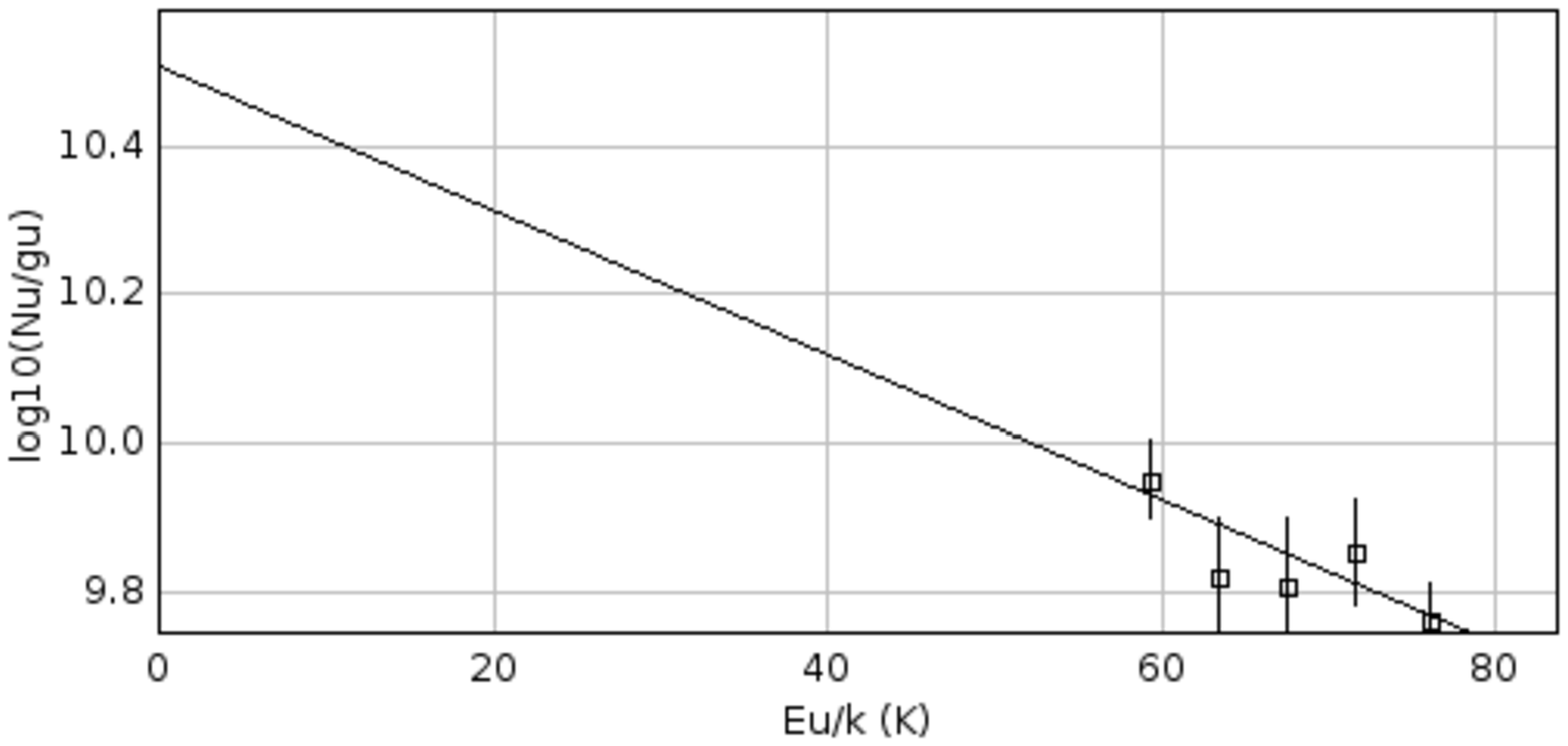}}

\subfloat[CH$_3$CHO-A, T$_{rot}$=5.4 (0.4) K]{\includegraphics[width=63mm]{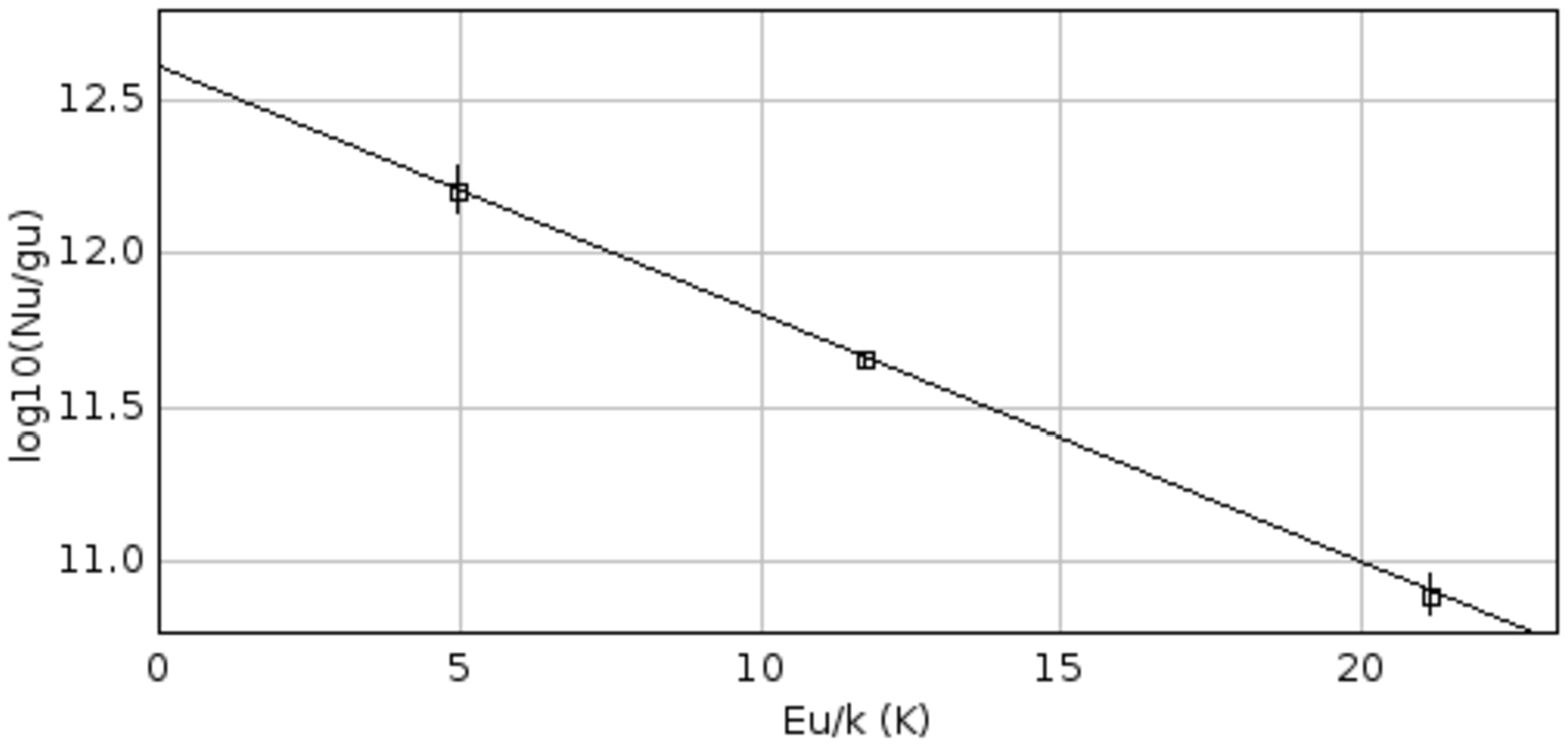}}\hfil
\subfloat[CH$_3$CHO-A, T$_{rot}$=4.5 (1.0) K]{\includegraphics[width=63mm]{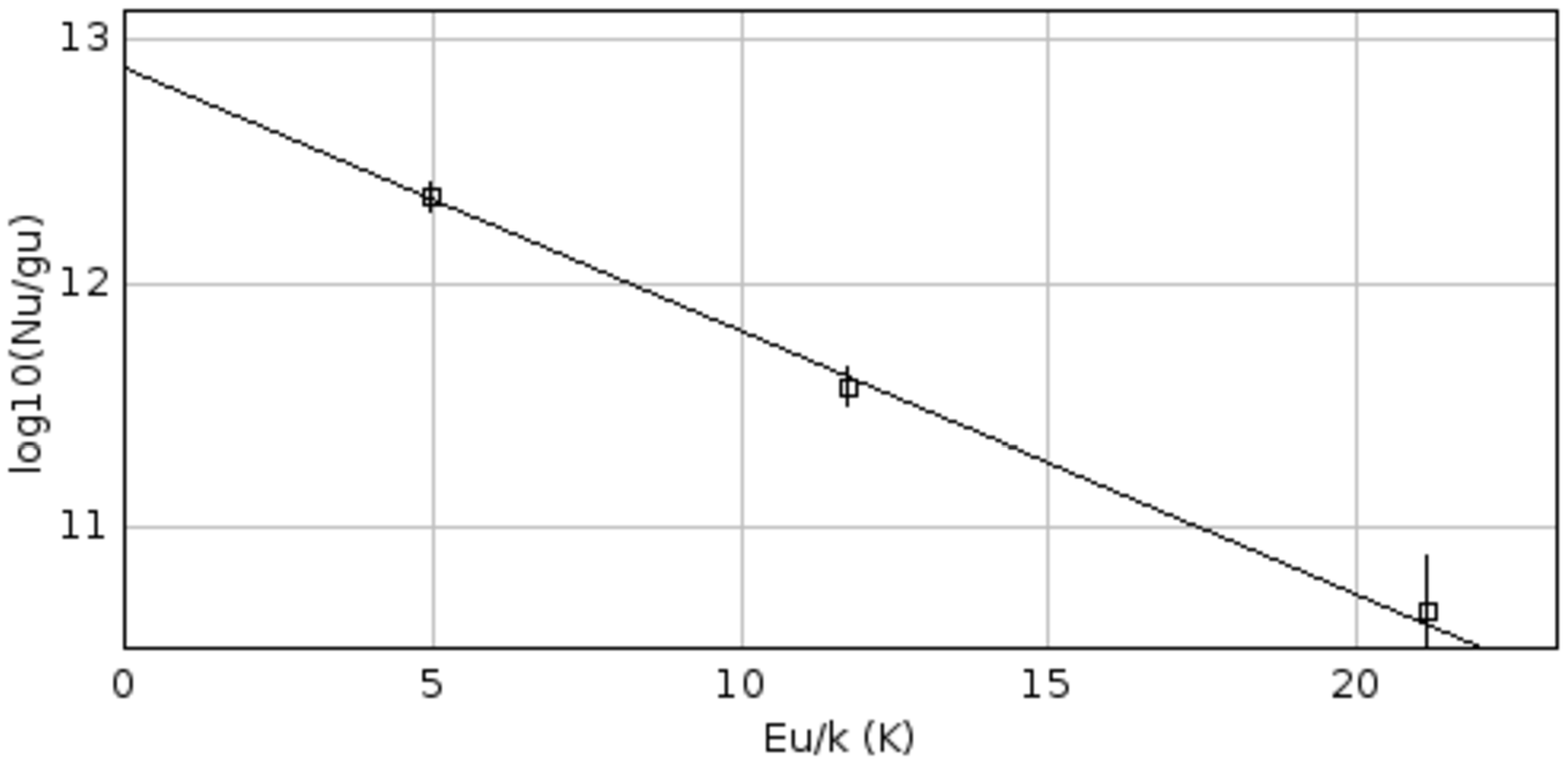}}

\subfloat[HOCO$^+$, T$_{rot}$=7.5 (0.4) K]{\includegraphics[width=63mm]{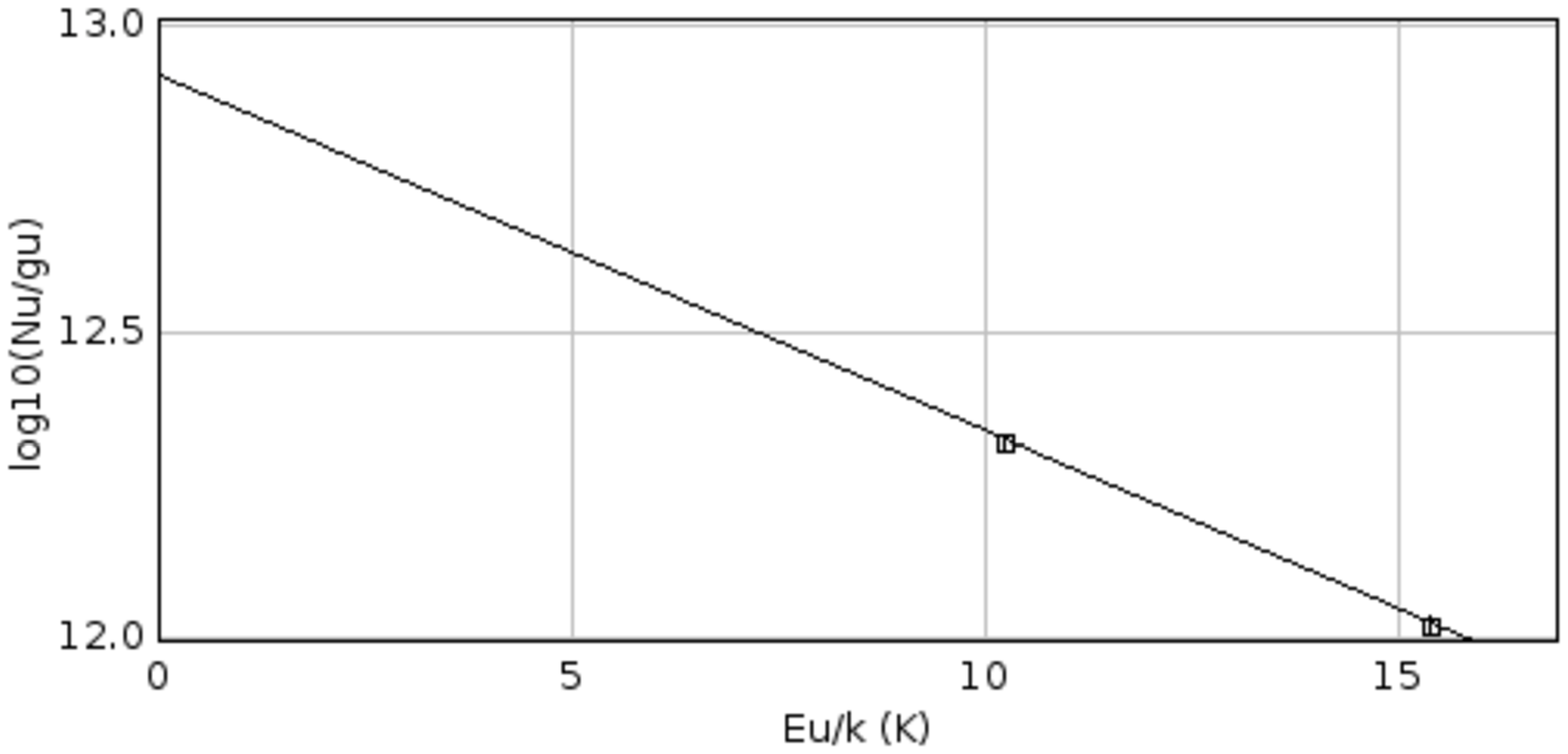}}\hfil
\subfloat[HOCO$^+$, T$_{rot}$=6.5 (0.8) K]{\includegraphics[width=63mm]{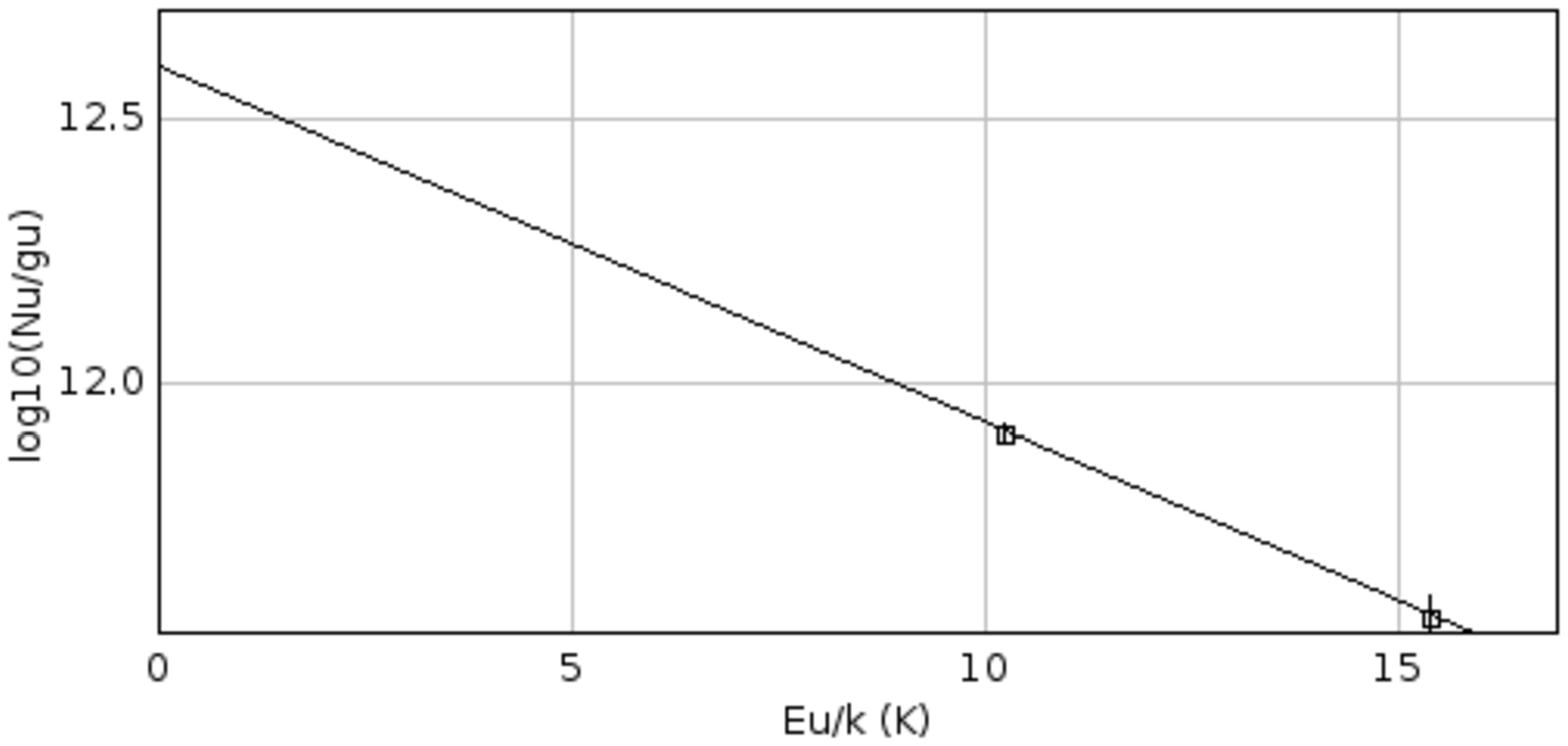}}





\contcaption{}
\end{figure*}


\begin{figure*}
\includegraphics[trim=2.5cm 0 0 0, clip, width=180mm]{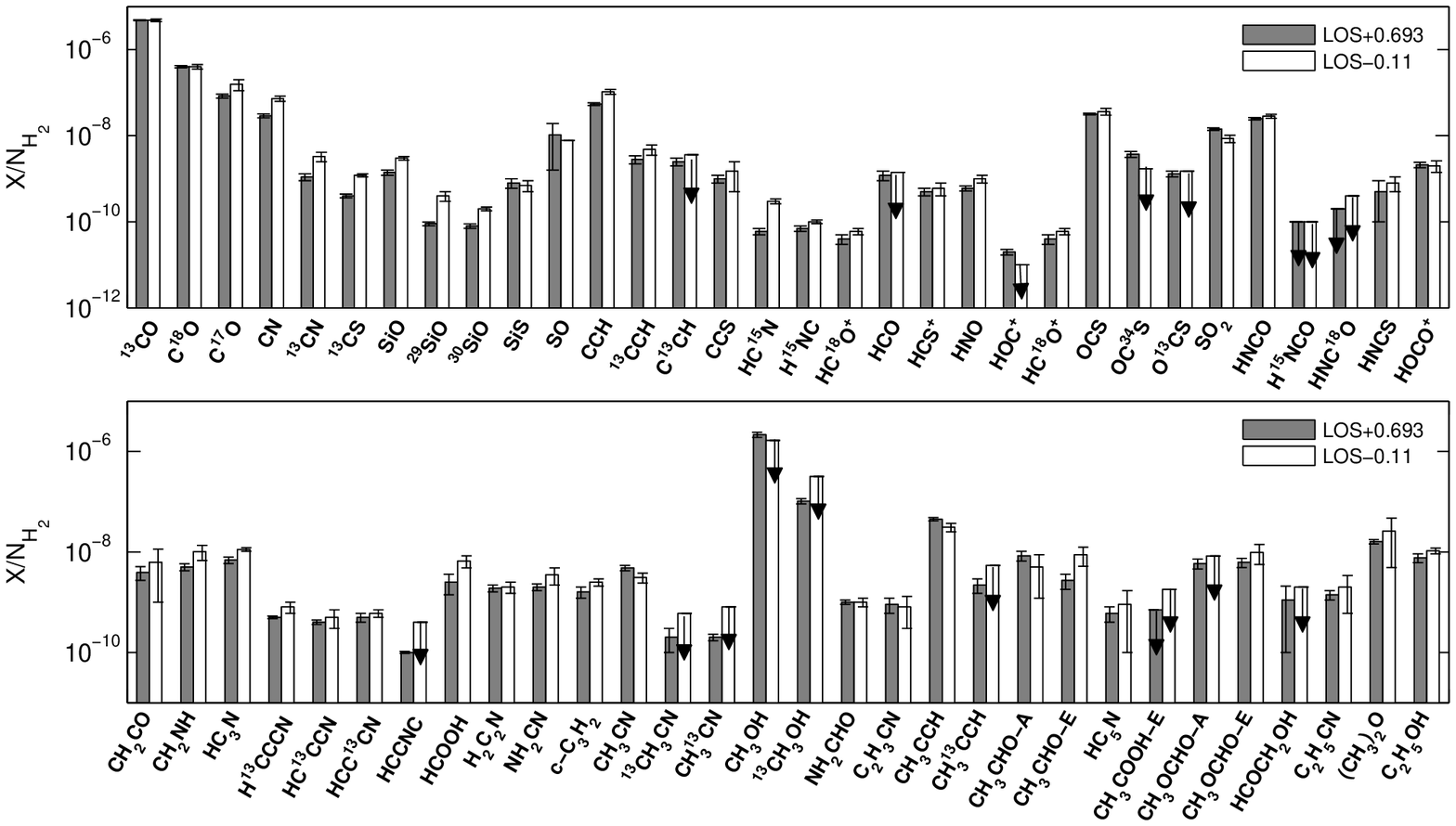}
\caption{Abundances relatives to H$_2$ for molecules detected in this survey. Arrows correspond to upper limits to the derived
abundances. CH$_3$OH lines are observed in \emph{LOS}$-$0.11, but we were only able to derive an upper limit to the
CH$_3$OH column density (see Sec. \ref{column_densities}).}
\label{fig6}
\end{figure*}

\subsubsection{H$_2$ densities}\label{densities}

We have used the non-LTE excitation and radiative transfer code RADEX with the Large Velocity Gradient (LVG) approximation \citep{vandertak}
to estimate the H$_2$ densities by using three HC$_3$N lines (see Tables \ref{table1} and \ref{table2}).
For a T$_{kin}$ of 100 K, a linewidth of 21 km s$^{-1}$, a background temperature of 2.73 K and a beam filling
factor of 1 (since the cloud size is much larger than the telescope beam size, see Fig.~\ref{fig1}), we have estimated H$_2$ densities of $\sim$2.3 (0.3)$\times$10$^4$ cm$^{-3}$ for both GC sources.

The HC$_3$N column densities and H$_2$ densities were considered free parameters for the modelling.  Our non-LTE analysis provided estimates of the HC$_3$N column densities of 6.0$\times$10$^{14}$ cm$^{-2}$ for \emph{LOS}+0.693 and 
4.9$\times$10$^{14}$ cm$^{-2}$ for \emph{LOS}$-$0.11, in good agreement within a factor of 2 with the LTE calculations of the HC$_3$N column densities (see 
Table \ref{table3}). 
Our estimated line optical depths are $\sim$0.5 for the three HC$_3$N lines. 
The predicted excitation temperatures of $\sim$14 K are also consistent with the LTE analysis and our assumption of subthermal excitation. Changes in 
the T$_{kin}$ to lower values of $\sim$50 K, will increase the H$_2$ densities by only a factor of 2, to $\sim$5$\times$10$^4$ cm$^{-3}$. The derived 
excitation temperatures are very sensitive to the H$_2$ density. Changes in a factor of 2 in the H$_2$ densities increase the excitation temperature 
from $\sim$14 K to $\sim$24 K. The latter predicted temperature is inconsistent with our measured T$_{rot}$ for HC$_3$N, ruling out higher H$_2$ densities 
in both GC sources.


\begin{table*}
\scriptsize
\begin{minipage}{200mm}

\caption{Line parameters for \emph{LOS}+0.693.}\label{table1}
\centering
\begin{tabular}{cccc||c||c||cr||r||r||rr||r||r||r||r||r||rr||r||r||rc}
\hline
Molecule  & Frequency  & Transition & \multicolumn{2}{r}{Area ($\sigma$)} & \multicolumn{4}{r}{V$_{\rm LSR}$ ($\sigma$)} & \multicolumn{5}{r}{$\Delta_{v_{1/2}}$ ($\sigma$)} & \multicolumn{4}{c}{T$^{*}$ $_{A}$ ($\sigma$)} & Notes\\
 & (MHz) &  & \multicolumn{2}{c}{(K km s$^{-1}$)} & \multicolumn{4}{c}{(km s$^{-1}$)} & \multicolumn{5}{c}{(km s$^{-1}$)} & \multicolumn{4}{c}{(mK)} & \\
\hline
C$_{2}$H$_{3}$CN & 77633.8 & 8$_{1,7}$-7$_{1,6}$ & \multicolumn{2}{r}{2.6 (0.5)} & \multicolumn{4}{r}{63.9 (1.4)} & \multicolumn{5}{r}{19.6 (3.2)} & \multicolumn{4}{r}{125.8 (15.6)} & \\
Unidentified & 77935.5 &  & \multicolumn{2}{r}{2.9 (1.1)} & \multicolumn{4}{r}{65.0 (5.0)} & \multicolumn{5}{r}{25.0 (5.0)} & \multicolumn{4}{r}{88.1 (18.2)} & \\
C$_{2}$H$_{5}$CN & 78183.6 & 9$_{1,9}$-8$_{1,8}$ & \multicolumn{2}{r}{3.6 (0.6)} & \multicolumn{4}{r}{64.2 (1.5)} & \multicolumn{5}{r}{30.5 (3.5)} & \multicolumn{4}{r}{111.0 (11.9)} & \\
CH$_{3}$OCHO$^{c}$ & 78481.3 & 7$_{1,7}$-6$_{1,6}$ A+E & \multicolumn{2}{r}{1.7 (0.4)} & \multicolumn{4}{r}{71.7 (1.4)} & \multicolumn{5}{r}{24.8 (2.7)} & \multicolumn{4}{r}{129.5 (7.6)} & bl\\
CH$_{3}$CHO & 79099.3 & 4$_{1,3}$-3$_{1,2}$ E  & \multicolumn{2}{r}{5.6 (0.4)} & \multicolumn{4}{r}{70.5 (0.4)} & \multicolumn{5}{r}{20.7 (1.0)} & \multicolumn{4}{r}{0.3 (11.6)} & \\
CH$_{3}$CHO & 79150.2 & 4$_{1,3}$-3$_{1,2}$ A & \multicolumn{2}{r}{10.0 (0.6)} & \multicolumn{4}{r}{70.9 (0.4)} & \multicolumn{5}{r}{22.5 (1.0)} & \multicolumn{4}{r}{418.0$^{a}$(16.1)} & \\
H$^{13}$CCCN & 79350.4 & 9$_K$-8$_K$, K=8-7, 9-8, 10-9 & \multicolumn{2}{r}{4.9 (0.3)} & \multicolumn{4}{r}{70.5 (0.5)} & \multicolumn{5}{r}{22.8 (1.3)} & \multicolumn{4}{r}{202.6 (8.8)} & hf\\
NH$_{2}$CN & 79449.7 & 4$_{1,4}$-3$_{1,3}$ & \multicolumn{2}{r}{10.9 (0.5)} & \multicolumn{4}{r}{68.4 (0.3)} & \multicolumn{5}{r}{19.2 (0.7)} & \multicolumn{4}{r}{534.9 (14.2)} & \\
C$_{2}$H$_{5}$CN & 79677.5 & 9$_{0,9}$-8$_{0,8}$ & \multicolumn{2}{r}{2.6 (0.5)} & \multicolumn{4}{r}{64.3 (1.9)} & \multicolumn{5}{r}{29.2 (4.5)} & \multicolumn{4}{r}{83.8 (11.5)} & \\
H$_{2}$C$_{2}$N & 79759.4 & 4$_{1,4}$-3$_{1,3}$, J=9/2-7/2 & \multicolumn{2}{r}{2.1 (0.5)} & \multicolumn{4}{r}{65.9$^{a}$(1.7)} & \multicolumn{5}{r}{19.1$^{a}$(3.9)} & \multicolumn{4}{r}{103.7 (7.3)} & hf$^a$\\
CH$_{3}$OCHO$^{c}$ & 79783.8 & 7$_{0,7}$-6$_{0,6}$ A+E & \multicolumn{2}{r}{1.6 (0.3)} & \multicolumn{4}{r}{75.7 (0.7)} & \multicolumn{5}{r}{25.4 (1.7)} & \multicolumn{4}{r}{115.9 (6.1)} & bl\\
HC$_{5}$N & 79876.9 & 30-29 & \multicolumn{2}{r}{2.8 (0.2)} & \multicolumn{4}{r}{67.4 (0.6)} & \multicolumn{5}{r}{22.4 (1.4)} & \multicolumn{4}{r}{115.5 (6.4)} & \\
NH$_{2}$CN & 79915.1 & 4$_{0,4}$-3$_{0,3}$ & \multicolumn{2}{r}{1.8 (0.2)} & \multicolumn{4}{r}{72.9 (0.8)} & \multicolumn{5}{r}{24.0 (2.0)} & \multicolumn{4}{r}{68.2 (5.1)} & \\
NH$_{2}$CN & 79963.2 & 4$_{2,3}$-3$_{2,2}$ & \multicolumn{2}{r}{1.1 (0.1)} & \multicolumn{4}{r}{63.7 (0.7)} & \multicolumn{5}{r}{19.0 (1.5)} & \multicolumn{4}{r}{109.9 (7.8)} & m\\
NH$_{2}$CN & 79979.5 & 4$_{0,4}$-3$_{0,3}$ & \multicolumn{2}{r}{4.1 (0.3)} & \multicolumn{4}{r}{65.1 (0.4)} & \multicolumn{5}{r}{19.0 (1.0)} & \multicolumn{4}{r}{200.9 (7.4)} & \\
CH$_{2}$CO & 80076.7 & 4$_{1,4}$-3$_{1,3}$ & \multicolumn{2}{r}{4.2 (0.3)} & \multicolumn{4}{r}{68.2 (0.6)} & \multicolumn{5}{r}{20.7 (1.3)} & \multicolumn{4}{r}{190.8 (9.0)} & \\
Unidentified & 80193.2 &  & \multicolumn{2}{r}{1.6 (0.4)} & \multicolumn{4}{r}{65.6 (1.5)} & \multicolumn{5}{r}{19.0 (3.5)} & \multicolumn{4}{r}{77.8 (10.4)} & \\
Unidentified & 80373.6 &  & \multicolumn{2}{r}{2.6 (1.0)} & \multicolumn{4}{r}{63.6 (3.2)} & \multicolumn{5}{r}{25.0 (7.8)} & \multicolumn{4}{r}{86.7 (20.3)} & \\
C$_{2}$H$_{5}$CN & 80404.9 & 9$_{2,8}$-8$_{2,7}$ & \multicolumn{2}{r}{3.2 (0.4)} & \multicolumn{4}{r}{68.8 (1.5)} & \multicolumn{5}{r}{35.0 (3.5)} & \multicolumn{4}{r}{84.9 (8.2)} & \\
H$_{2}$C$_{2}$N & 80480.9 & 4$_{0,4}$-3$_{0,3}$, J=9/2-7/2 & \multicolumn{2}{r}{3.9 (0.5)} & \multicolumn{4}{r}{61.1$^{a}$(1.1)} & \multicolumn{5}{r}{21.4$^{a}$(2.2)} & \multicolumn{4}{r}{170.7 (10.9)} & hf$^a$\\
H$_{2}$C$_{2}$N & 80489.9 & 4$_{0,4}$-3$_{0,3}$, J=7/2-5/2 & \multicolumn{2}{r}{2.4 (1.6)} & \multicolumn{4}{r}{68.5$^{a}$(9.2)} & \multicolumn{5}{r}{18.9$^{a}$(10.4)} & \multicolumn{4}{r}{119.0$^{a}$(45.5)} & hf$^a$\\
NH$_{2}$CN & 80504.6 & 4$_{1,3}$-3$_{1,2}$ & \multicolumn{2}{r}{9.1 (0.3)} & \multicolumn{4}{r}{65.7 (0.2)} & \multicolumn{5}{r}{18.9 (0.4)} & \multicolumn{4}{r}{451.4 (7.4)} & \\
(CH$_{3}$)$_{2}$O$^{b}$ & 80538.5 & 5$_{2,3}$-5$_{1,4}$ & \multicolumn{2}{r}{2.6 (0.4)} & \multicolumn{4}{r}{75.5$^{a}$(2.4)} & \multicolumn{5}{r}{28.4$^{a}$(4.1)} & \multicolumn{4}{r}{86.0 (6.7)} & \\
CH$_{2}$CO & 80832.1 & 4$_{0,4}$-3$_{0,3}$ & \multicolumn{2}{r}{2.0 (0.2)} & \multicolumn{4}{r}{65.6 (0.8)} & \multicolumn{5}{r}{23.8 (1.8)} & \multicolumn{4}{r}{80.1 (4.9)} & \\
CH$_{3}$OH & 80993.2 & 7$_{2,6}$-8$_{1,7}$ A- & \multicolumn{2}{r}{1.0 (0.2)} & \multicolumn{4}{r}{62.5 (1.5)} & \multicolumn{5}{r}{19.0 (3.4)} & \multicolumn{4}{r}{49.0 (7.0)} & cd\\
H$_{2}$C$_{2}$N & 81207.3 & 4$_{1,3}$-3$_{1,2}$, J=7/2-5/2 & \multicolumn{2}{r}{1.2 (0.2)} & \multicolumn{4}{r}{67.2 (0.8)} & \multicolumn{5}{r}{18.7 (1.9)} & \multicolumn{4}{r}{60.4 (4.8)} & hf$^a$\\
H$_{2}$C$_{2}$N & 81232.5 & 4$_{1,3}$-3$_{1,2}$, J=9/2-7/2 & \multicolumn{2}{r}{1.3 (0.2)} & \multicolumn{4}{r}{70.4 (0.7)} & \multicolumn{5}{r}{18.7 (1.7)} & \multicolumn{4}{r}{64.2 (4.6)} & hf$^a$, cd\\
C$_{2}$H$_{5}$CN & 81261.3 & 9$_{2,7}$-8$_{2,6}$ & \multicolumn{2}{r}{2.4 (0.5)} & \multicolumn{4}{r}{65.1 (1.8)} & \multicolumn{5}{r}{29.2 (4.3)} & \multicolumn{4}{r}{76.0$^{a}$(9.2)} & \\
Unidentified & 81337.8 &  & \multicolumn{2}{r}{1.7 (0.3)} & \multicolumn{4}{r}{68.3 (1.5)} & \multicolumn{5}{r}{28.1 (3.4)} & \multicolumn{4}{r}{57.0 (5.7)} & \\
HNO & 81477.4 & 1$_{0,1}$-0$_{0,0}$ & \multicolumn{2}{r}{1.5 (0.2)} & \multicolumn{4}{r}{67.5$^{a}$(0.8)} & \multicolumn{5}{r}{18.7$^{a}$(2.0)} & \multicolumn{4}{r}{77.0 (6.1)} & hf\\
CCS & 81505.1 & 6$_{7}$-5$_{6}$ & \multicolumn{2}{r}{4.8 (0.3)} & \multicolumn{4}{r}{68.5 (0.4)} & \multicolumn{5}{r}{19.6$^{a}$(0.8)} & \multicolumn{4}{r}{232.1 (7.8)} & \\
HC$^{13}$CCN & 81534.1 & 9$_K$-8$_K$, K=8-7, 9-8, 10-9 & \multicolumn{2}{r}{4.1 (0.4)} & \multicolumn{4}{r}{64.5$^{a}$(1.0)} & \multicolumn{5}{r}{28.0 (2.0)} & \multicolumn{4}{r}{137.3 (8.7)} & hf\\
HCC$^{13}$CN & 81541.9 & 9$_K$-8$_K$, K=8-7, 9-8, 10-9 & \multicolumn{2}{r}{3.8 (0.3)} & \multicolumn{4}{r}{67.8 (0.6)} & \multicolumn{5}{r}{23.3 (1.6)} & \multicolumn{4}{r}{154.2 (5.6)} & hf\\
CH$_2$CO & 81586.2 & 4$_{1,3}$-3$_{1,2}$ & \multicolumn{2}{r}{3.5 (0.2)} & \multicolumn{4}{r}{66.7 (0.4)} & \multicolumn{5}{r}{21.7 (0.8)} & \multicolumn{4}{r}{151.1 (4.4)} & \\
NH$_{2}$CHO & 81693.5 & 4$_{1,4}$-3$_{1,3}$ & \multicolumn{2}{r}{6.3 (0.3)} & \multicolumn{4}{r}{68.0 (0.4)} & \multicolumn{5}{r}{24.3 (1.0)} & \multicolumn{4}{r}{245.4 (8.1)} & hf\\
HC$_{3}$N & 81881.4 & 9-8 & \multicolumn{2}{r}{67.2 (1.8)} & \multicolumn{4}{r}{66.7 (0.2)} & \multicolumn{5}{r}{24.9 (0.5)} & \multicolumn{4}{r}{2534.8 (43.0)} & \\
c-C$_{3}$H$_{2}$ & 82093.5 & 2$_{0,2}$-1$_{1,1}$ & \multicolumn{2}{r}{3.8 (0.4)} & \multicolumn{4}{r}{71.0 (0.6)} & \multicolumn{5}{r}{18.5 (1.5)} & \multicolumn{4}{r}{194.8 (12.1)} & \\
HNCS, a-type & 82101.8 & 7$_{0,7}$-6$_{0,6}$ & \multicolumn{2}{r}{1.8 (0.4)} & \multicolumn{4}{r}{67.8 (1.6)} & \multicolumn{5}{r}{23.9 (3.9)} & \multicolumn{4}{r}{70.1 (8.5)} & \\
C$_{2}$H$_{5}$OH & 82115.7 & 3$_{2,2}$-3$_{1,3}$ & \multicolumn{2}{r}{1.2 (0.3)} & \multicolumn{4}{r}{71.8 (1.3)} & \multicolumn{5}{r}{18.7 (3.0)} & \multicolumn{4}{r}{60.6 (8.4)} & \\
Unidentified & 82198.8 &  & \multicolumn{2}{r}{2.7 (0.2)} & \multicolumn{4}{r}{67.9 (0.9)} & \multicolumn{5}{r}{27.8 (2.0)} & \multicolumn{4}{r}{90.0 (5.0)} & \\
C$_{2}$H$_{5}$CN & 82458.5 & 9$_{1,8}$-8$_{1,7}$ & \multicolumn{2}{r}{3.7 (0.3)} & \multicolumn{4}{r}{67.6 (0.7)} & \multicolumn{5}{r}{29.2 (1.8)} & \multicolumn{4}{r}{20.4 (4.7)} & \\
HCOCH$_{2}$OH & 82470.6 & 8$_{0,8}$-7$_{1,7}$ & \multicolumn{2}{r}{1.0 (0.2)} & \multicolumn{4}{r}{73.8 (0.9)} & \multicolumn{5}{r}{19.6 (2.3)} & \multicolumn{4}{r}{47.3 (4.6)} & \\
HC$_{5}$N & 82539.2 & 31-30 & \multicolumn{2}{r}{4.0 (0.6)} & \multicolumn{4}{r}{64.0 (1.4)} & \multicolumn{5}{r}{27.7 (3.3)} & \multicolumn{4}{r}{141.5 (14.0)} & \\
NH$_{2}$CHO & 82549.5 & 1$_{1,1}$-0$_{0,0}$ & \multicolumn{2}{r}{2.9 (0.2)} & \multicolumn{4}{r}{67.0 (0.7)} & \multicolumn{5}{r}{25.6 (1.8)} & \multicolumn{4}{r}{108.9 (4.6)} & hf, cl\\
(CH$_{3}$)$_{2}$O$^{b}$ & 82650.1 & 3$_{1,3}$-2$_{0,2}$ & \multicolumn{2}{r}{3.6 (0.2)} & \multicolumn{4}{r}{65.6 (0.6)} & \multicolumn{5}{r}{27.6 (1.4)} & \multicolumn{4}{r}{123.0$^{a}$(4.3)} & \\
O$^{13}$C$^{34}$S & 82762.5 & 7-6 & \multicolumn{2}{r}{$\la$0.3} & \multicolumn{4}{c}{...} & \multicolumn{5}{c}{...} & \multicolumn{4}{r}{$\la$33} & \\
c-C$_{3}$H$_{2}$ & 82966.2 & 3$_{1,2}$-3$_{0,3}$ & \multicolumn{2}{r}{4.2 (0.2)} & \multicolumn{4}{r}{67.4 (0.3)} & \multicolumn{5}{r}{18.4 (0.8)} & \multicolumn{4}{r}{212.5 (7.5)} & \\
OC$^{34}$S & 83057.9 & 7-6 & \multicolumn{2}{r}{2.1 (0.3)} & \multicolumn{4}{r}{65.0 (1.3)} & \multicolumn{5}{r}{27.5 (3.0)} & \multicolumn{4}{r}{72.4 (6.6)} & \\
HNC$^{18}$O & 83191.5 & 4$_{0,4}$-3$_{0,3}$ & \multicolumn{2}{r}{$\la$0.1} & \multicolumn{4}{c}{...} & \multicolumn{5}{c}{...} & \multicolumn{4}{r}{$\la$39} & hf \\
C$_{2}$H$_{3}$CN & 83207.5 & 9$_{1,9}$-8$_{1,8}$ & \multicolumn{2}{r}{2.2 (0.1)} & \multicolumn{4}{r}{70.1 (0.5)} & \multicolumn{5}{r}{20.4 (1.1)} & \multicolumn{4}{r}{98.7 (3.9)} & \\
Unidentified & 83404.9 &  & \multicolumn{2}{r}{2.0 (0.3)} & \multicolumn{4}{r}{68.7 (1.2)} & \multicolumn{5}{r}{27.4 (2.7)} & \multicolumn{4}{r}{69.0 (5.4)} & \\
CH$_{3}$CHO & 83584.2 & 2$_{-1,2}$-1$_{0,1}$ E & \multicolumn{2}{r}{2.1 (0.4)} & \multicolumn{4}{r}{68.9 (1.2)} & \multicolumn{5}{r}{18.2 (3.1)} & \multicolumn{4}{r}{0.1$^{a}$(10.6)} & \\
SO$_{2}$ & 83688.0 & 8$_{1,7}$-8$_{0,8}$ & \multicolumn{2}{r}{4.2 (0.2)} & \multicolumn{4}{r}{65.0 (0.4)} & \multicolumn{5}{r}{25.9 (0.8)} & \multicolumn{4}{r}{153.0 (4.1)} & \\
Unidentified & 83900.3 &  & \multicolumn{2}{r}{5.2 (0.4)} & \multicolumn{4}{r}{65.4 (0.6)} & \multicolumn{5}{r}{24.2 (1.4)} & \multicolumn{4}{r}{202.7 (9.3)} & \\
$^{13}$CCH & 84119.3 & N=1-0, F$_{1}$=2-1, F=5/2-3/2 & \multicolumn{2}{r}{1.2 (0.3)} & \multicolumn{4}{r}{67.6 (2.4)} & \multicolumn{5}{r}{36.2 (5.7)} & \multicolumn{4}{r}{32.0 (4.1)} & hf, cd\\
Unidentified & 84139.0 &  & \multicolumn{2}{r}{0.7 (0.3)} & \multicolumn{4}{r}{67.6 (2.4)} & \multicolumn{5}{r}{18.1 (5.8)} & \multicolumn{4}{r}{37.3 (8.2)} & \\
$^{13}$CCH & 84153.3 & N=1-0, F$_{1}$=1-0, F=3/2-1/2 & \multicolumn{2}{r}{2.1 (0.5)} & \multicolumn{4}{r}{84.5 (2.6)} & \multicolumn{5}{r}{36.2 (6.1)} & \multicolumn{4}{r}{55.1 (7.2)} & hf\\
Unidentified & 84183.9 &  & \multicolumn{2}{r}{1.2 (0.3)} & \multicolumn{4}{r}{65.2 (1.4)} & \multicolumn{5}{r}{18.1 (3.3)} & \multicolumn{4}{r}{60.0 (8.9)} & \\
CH$_{3}$CHO & 84219.7 & 2$_{1,2}$-1$_{0,1}$ A & \multicolumn{2}{r}{2.7 (0.5)} & \multicolumn{4}{r}{74.8 (1.5)} & \multicolumn{5}{r}{29.0 (3.6)} & \multicolumn{4}{r}{88.6 (10.1)} & \\
CH$_{3}$OCHO & 84449.1 & 7$_{2,6}$-6$_{2,5}$ E & \multicolumn{2}{r}{0.9 (0.3)} & \multicolumn{4}{r}{69.5$^{a}$(1.7)} & \multicolumn{5}{r}{18.0 (4.0)} & \multicolumn{4}{r}{48.0 (8.4)} & \\
CH$_{3}$OCHO & 84454.7 & 7$_{2,6}$-6$_{2,5}$ A & \multicolumn{2}{r}{1.3 (0.3)} & \multicolumn{4}{r}{66.0 (1.3)} & \multicolumn{5}{r}{18.0 (3.3)} & \multicolumn{4}{r}{69.0 (6.4)} & \\
CH$_{3}$OH & 84521.1 & 5$_{-1,5}$-4$_{0,4}$ E & \multicolumn{2}{r}{128.0 (5.1)} & \multicolumn{4}{r}{68.5 (0.2)} & \multicolumn{5}{r}{18.0 (0.6)} & \multicolumn{4}{r}{6678.0$^{a}$(164.6)} & \\
NH$_{2}$CHO & 84542.4 & 4$_{0,4}$-3$_{0,3}$ & \multicolumn{2}{r}{7.3 (0.5)} & \multicolumn{4}{r}{71.4 (0.6)} & \multicolumn{5}{r}{24.4 (1.3)} & \multicolumn{4}{r}{280.2 (11.8)} & hf\\
c-C$_{3}$H$_{2}$ & 84727.6 & 3$_{2,2}$-3$_{1,3}$ & \multicolumn{2}{r}{0.8 (0.2)} & \multicolumn{4}{r}{67.3 (1.4)} & \multicolumn{5}{r}{19.8 (3.2)} & \multicolumn{4}{r}{38.8 (4.8)} & \\
$^{30}$SiO & 84745.9 & 2-1 & \multicolumn{2}{r}{2.2 (0.2)} & \multicolumn{4}{r}{71.8 (0.6)} & \multicolumn{5}{r}{18.0 (1.4)} & \multicolumn{4}{r}{116.0$^{a}$(6.0)} & \\
NH$_{2}$CHO & 84807.9 & 4$_{2,3}$-3$_{2,2}$ & \multicolumn{2}{r}{2.6 (0.2)} & \multicolumn{4}{r}{69.0 (0.4)} & \multicolumn{5}{r}{20.9 (1.0)} & \multicolumn{4}{r}{116.7 (4.3)} & hf\\
O$^{13}$CS & 84865.1 & 7-6 & \multicolumn{2}{r}{0.9 (0.1)} & \multicolumn{4}{r}{65.8 (0.9)} & \multicolumn{5}{r}{17.9 (2.1)} & \multicolumn{4}{r}{45.0$^{a}$(3.8)} & \\
NH$_{2}$CHO & 84888.9 & 4$_{3,2}$-3$_{3,1}$ & \multicolumn{2}{r}{1.1 (0.1)} & \multicolumn{4}{r}{67.5 (0.8)} & \multicolumn{5}{r}{17.9 (1.9)} & \multicolumn{4}{r}{56.3 (4.6)} & hf, bl\\
C$_{2}$H$_{3}$CN & 84946.0 & 9$_{0,9}$-8$_{0,8}$ & \multicolumn{2}{r}{2.6 (0.2)} & \multicolumn{4}{r}{67.6 (0.7)} & \multicolumn{5}{r}{25.8 (1.7)} & \multicolumn{4}{r}{93.7 (4.6)} & \\
$^{13}$CH$_{3}$OH & 84970.2 & 8$_{0,8}$-7$_{1,7}$ A+ & \multicolumn{2}{r}{1.5 (0.3)} & \multicolumn{4}{r}{70.4 (1.9)} & \multicolumn{5}{r}{26.9 (4.3)} & \multicolumn{4}{r}{52.1 (4.6)} & \\
Unidentified &      84980.0 &  & \multicolumn{2}{r}{1.0 (0.2)} & \multicolumn{4}{r}{67.9 (2.5)} & \multicolumn{5}{r}{23.7 (5.0)} & \multicolumn{4}{r}{37.0 (6.0)} & \\
NH$_{2}$CHO & 85093.3 & 4$_{2,2}$-3$_{2,1}$ & \multicolumn{2}{r}{3.3 (0.3)} & \multicolumn{4}{r}{67.6 (0.7)} & \multicolumn{5}{r}{23.0 (1.5)} & \multicolumn{4}{r}{136(7.6)} & hf\\
OCS & 85139.1 & 7-6 & \multicolumn{2}{r}{20.7 (0.5)} & \multicolumn{4}{r}{64.6 (0.2)} & \multicolumn{5}{r}{22.2(0.4)} & \multicolumn{4}{r}{876.7(12.7)} & \\
HC$^{18}$O$^+$ & 85162.1 & 1-0 & \multicolumn{2}{r}{2.0 (0.8)} & \multicolumn{4}{r}{74.5 (2.8)} & \multicolumn{5}{r}{17.9 (7.1)} & \multicolumn{4}{r}{104.0 (7.6)} & \\
HC$_{5}$N & 85201.6 & 32-31 & \multicolumn{2}{r}{2.4 (0.4)} & \multicolumn{4}{r}{67.3 (1.1)} & \multicolumn{5}{r}{21.7 (2.6)} & \multicolumn{4}{r}{105.3 (10.0)} & \\
C$^{13}$CH & 85229.3 & N=1-0, F1=2-1 F=5/2-3/2 & \multicolumn{2}{r}{2.0 (0.3)} & \multicolumn{4}{r}{60.6 (1.5)} & \multicolumn{5}{r}{26.8 (3.6)} & \multicolumn{4}{r}{68.9 (7.3)} & hf\\
C$_{2}$H$_{5}$OH & 85265.5 & 6$_{0,6}$-5$_{1,5}$ & \multicolumn{2}{r}{7.2 (0.3)} & \multicolumn{4}{r}{64.5 (0.3)} & \multicolumn{5}{r}{23.9 (0.8)} & \multicolumn{4}{r}{282.9 (7.3)} & cl\\
H$^{15}$NCO & 85292.1 & 4$_{0,4}$-3$_{0,3}$ & \multicolumn{2}{r}{$\la$0.3} & \multicolumn{4}{c}{...} & \multicolumn{5}{c}{...} & \multicolumn{4}{r}{$\la$36} & \\

\hline
\end{tabular}
\end{minipage}
\end{table*}

\begin{table*}
\scriptsize
\begin{minipage}{200mm}
\contcaption{}
\centering
\begin{tabular}{cccc||c||c||cr||r||r||rr||r||r||r||r||r||rr||r||r||rc}
\hline
Molecule  & Frequency  & Transition & \multicolumn{2}{r}{Area ($\sigma$)} & \multicolumn{4}{r}{V$_{\rm LSR}$ ($\sigma$)} & \multicolumn{5}{r}{$\Delta_{v_{1/2}}$ ($\sigma$)} & \multicolumn{4}{c}{T$^{*}$ $_{A}$ ($\sigma$)} & Notes\\
 & (MHz) &  & \multicolumn{2}{c}{(K km s$^{-1}$)} & \multicolumn{4}{c}{(km s$^{-1}$)} & \multicolumn{5}{c}{(km s$^{-1}$)} & \multicolumn{4}{c}{(mK)} & \\
\hline
C$_{2}$H$_{3}$CN & 85302.6 & 9$_{2,8}$-8$_{2,7}$ & \multicolumn{2}{r}{1.4 (0.3)} & \multicolumn{4}{r}{68.5$^{a}$(1.8)} & \multicolumn{5}{r}{17.8 (3.6)} & \multicolumn{4}{r}{72.1 (5.6)} & \\
c-C$_{3}$H$_{2}$ & 85338.8 & 2$_{1,2}$-1$_{0,1}$ & \multicolumn{2}{r}{8.9 (0.4)} & \multicolumn{4}{r}{[40, 80]}  & \multicolumn{5}{r}{...} & \multicolumn{4}{r}{...} & \\
                                           &  &  & \multicolumn{2}{r}{1.6 (0.4)} & \multicolumn{4}{r}{[80, 110]} & \multicolumn{5}{r}{...} & \multicolumn{4}{r}{...} & \\
HCS$^+$ & 85347.9 & 2-1 & \multicolumn{2}{r}{5.2 (0.6)} & \multicolumn{4}{r}{70.4 (0.7)} & \multicolumn{5}{r}{17.8 (1.8)} & \multicolumn{4}{r}{275.0$^{a}$(10.4)} & bl\\
CH$_{3}^{13}$CCH & 85407.2 & 5$_{3}$-4$_{3}$ & \multicolumn{2}{r}{0.4 (0.2)} & \multicolumn{4}{r}{65.8 (1.7)} & \multicolumn{5}{r}{8.9 (4.1)} & \multicolumn{4}{r}{41.8 (9.0)} & \\
CH$_{3}^{13}$CCH & 85421.8 & 5$_K$-4$_K$, K=0, 1 & \multicolumn{2}{r}{0.7 (0.2)} & \multicolumn{4}{r}{80.9 (2.2)} & \multicolumn{5}{r}{17.8 (5.2)} & \multicolumn{4}{r}{36.6 (6.9)} & m, cd\\
CH$_{3}$CCH & 85442.5 & 5$_{3}$-4$_{3}$ & \multicolumn{2}{r}{2.6 (0.5)} & \multicolumn{4}{r}{68.5$^{a}$(1.1)} & \multicolumn{5}{r}{17.8 (2.7)} & \multicolumn{4}{r}{137.3 (13.7)} & \\
CH$_{3}$CCH & 85450.7 & 5$_{2}$-4$_{2}$ & \multicolumn{2}{r}{3.5 (0.6)} & \multicolumn{4}{r}{69.5$^{a}$(1.0)} & \multicolumn{5}{r}{17.8 (2.7)} & \multicolumn{4}{r}{183.0$^{a}$(16.1)} & bl\\
CH$_{3}$CCH & 85457.2 & 5$_K$-4$_K$, K=0, 1 & \multicolumn{2}{r}{11.2 (0.9)} & \multicolumn{4}{r}{66.5 (0.5)} & \multicolumn{5}{r}{20.8 (1.4)} & \multicolumn{4}{r}{508.6 (25.1)} & m\\
HOCO$^+$ & 85531.5 & 4$_{0,4}$-3$_{0,3}$ & \multicolumn{2}{r}{17.3 (0.4)} & \multicolumn{4}{r}{64.8 (0.2)} & \multicolumn{5}{r}{22.1 (0.4)} & \multicolumn{4}{r}{737.4 (10.4)} & \\
C$_{2}$H$_{3}$CN & 85715.4 & 9$_{2,7}$-8$_{2,6}$ & \multicolumn{2}{r}{1.0 (0.2)} & \multicolumn{4}{r}{67.4 (1.5)} & \multicolumn{5}{r}{15.7 (3.6)} & \multicolumn{4}{r}{56.4 (6.1)} & \\
$^{29}$SiO & 85759.0 & 2-1 & \multicolumn{2}{r}{2.4 (0.2)} & \multicolumn{4}{r}{65.6 (0.6)} & \multicolumn{5}{r}{17.8 (1.3)} & \multicolumn{4}{r}{129.0$^{a}$(6.9)} & \\
HC$^{15}$N & 86054.9 & 1-0 & \multicolumn{2}{r}{1.8 (0.2)} & \multicolumn{4}{r}{[40, 80]}  & \multicolumn{5}{c}{...} & \multicolumn{4}{c}{...} & \\
 &  &                      & \multicolumn{2}{r}{1.0 (0.2)} & \multicolumn{4}{r}{[80, 110]} & \multicolumn{5}{c}{...} & \multicolumn{4}{c}{...} & \\
SO & 86093.9 & 2$_{2}$-1$_{1}$ & \multicolumn{2}{r}{8.0 (0.4)} & \multicolumn{4}{r}{67.6 (0.5)} & \multicolumn{5}{r}{26.6 (1.1)} & \multicolumn{4}{r}{288.0 (10.1)} & \\
CCS & 86181.3 & 7$_{6}$-6$_{5}$ & \multicolumn{2}{r}{1.1 (0.2)} & \multicolumn{4}{r}{71.9 (1.4)} & \multicolumn{5}{r}{19.6$^{a}$(3.3)} & \multicolumn{4}{r}{54.0$^{a}$(5.7)} & \\
(CH$_{3}$)$_{2}$O$^{b}$ & 86226.7 & 2$_{2,0}$-2$_{1,1}$ & \multicolumn{2}{r}{1.7 (0.3)} & \multicolumn{4}{r}{76.7 (1.2)} & \multicolumn{5}{r}{26.5 (3.5)} & \multicolumn{4}{r}{60.6 (5.1)} & bl\\
CH$_{3}$OCHO$^{c}$ & 86265.8 & 7$_{3,5}$-6$_{3,4}$ A+E & \multicolumn{2}{r}{0.8 (0.2)} & \multicolumn{4}{r}{62.9 (1.0)} & \multicolumn{5}{r}{17.6 (2.2)} & \multicolumn{4}{r}{84.0$^{a}$(7.2)} & bl\\
H$^{13}$CN & 86340.1 & 1-0, F=1-1, 2-1, 0-1 & \multicolumn{2}{r}{9.1 (0.1)} & \multicolumn{4}{r}{[40,80]} & \multicolumn{5}{c}{...} & \multicolumn{4}{c}{...} & hf\\
 &  &                                       & \multicolumn{2}{r}{6.3 (0.1)} & \multicolumn{4}{r}{[80,110]} & \multicolumn{5}{c}{...} & \multicolumn{4}{c}{...} & \\
HCOOH & 86546.1 & 4$_{1,4}$-3$_{1,3}$ & \multicolumn{2}{r}{1.9 (0.3)} & \multicolumn{4}{r}{68.5$^{a}$(0.8)} & \multicolumn{5}{r}{17.6 (2.0)} & \multicolumn{4}{r}{102.9 (7.4)} & \\
HCO & 86670.7 & 1$_{0,1}$-0$_{0,0}$, J=3/2-1/2, F=2-1 & \multicolumn{2}{r}{2.0 (0.3)} & \multicolumn{4}{r}{69.4 (0.9)} & \multicolumn{5}{r}{17.6$^{a}$(2.2)} & \multicolumn{4}{r}{108.4 (9.8)} & hf$^{a}$\\
HCO & 86708.3 & 1$_{0,1}$-0$_{0,0}$, J=3/2-1/2, F=1-0 & \multicolumn{2}{r}{1.7 (0.3)} & \multicolumn{4}{r}{68.5$^{a}$(1.3)} & \multicolumn{5}{r}{17.6$^{a}$(3.0)} & \multicolumn{4}{r}{89.0$^{a}$(10.0)} & hf$^{a}$, cd\\
H$^{13}$CO$^+$ & 86754.2 & 1-0 & \multicolumn{2}{r}{7.3 (0.1)} & \multicolumn{4}{r}{[40, 80]}  & \multicolumn{5}{c}{...} & \multicolumn{4}{c}{...} & \\
 &  &                       & \multicolumn{2}{r}{3.5 (0.1)} & \multicolumn{4}{r}{[80, 110]} & \multicolumn{5}{c}{...} & \multicolumn{4}{c}{...} & \\
HCO & 86777.4 & 1$_{0,1}$-0$_{0,0}$, J=1/2-1/2, F=1-1 & \multicolumn{2}{r}{1.0 (0.3)} & \multicolumn{4}{r}{75.0 (1.6)} & \multicolumn{5}{r}{17.5 (3.8)} & \multicolumn{4}{r}{54.7 (9.5)} & hf$^{a}$\\
HCO & 86805.7 & 1$_{0,1}$-0$_{0,0}$, J=1/2-1/2, F=0-1 & \multicolumn{2}{r}{0.7 (0.2)} & \multicolumn{4}{r}{67.9 (1.4)} & \multicolumn{5}{r}{14.9 (3.7)} & \multicolumn{4}{r}{45.5 (7.0)} & hf$^{a}$\\
SiO & 86846.9 & 2-1 & \multicolumn{2}{r}{13.9 (0.4)} & \multicolumn{4}{r}{[40,80]}  & \multicolumn{5}{c}{...} & \multicolumn{4}{c}{...} & \\
 &  &               & \multicolumn{2}{r}{6.4  (0.4)} & \multicolumn{4}{r}{[80,110]} & \multicolumn{5}{c}{...} & \multicolumn{4}{c}{...} & \\
HN$^{13}$C & 87090.8 & 1-0 & \multicolumn{2}{r}{11.1 (0.2)} & \multicolumn{4}{r}{[40, 80]}  & \multicolumn{5}{c}{...} & \multicolumn{4}{c}{...} & \\
 &  &                      & \multicolumn{2}{r}{0.1 (0.2)}  & \multicolumn{4}{r}{[80, 110]} & \multicolumn{5}{c}{...} & \multicolumn{4}{c}{...} & \\
CCH & 87284.1 & N=1-0, J=3/2-1/2, F=1-1 & \multicolumn{2}{r}{3.4 (0.5)} & \multicolumn{4}{r}{67.2 (0.9)} & \multicolumn{5}{r}{17.8 (2.1)} & \multicolumn{4}{r}{181.4 (18.1)} & hf$^{a}$\\
CCH & 87316.9 & N=1-0, J=3/2-1/2, F=2-1 & \multicolumn{2}{r}{19.1 (0.8)} & \multicolumn{4}{r}{69.1 (0.4)} & \multicolumn{5}{r}{26.0 (0.9)} & \multicolumn{4}{r}{689.0$^{a}$(14.9)} & hf$^{a}$\\
CCH & 87328.6 & N=1-0, J=3/2-1/2, F=1-0 & \multicolumn{2}{r}{6.8 (0.8)} & \multicolumn{4}{r}{70.3 (0.8)} & \multicolumn{5}{r}{17.4$^{a}$(2.0)} & \multicolumn{4}{r}{367.3 (15.0)} & hf$^{a}$\\
CCH & 87402.0 & N=1-0, J=1/2-1/2, F=1-1 & \multicolumn{2}{r}{7.8 (0.5)} & \multicolumn{4}{r}{69.5 (0.5)} & \multicolumn{5}{r}{16.1 (0.9)} & \multicolumn{4}{r}{455.2 (19.4)} & hf$^{a}$, bl\\
CCH & 87407.1 & N=1-0, J=1/2-1/2, F=0-1 & \multicolumn{2}{r}{6.6 (0.5)} & \multicolumn{4}{r}{66.8 (0.7)} & \multicolumn{5}{r}{23.2 (1.5)} & \multicolumn{4}{r}{266.5 (9.2)} & hf$^{a}$, bl\\
CCH & 87446.5 & N=1-0, J=1/2-1/2, F=1-0 & \multicolumn{2}{r}{4.5 (0.3)} & \multicolumn{4}{r}{65.4 (0.4)} & \multicolumn{5}{r}{20.1 (0.9)} & \multicolumn{4}{r}{211.0 (8.3)} & hf$^{a}$, cd\\
HNCO & 87597.3 & 4$_{1,4}$-3$_{1,3}$ & \multicolumn{2}{r}{2.1 (0.4)} & \multicolumn{4}{r}{66.6 (1.0)} & \multicolumn{5}{r}{17.4 (2.4)} & \multicolumn{4}{r}{113.0$^{a}$(10.9)} & hf\\
C$_{2}$H$_{5}$OH & 87716.1 & 5$_{2,4}$-5$_{1,5}$ & \multicolumn{2}{r}{1.1 (0.2)} & \multicolumn{4}{r}{66.6 (1.2)} & \multicolumn{5}{r}{18.8 (2.9)} & \multicolumn{4}{r}{53.7 (7.1)} & \\
NH$_{2}$CHO & 87848.9 & 4$_{1,3}$-3$_{1,2}$ & \multicolumn{2}{r}{11.5 (1.1)} & \multicolumn{4}{r}{67.8 (0.8)} & \multicolumn{5}{r}{26.7 (1.9)} & \multicolumn{4}{r}{405.4 (26.2)} & hf\\
HC$_{5}$N & 87863.9 & 33-32 & \multicolumn{2}{r}{2.6 (0.3)} & \multicolumn{4}{r}{69.5 (1.0)} & \multicolumn{5}{r}{22.2 (2.3)} & \multicolumn{4}{r}{109.0 (8.2)} & \\
HNCO & 87925.2 & 4$_{0,4}$-3$_{0,3}$ & \multicolumn{2}{r}{140.2 (3.2)} & \multicolumn{4}{r}{66.1 (0.2)} & \multicolumn{5}{r}{24.6 (0.4)} & \multicolumn{4}{r}{5355.9 (78.9)} & hf\\
H$^{13}$CCCN & 88166.8 & 10$_K$-9$_K$, K=10-9, 11-10, 9-8 & \multicolumn{2}{r}{4.1 (0.3)} & \multicolumn{4}{r}{67.5 (0.5)} & \multicolumn{5}{r}{23.6 (1.1)} & \multicolumn{4}{r}{162.8 (6.3)} & hf\\
HNCO & 88239.0 & 4$_{1,3}$-3$_{1,2}$ & \multicolumn{2}{r}{3.1 (0.4)} & \multicolumn{4}{r}{65.9 (1.2)} & \multicolumn{5}{r}{25.9 (3.1)} & \multicolumn{4}{r}{113.6 (8.5)} & hf\\
C$_{2}$H$_{5}$CN & 88323.7 & 10$_{0,10}$-9$_{0,9}$ & \multicolumn{2}{r}{1.7 (0.3)} & \multicolumn{4}{r}{68.7 (1.1)} & \multicolumn{5}{r}{22.9 (2.6)} & \multicolumn{4}{r}{69.0$^{a}$(7.1)} & \\
HCN & 88631.8 & 1-0, F=0-1, 1-1, 2-1    & \multicolumn{2}{r}{53.9 (0.2)} & \multicolumn{4}{r}{[40,80]}  & \multicolumn{5}{c}{...} & \multicolumn{4}{c}{...} & hf\\
 &  &                                   & \multicolumn{2}{r}{34.2 (0.2)} & \multicolumn{4}{r}{[80,110]} & \multicolumn{5}{c}{...} & \multicolumn{4}{c}{...} & \\
CH$_{3}$OCHO & 88843.2 & 7$_{1,6}$-6$_{1,5}$ E & \multicolumn{2}{r}{0.6 (0.2)} & \multicolumn{4}{r}{70.1 (2.5)} & \multicolumn{5}{r}{17.1 (6.1)} & \multicolumn{4}{r}{30.0$^{a}$(7.7)} & \\
CH$_{3}$OCHO & 88851.6 & 7$_{1,6}$-6$_{1,5}$ A & \multicolumn{2}{r}{1.0 (0.4)} & \multicolumn{4}{r}{67.7 (2.6)} & \multicolumn{5}{r}{17.1 (5.9)} & \multicolumn{4}{r}{56.8 (14.0)} & \\
H$^{15}$NC & 88865.7 & 1-0 & \multicolumn{2}{r}{1.7 (0.2)} & \multicolumn{4}{r}{[40, 80]} & \multicolumn{5}{c}{...} & \multicolumn{4}{c}{...} & \\
HCO$^+$ & 89188.5 & 1-0 & \multicolumn{2}{r}{34.6 (0.2)} & \multicolumn{4}{r}{[40, 80]}  & \multicolumn{5}{c}{...} & \multicolumn{4}{c}{...} & \\
 &  &                & \multicolumn{2}{r}{24.6 (0.2)} & \multicolumn{4}{r}{[80, 110]} & \multicolumn{5}{c}{...} & \multicolumn{4}{c}{...} & \\
CH$_{3}$OCHO$^{c}$ & 89316.6 & 8$_{1,8}$-7$_{1,7}$ A+E & \multicolumn{2}{r}{1.4 (0.2)} & \multicolumn{4}{r}{66.6 (0.6)} & \multicolumn{5}{r}{25.7 (1.6)} & \multicolumn{4}{r}{101.4 (5.1)} & bl\\
$^{13}$CH$_{3}$CN & 89331.3 & 5$_K$-4$_K$ K=0, 1 & \multicolumn{2}{r}{2.0 (0.2)} & \multicolumn{4}{r}{69.7 (1.0)} & \multicolumn{5}{r}{25.6 (2.5)} & \multicolumn{4}{r}{74.9 (4.9)} & hf, bl\\
HCCNC & 89419.3 & 9-8 & \multicolumn{2}{r}{0.7 (0.3)} & \multicolumn{4}{r}{70.1 (3.1)} & \multicolumn{5}{r}{21.5 (8.1)} & \multicolumn{4}{r}{28.8 (6.9)} & hf \\
HOC$^+$ & 89487.4 & 1-0 & \multicolumn{2}{r}{1.1 (0.1)} & \multicolumn{4}{r}{68.0 (0.8)} & \multicolumn{5}{r}{19.4 (2.0)} & \multicolumn{4}{r}{51.8 (4.3)} & \\
HCOOH & 89579.1 & 4$_{0,4}$-3$_{0,3}$ & \multicolumn{2}{r}{2.4 (0.2)} & \multicolumn{4}{r}{66.5$^{a}$(0.8)} & \multicolumn{5}{r}{20.0 (1.9)} & \multicolumn{4}{r}{91.0$^{a}$(6.6)} & \\
HCOOH & 89861.4 & 4$_{2,3}$-3$_{2,2}$ & \multicolumn{2}{r}{1.0 (0.3)} & \multicolumn{4}{r}{60.3 (2.7)} & \multicolumn{5}{r}{25.4 (7.1)} & \multicolumn{4}{r}{36.8 (4.6)} & \\
C$_{2}$H$_{5}$OH & 90117.6 & 4$_{1,4}$-3$_{0,3}$ & \multicolumn{2}{r}{6.7 (0.3)} & \multicolumn{4}{r}{65.0 (0.3)} & \multicolumn{5}{r}{21.6 (0.8)} & \multicolumn{4}{r}{290.8 (7.9)} & \\
CH$_{3}$OCHO & 90145.6 & 7$_{2,5}$-6$_{2,4}$ E & \multicolumn{2}{r}{1.1 (0.4)} & \multicolumn{4}{r}{64.7 (2.5)} & \multicolumn{5}{r}{16.9 (6.3)} & \multicolumn{4}{r}{58.3 (9.7)} & \\
CH$_{3}$OCHO & 90156.4 & 7$_{2,5}$-6$_{2,4}$ A & \multicolumn{2}{r}{1.1 (0.3)} & \multicolumn{4}{r}{66.0 (1.5)} & \multicolumn{5}{r}{18.1 (3.9)} & \multicolumn{4}{r}{57.4 (7.5)} & \\
HCOOH & 90164.6 & 4$_{2,2}$-3$_{2,1}$ & \multicolumn{2}{r}{1.5 (0.5)} & \multicolumn{4}{r}{65.8 (1.9)} & \multicolumn{5}{r}{17.3 (4.6)} & \multicolumn{4}{r}{81.0 (14.6)} & cl\\
CH$_{3}$COOH & 90203.3 & 8$_{*,8}$-7$_{*,7}$ E & \multicolumn{2}{r}{$\la$0.3} & \multicolumn{4}{c}{...} & \multicolumn{5}{c}{\emph{...}} & \multicolumn{4}{r}{$\la$33} & \\
CH$_{3}$OCHO$^{c}$ & 90229.6 & 8$_{0,8}$-7$_{0,7}$ A+E & \multicolumn{2}{r}{1.1 (0.3)} & \multicolumn{4}{r}{69.9 (0.9)} & \multicolumn{5}{r}{18.8 (2.1)} & \multicolumn{4}{r}{88.9 (9.0)} & bl\\
C$_{2}$H$_{5}$CN & 90453.2 & 10$_{2,8}$-9$_{2,7}$ & \multicolumn{2}{r}{1.6 (0.3)} & \multicolumn{4}{r}{70.2 (1.4)} & \multicolumn{5}{r}{25.2 (3.4)} & \multicolumn{4}{r}{58.6 (5.3)} & \\
HC$_{5}$N & 90526.2 & 34-33 & \multicolumn{2}{r}{2.0 (0.2)} & \multicolumn{4}{r}{67.1 (0.9)} & \multicolumn{5}{r}{23.6 (2.2)} & \multicolumn{4}{r}{78.3 (6.2)} & \\
HC$^{13}$CCN & 90593.0 & 10$_K$-9$_K$, K=9-8, 10-9, 11-10 & \multicolumn{2}{r}{5.5 (0.5)} & \multicolumn{4}{r}{65.7 (1.1)} & \multicolumn{5}{r}{25.2 (2.1)} & \multicolumn{4}{r}{203.6 (9.2)} & hf\\
HCC$^{13}$CN & 90601.7 & 10$_K$-9$_K$, K=9-8, 10-9, 11-10 & \multicolumn{2}{r}{3.8 (0.4)} & \multicolumn{4}{r}{68.8 (0.9)} & \multicolumn{5}{r}{22.6 (1.9)} & \multicolumn{4}{r}{158.3 (6.6)} & hf\\
HNC & 90663.5 & 1-0 & \multicolumn{2}{r}{39.4 (0.3)} & \multicolumn{4}{r}{[40, 80]}  & \multicolumn{5}{c}{...} & \multicolumn{4}{c}{...} & \\
 &  &               & \multicolumn{2}{r}{17.1 (0.3)} & \multicolumn{4}{r}{[80, 110]} & \multicolumn{5}{c}{...} & \multicolumn{4}{c}{...} & \\
SiS & 90771.5 & 5-4 & \multicolumn{2}{r}{3.6 (1.0)} & \multicolumn{4}{r}{69.5$^{a}$(3.7)} & \multicolumn{5}{r}{25.2 (6.6)} & \multicolumn{4}{r}{136.1 (12.3)} & \\
(CH$_{3}$)$_{2}$O$^{b}$ & 90938.0 & 6$_{0,6}$-5$_{1,5}$ & \multicolumn{2}{r}{4.2 (0.4)} & \multicolumn{4}{r}{66.9 (0.6)} & \multicolumn{5}{r}{21.9 (1.4)} & \multicolumn{4}{r}{179.1 (11.1)} & \\
HC$_{3}$N & 90979.0 & 10-9 & \multicolumn{2}{r}{64.3 (1.8)} & \multicolumn{4}{r}{67.4 (0.2)} & \multicolumn{5}{r}{23.6 (0.5)} & \multicolumn{4}{r}{2557.7 (46.4)} & \\
C$_{2}$H$_{5}$CN & 91549.1 & 10$_{1,9}$-9$_{1,8}$ & \multicolumn{2}{r}{2.2 (0.5)} & \multicolumn{4}{r}{64.1 (2.1)} & \multicolumn{5}{r}{27.1 (4.5)} & \multicolumn{4}{r}{75.4 (7.4)} & \\
Unidentified & 91750.0 &  & \multicolumn{2}{r}{1.6 (0.2)} & \multicolumn{4}{r}{70.3 (0.8)} & \multicolumn{5}{r}{18.6 (2.0)} & \multicolumn{4}{r}{78.5 (7.8)} & \\
Unidentified & 91848.0 &  & \multicolumn{2}{r}{3.7 (0.3)} & \multicolumn{4}{r}{67.4 (0.7)} & \multicolumn{5}{r}{21.4 (1.6)} & \multicolumn{4}{r}{160.3 (8.3)} & \\

\hline
\end{tabular}
\end{minipage}
\end{table*}
\begin{table*}
\begin{threeparttable}
\scriptsize
\begin{minipage}{200mm}
\contcaption{}
\centering
\begin{tabular}{cccc||c||c||cr||r||r||rr||r||r||r||r||r||rr||r||r||rc}
\hline
Molecule  & Frequency  & Transition & \multicolumn{2}{r}{Area ($\sigma$)} & \multicolumn{4}{r}{V$_{\rm LSR}$ ($\sigma$)} & \multicolumn{5}{r}{$\Delta_{v_{1/2}}$ ($\sigma$)} & \multicolumn{4}{c}{T$^{*}$ $_{A}$ ($\sigma$)} & Notes\\
 & (MHz) &  & \multicolumn{2}{c}{(K km s$^{-1}$)} & \multicolumn{4}{c}{(km s$^{-1}$)} & \multicolumn{5}{c}{(km s$^{-1}$)} & \multicolumn{4}{c}{(mK)} & \\
\hline
CH$_{3}^{13}$CN & 91941.5 & 5$_K$-4$_K$, K=0, 1 & \multicolumn{2}{r}{1.7 (0.4)} & \multicolumn{4}{r}{67.7 (1.6)} & \multicolumn{5}{r}{24.8 (3.9)} & \multicolumn{4}{r}{64.9 (8.5)} & hf, m\\
CH$_{3}$CN, v8=0,1 & 91959.2 & 5$_{4}$-4$_{4}$ & \multicolumn{2}{r}{1.9 (0.6)} & \multicolumn{4}{r}{71.7 (1.7)} & \multicolumn{5}{r}{17.4 (3.9)} & \multicolumn{4}{r}{102.2 (19.2)} & hf\\
CH$_{3}$CN, v8=0,1 & 91971.3 & 5$_{3}$-4$_{3}$ & \multicolumn{2}{r}{6.4 (0.8)} & \multicolumn{4}{r}{70.4 (0.8)} & \multicolumn{5}{r}{21.8 (2.0)} & \multicolumn{4}{r}{275.2 (19.4)} & hf\\
CH$_{3}$CN, v8=0,1 & 91980.0 & 5$_{2}$-4$_{2}$ & \multicolumn{2}{r}{6.5 (1.3)} & \multicolumn{4}{r}{69.3 (1.4)} & \multicolumn{5}{r}{16.6$^{a}$(2.7)} & \multicolumn{4}{r}{370.0$^{a}$(44.0)} & hf\\
CH$_{3}$CN, v8=0,1 & 91987.0 & 5$_K$-4$_K$, K=0, 1 & \multicolumn{2}{r}{19.1 (0.8)} & \multicolumn{4}{r}{73.6 (0.4)} & \multicolumn{5}{r}{24.8 (0.9)} & \multicolumn{4}{r}{722.4 (15.8)} & hf, m\\
C$_{2}$H$_{3}$CN & 92426.2 & 10$_{1,10}$-9$_{1,9}$ & \multicolumn{2}{r}{1.2 (0.2)} & \multicolumn{4}{r}{67.9 (0.8)} & \multicolumn{5}{r}{18.3 (1.8)} & \multicolumn{4}{r}{62.0 (4.6)} & \\
$^{13}$CS & 92494.3 & 2-1 & \multicolumn{2}{r}{4.7 (0.5)} & \multicolumn{4}{r}{[45,80]} & \multicolumn{5}{r}{...} & \multicolumn{4}{r}{...} & \\
Unidentified & 92724.8 &  & \multicolumn{2}{r}{2.5 (0.3)} & \multicolumn{4}{r}{66.3 (1.0)} & \multicolumn{5}{r}{29.0 (2.3)} & \multicolumn{4}{r}{81.0 (6.0)} & \\
NH$_{2}$CHO & 105464.2 & 5$_{0,5}$-4$_{0,4}$ & \multicolumn{2}{r}{7.7 (0.3)} & \multicolumn{4}{r}{64.6 (0.3)} & \multicolumn{5}{r}{24.1 (0.8)} & \multicolumn{4}{r}{299.6 (8.9)} & hf\\
HNCS, a-type & 105558.0 & 9$_{0,9}$-8$_{0,8}$ & \multicolumn{2}{r}{1.6 (0.2)} & \multicolumn{4}{r}{64.4 (0.8)} & \multicolumn{5}{r}{21.6 (1.8)} & \multicolumn{4}{r}{68.3 (4.6)} & \\
$^{13}$C$^{15}$N & 105747.7 & 1$_{2,1}$-0$_{1,0}$ & \multicolumn{2}{r}{1.9 (0.2)} & \multicolumn{4}{r}{67.9 (1.0)} & \multicolumn{5}{r}{23.7 (2.3)} & \multicolumn{4}{r}{73.5 (6.2)} & hf, cl\\
CH$_{2}$NH & 105793.9 & 4$_{0,4}$-3$_{1,3}$ & \multicolumn{2}{r}{2.4 (0.4)} & \multicolumn{4}{r}{71.5$^{a}$(1.0)} & \multicolumn{5}{r}{14.4 (1.7)} & \multicolumn{4}{r}{157.0 (18.8)} & hf, bl\\
H$^{13}$CCCN & 105799.0 & 12$_K$-11$_K$, K=12-11, 13-12, 11-10 & \multicolumn{2}{r}{2.7 (0.4)} & \multicolumn{4}{r}{70.5 (1.0)} & \multicolumn{5}{r}{14.4 (1.9)} & \multicolumn{4}{r}{173.2 (7.7)} & hf, bl\\
NH$_{2}$CHO & 105972.6 & 5$_{2,4}$-4$_{2,3}$ & \multicolumn{2}{r}{2.5 (0.2)} & \multicolumn{4}{r}{67 (0.6)} & \multicolumn{5}{r}{21.9 (1.4)} & \multicolumn{4}{r}{105.4 (6.0)} & hf\\
Unidentified & 106302.5 &  & \multicolumn{2}{r}{2.8 (0.8)} & \multicolumn{4}{r}{68.5 (3.0)} & \multicolumn{5}{r}{21.5 (5.6)} & \multicolumn{4}{r}{124.1 (9.1)} & \\
Unidentified & 106311.8 &  & \multicolumn{2}{r}{4.9 (0.5)} & \multicolumn{4}{r}{70.0 (0.8)} & \multicolumn{5}{r}{21.5 (1.9)} & \multicolumn{4}{r}{214.0 (12.0)} & \\
CCS & 106347.7 & 8$_{9}$-7$_{8}$ & \multicolumn{2}{r}{2.2 (0.1)} & \multicolumn{4}{r}{66.9 (0.4)} & \multicolumn{5}{r}{19.6$^{a}$(0.9)} & \multicolumn{4}{r}{104.8 (4.6)} & \\
HC$_{5}$N & 106499.4 & 40-39 & \multicolumn{2}{r}{1.3 (0.4)} & \multicolumn{4}{r}{64.5$^{a}$(3.0)} & \multicolumn{5}{r}{21.4 (5.3)} & \multicolumn{4}{r}{56.5 (9.0)} & \\
NH$_{2}$CHO & 106541.7 & 5$_{2,3}$-4$_{2,2}$ & \multicolumn{2}{r}{2.2 (0.2)} & \multicolumn{4}{r}{68.6 (0.8)} & \multicolumn{5}{r}{21.4 (1.8)} & \multicolumn{4}{r}{95.6 (6.6)} & hf\\
C$_{2}$H$_{3}$CN & 106641.3 & 11$_{1,10}$-10$_{1,9}$ & \multicolumn{2}{r}{1.0 (0.2)} & \multicolumn{4}{r}{74.2 (0.9)} & \multicolumn{5}{r}{16.7 (2.2)} & \multicolumn{4}{r}{56.0 (6.9)} & \\
(CH$_{3}$)$_{2}$O$^{b}$ & 106777.3 & 9$_{1,8}$-8$_{2,7}$ & \multicolumn{2}{r}{0.9 (0.3)} & \multicolumn{4}{r}{72.2 (1.9)} & \multicolumn{5}{r}{14.3 (4.3)} & \multicolumn{4}{r}{60.0 (11.2)} & \\
OC$^{34}$S & 106787.3 & 9-8 & \multicolumn{2}{r}{1.6 (0.2)} & \multicolumn{4}{r}{67.5$^{a}$(0.9)} & \multicolumn{5}{r}{21.4 (2.3)} & \multicolumn{4}{r}{69.3 (4.9)} & \\
HOCO$^+$ & 106913.5 & 5$_{0,5}$-4$_{0,4}$ & \multicolumn{2}{r}{13.6 (0.4)} & \multicolumn{4}{r}{68.0 (0.2)} & \multicolumn{5}{r}{21.1 (0.5)} & \multicolumn{4}{r}{606.0 (11.8)} & \\
CH$_{3}$OH & 107013.8 & 3$_{1,3}$-4$_{0,4}$ A+ & \multicolumn{2}{r}{-2.7 (...)} & \multicolumn{4}{r}{66.1 (3.1)} & \multicolumn{5}{r}{40.0 (7.0)} & \multicolumn{4}{r}{-64.1 (...)} & al\\
C$_{2}$H$_{5}$CN & 107043.5 & 12$_{2,11}$-11$_{2,10}$ & \multicolumn{2}{r}{1.8 (0.3)} & \multicolumn{4}{r}{68.7 (1.3)} & \multicolumn{5}{r}{21.3 (3.2)} & \multicolumn{4}{r}{77.8 (9.1)} & \\
Unidentified & 107100.0 &  & \multicolumn{2}{r}{0.9 (0.2)} & \multicolumn{4}{r}{68.5 (1.3)} & \multicolumn{5}{r}{18.6 (3.1)} & \multicolumn{4}{r}{47.2 (6.5)} & \\
Unidentified & 107134.6 &  & \multicolumn{2}{r}{3.2 (0.2)} & \multicolumn{4}{r}{70.6 (0.5)} & \multicolumn{5}{r}{21.4 (1.1)} & \multicolumn{4}{r}{139.6 (6.2)} & \\
$^{13}$CH$_{3}$CN & 107178.5 & 6$_3$-5$_3$ & \multicolumn{2}{r}{0.5 (0.2)} & \multicolumn{4}{r}{72.4 (1.8)} & \multicolumn{5}{r}{16.4 (4.2)} & \multicolumn{4}{r}{28.7 (7.2)} & hf\\
$^{13}$CH$_{3}$CN & 107196.5 & 6$_K$-5$_K$, K=0, 1 & \multicolumn{2}{r}{1.5 (0.2) } & \multicolumn{4}{r}{74.6 (1.0)} & \multicolumn{5}{r}{21.3 (2.23)} & \multicolumn{4}{r}{66.1 (5.68)} & hf, bl\\
HCOOH & 108126.7 & 5$_{1,5}$-4$_{1,4}$ & \multicolumn{2}{r}{2.2 (0.3)} & \multicolumn{4}{r}{67.5$^{a}$(1.0)} & \multicolumn{5}{r}{21.1$^{a}$(2.5)} & \multicolumn{4}{r}{97.8 (5.2)} & \\
$^{13}$CN & 108636.9 & N=1-0, F1=0, F2=1-0, F=1-1 & \multicolumn{2}{r}{2.5 (0.6)} & \multicolumn{4}{r}{57.6 (4.1)} & \multicolumn{5}{r}{35.0 (7.3)} & \multicolumn{4}{r}{67.2 (6.3)} & hf\\
$^{13}$CN & 108651.2 & N=1-0, F1=0, F2=1-0, F=2-1 & \multicolumn{2}{r}{2.0 (0.4)} & \multicolumn{4}{r}{56.8 (2.3)} & \multicolumn{5}{r}{32.6 (4.8)} & \multicolumn{4}{r}{56.7 (5.6)} & hf\\
HC$^{13}$CCN & 108710.5 & 12$_K$-11$_K$, K=12-11, 13-12, 11-10 & \multicolumn{2}{r}{2.3 (0.4)} & \multicolumn{4}{r}{69.0 (1.9)} & \multicolumn{5}{r}{25.4 (3.8)} & \multicolumn{4}{r}{86.1 (6.5)} & hf\\
HCC$^{13}$CN & 108721.0 & 12$_K$-11$_K$, K=12-11, 13-12, 11-10 & \multicolumn{2}{r}{1.6 (0.3)} & \multicolumn{4}{r}{68.7 (1.7)} & \multicolumn{5}{r}{22.0 (3.8)} & \multicolumn{4}{r}{68.7 (6.4)} & hf\\
$^{13}$CN & 108780.2 & N=1-0, F1=1, F2=2-1, F=3-2 & \multicolumn{2}{r}{1.6 (0.2)} & \multicolumn{4}{r}{69.4 (0.9)} & \multicolumn{5}{r}{21.0 (2.1)} & \multicolumn{4}{r}{71.8 (5.9)} & hf\\
$^{13}$CN & 108786.9 & N=1-0, F1=1, F2=2-1, F=1-0 & \multicolumn{2}{r}{0.7 (0.2)} & \multicolumn{4}{r}{69.8 (1.0)} & \multicolumn{5}{r}{13.3 (2.7)} & \multicolumn{4}{r}{51.2 (6.1)} & hf$^{a}$\\
$^{13}$CN & 108796.4 & N=1-0, F1=1, F2=2-1, F=2-2 & \multicolumn{2}{r}{0.9 (0.2)} & \multicolumn{4}{r}{71.3 (1.0)} & \multicolumn{5}{r}{17.8 (2.4)} & \multicolumn{4}{r}{48.9 (5.9)} & hf, cd\\
CH$_{3}$OH & 108893.9 & 0$_{0,0}$-1$_{-1,1}$ E & \multicolumn{2}{r}{12.1 (0.3)} & \multicolumn{4}{r}{[40,80]}  & \multicolumn{5}{r}{...} & \multicolumn{4}{r}{...} & \\
 &  &                                          & \multicolumn{2}{r}{2.4 (0.3)}  & \multicolumn{4}{r}{[80,110]} & \multicolumn{5}{r}{...} & \multicolumn{4}{r}{...} & \\

O$^{13}$CS & 109110.8 & 9-8 & \multicolumn{2}{r}{0.6 (0.2)} & \multicolumn{4}{r}{66.4 (1.7)} & \multicolumn{5}{r}{17.1 (3.9)} & \multicolumn{4}{r}{32.4 (7.8)} & \\
HC$_{3}$N & 109173.6 & 12-11 & \multicolumn{2}{r}{47.3 (1.2)} & \multicolumn{4}{r}{66.9 (0.2)} & \multicolumn{5}{r}{23.0 (0.4)} & \multicolumn{4}{r}{1932.8 (31.5)} & \\
SO & 109252.2 & 3$_{2}$-2$_{1}$ & \multicolumn{2}{r}{8.2 (0.3)} & \multicolumn{4}{r}{67.6 (0.2)} & \multicolumn{5}{r}{20.3 (0.5)} & \multicolumn{4}{r}{380.4 (7.9)} & \\
OCS & 109463.0 & 9-8 & \multicolumn{2}{r}{19.5 (0.5)} & \multicolumn{4}{r}{65.8 (0.2)} & \multicolumn{5}{r}{22.4 (0.4)} & \multicolumn{4}{r}{819.0 (13.4)} & \\
HNCO & 109496.0 & 5$_{1,5}$-4$_{1,4}$ & \multicolumn{2}{r}{2.8 (0.4)} & \multicolumn{4}{r}{64.3 (1.0)} & \multicolumn{5}{r}{20.7 (2.4)} & \multicolumn{4}{r}{127.1 (12.5)} & hf\\
C$_{2}$H$_{5}$CN & 109650.2 & 12$_{1,11}$-11$_{1,10}$ & \multicolumn{2}{r}{2.0 (0.3)} & \multicolumn{4}{r}{64.6 (1.4)} & \multicolumn{5}{r}{27.9 (3.3)} & \multicolumn{4}{r}{67.3 (7.7)} & \\
NH$_{2}$CHO & 109753.5 & 5$_{1,4}$-4$_{1,3}$ & \multicolumn{2}{r}{5.0 (0.4)} & \multicolumn{4}{r}{66.9 (0.5)} & \multicolumn{5}{r}{21.5 (1.2)} & \multicolumn{4}{r}{220.2 (10.4)} & hf\\
C$^{18}$O & 109782.1 & 1-0 & \multicolumn{2}{r}{23.4 (0.9)} & \multicolumn{4}{r}{66.8 (0.3)} & \multicolumn{5}{r}{25.7 (0.8)} & \multicolumn{4}{r}{855.2 (21.5)} & \\
HNCO & 109905.7 & 5$_{0,5}$-4$_{0,4}$ & \multicolumn{2}{r}{111.9 (2.7)} & \multicolumn{4}{r}{67.6 (0.2)} & \multicolumn{5}{r}{23.8 (0.4)} & \multicolumn{4}{r}{4420.4 (68.9)} & hf\\
$^{13}$CO & 110201.3 & 1-0 & \multicolumn{2}{r}{148.6 (5.5)} & \multicolumn{4}{r}{67.5$^{a}$(0.3)} & \multicolumn{5}{r}{27.6 (0.9)} & \multicolumn{4}{r}{5052.1 (95.0)} & \\
HNCO & 110298.0 & 5$_{1,4}$-4$_{1,3}$ & \multicolumn{2}{r}{3.5 (0.6)} & \multicolumn{4}{r}{63.1 (1.3)} & \multicolumn{5}{r}{20.7 (3.1)} & \multicolumn{4}{r}{157.0$^{a}$(13.9)} & hf\\
CH$_{3}^{13}$CN & 110328.8 & 6$_K$-5$_K$, K=0, 1 & \multicolumn{2}{r}{2.0 (0.3)} & \multicolumn{4}{r}{69.7 (0.8)} & \multicolumn{5}{r}{20.7 (2.0)} & \multicolumn{4}{r}{89.3 (7.8)} & hf, m, cd\\
CH$_{3}$CN, v8=0,1 & 110349.7 & 6$_{4}$-5$_{4}$ & \multicolumn{2}{r}{1.4 (0.6)} & \multicolumn{4}{r}{70.0 (2.2)} & \multicolumn{5}{r}{15.6 (5.3)} & \multicolumn{4}{r}{82.0 (23.0)} & hf\\
CH$_{3}$CN, v8=0,1 & 110364.4 & 6$_{3}$-5$_{3}$ & \multicolumn{2}{r}{7.5 (0.8)} & \multicolumn{4}{r}{68.5 (0.7)} & \multicolumn{5}{r}{20.7 (1.8)} & \multicolumn{4}{r}{338.1 (18.4)} & hf\\
CH$_{3}$CN, v8=0,1 & 110374.9 & 6$_{2}$-5$_{2}$ & \multicolumn{2}{r}{8.6 (2.6)} & \multicolumn{4}{r}{70.5$^{a}$(1.8)} & \multicolumn{5}{r}{20.7 (6.1)} & \multicolumn{4}{r}{390.0$^{a}$(26.9)} & hf\\
CH$_{3}$CN, v8=0,1 & 110383.4 & 6$_K$-5$_K$, K=0, 1 & \multicolumn{2}{r}{25.1 (0.9)} & \multicolumn{4}{r}{74.2 (0.3)} & \multicolumn{5}{r}{27.6 (0.8)} & \multicolumn{4}{r}{853.4 (16.4)} & hf, m\\
CH$_{3}$OCHO$^{c}$ & 110790.5 & 10$_{1,10}$-9$_{1,9}$ A+E & \multicolumn{2}{r}{0.8 (0.2)} & \multicolumn{4}{r}{70.7 (0.7)} & \multicolumn{5}{r}{20.6 (1.8)} & \multicolumn{4}{r}{71.7 (5.4)} & bl\\
CH$_{3}$OCHO$^{c}$ & 111682.1 & 9$_{1,8}$-8$_{1,7}$ A+E & \multicolumn{2}{r}{1.4 (0.6)} & \multicolumn{4}{r}{76.6 (2.1)} & \multicolumn{5}{r}{27.3 (5.0)} & \multicolumn{4}{r}{96.0$^{a}$(7.1)} & bl\\
HCOOH & 111746.7 & 5$_{0,5}$-4$_{0,4}$ & \multicolumn{2}{r}{2.9 (0.5)} & \multicolumn{4}{r}{70.4$^{a}$(1.6)} & \multicolumn{5}{r}{20.4$^{a}$(3.3)} & \multicolumn{4}{r}{134.5 (8.4)} & \\
(CH$_{3}$)$_{2}$O$^{b}$ & 111783.0 & 7$_{0,7}$-6$_{1,6}$ & \multicolumn{2}{r}{3.5 (0.4)} & \multicolumn{4}{r}{65.3 (0.8)} & \multicolumn{5}{r}{20.4 (1.9)} & \multicolumn{4}{r}{159.4 (11.5)} & \\
CH$_{3}$CHO & 112248.7 & 6$_{1,6}$-5$_{1,5}$ A & \multicolumn{2}{r}{3.8 (0.6)} & \multicolumn{4}{r}{61.8 (0.8)} & \multicolumn{5}{r}{31.7 (1.8)} & \multicolumn{4}{r}{226.9 (12.2)} & bl\\
CH$_{3}$CHO & 112254.5 & 6$_{-1,6}$-5$_{-1,5}$ E & \multicolumn{2}{r}{3.8 (0.5)} & \multicolumn{4}{r}{77.3 (0.7)} & \multicolumn{5}{r}{31.7 (1.6)} & \multicolumn{4}{r}{0.2 (10.4)} & bl\\
C$^{17}$O & 112358.9 & 1-0 & \multicolumn{2}{r}{5.0 (0.5)} & \multicolumn{4}{r}{65.7 (0.7)} & \multicolumn{5}{r}{23.6 (1.7)} & \multicolumn{4}{r}{200.7 (11.8)} & \\
Unidentified & 112464.0 &  & \multicolumn{2}{r}{1.9 (1.0)} & \multicolumn{4}{r}{68.6 (4.0)} & \multicolumn{5}{r}{26.2 (8.0)} & \multicolumn{4}{r}{68.8 (20.0)} & \\

C$_{2}$H$_{5}$OH & 112807.1 & 2$_{2,1}$-1$_{1,0}$ & \multicolumn{2}{r}{4.2 (0.4)} & \multicolumn{4}{r}{67.2 (0.9)} & \multicolumn{5}{r}{30.3 (2.4)} & \multicolumn{4}{r}{130.9 (6.2)} & \\
CN & 113123.3 & N=1-0, J=1/2-1/2, F=1/2-1/2 & \multicolumn{2}{r}{3.8 (0.4)} & \multicolumn{4}{r}{70.5$^{a}$(0.5)} & \multicolumn{5}{r}{13.5 (1.1)} & \multicolumn{4}{r}{264.0$^{a}$(14.6)} & hf$^{a}$, ot, cd\\
CN & 113144.1 & N=1-0, J=1/2-1/2, F=1/2-3/2 & \multicolumn{2}{r}{0.8 (0.2)} & \multicolumn{4}{r}{{[}49.5, 79.5{]}} & \multicolumn{5}{c}{...} & \multicolumn{4}{c}{...} & hf$^{a}$\\
CN & 113170.5 & N=1-0, J=1/2-1/2, F=3/2-1/2 & \multicolumn{2}{r}{2.5 (0.2)} & \multicolumn{4}{r}{{[}49.5, 79.5{]}} & \multicolumn{5}{c}{...} & \multicolumn{4}{c}{...} & hf$^{a}$\\
CN & 113191.3 & N=1-0, J=1/2-1/2, F=3/2-3/2 & \multicolumn{2}{r}{1.9 (0.2)} & \multicolumn{4}{r}{{[}49.5, 79.5{]}} & \multicolumn{5}{c}{...} & \multicolumn{4}{c}{...} & hf$^{a}$\\
CN & 113490.9 & N=1-0, J=3/2-1/2, F=5/2-3/2 & \multicolumn{2}{r}{7.0 (0.2)} & \multicolumn{4}{r}{{[}49.5, 79.5{]}} & \multicolumn{5}{c}{...} & \multicolumn{4}{c}{...} & hf\\
CN & 113508.9 & N=1-0, J=3/2-1/2, F=3/2-3/2 & \multicolumn{2}{r}{2.3 (0.2)} & \multicolumn{4}{r}{{[}49.5, 79.5{]}} & \multicolumn{5}{c}{...} & \multicolumn{4}{c}{...} & hf$^{a}$\\

\hline
\end{tabular}
\begin{tablenotes}
\item Notes: (bl) blended line; (m) multitransition line (frequency refers to the main component of the group); (hf) hyperfine 
structure (frequency refers to the main component of the group); (hf$^{a}$) hyperfine component, it is possible to resolve this hyperfine 
component since its frequency is sufficiently far from the frequencies of the other hyperfine components; (ot) transition less affected by 
opacity; (cl) this line is contaminated by the emission from an unknown molecular species; (al) absorption line; (cd) this transition is used to derive the column density (although several 
transitions of this molecule are detected, there is an insufficient dynamical range in $E_u$ to derive the column density by using a RD).
\item  $^a$ Parameter fixed in the Gaussian fit.
\item  $^b$ Substates EE, AA, EA, AE blended, we show just the most intense transition.
\item  $^c$ Frequency refers to species A.
\end{tablenotes}
\end{minipage}
\end{threeparttable}
\end{table*}


\begin{table*}
\scriptsize
\begin{minipage}{200mm}

\caption{Line parameters for \emph{LOS}$-$0.11.}\label{table2}
\centering
\begin{tabular}{cccc||c||c||cr||r||r||rr||r||r||r||r||r||rr||r||r||rc}
\hline
Molecule  & Frequency & Transition & \multicolumn{4}{c}{Area ($\sigma$)} & \multicolumn{4}{c}{V$_{\rm LSR}$ ($\sigma$)} & \multicolumn{7}{c}{$\Delta_{v_{1/2}}$ ($\sigma$)} & \multicolumn{4}{c}{T$^*_a$ ($\sigma$)} & Notes\\
 & (MHz) &  & \multicolumn{4}{c}{(K km s$^{-1}$)} & \multicolumn{4}{c}{(km s$^{-1}$)} & \multicolumn{7}{c}{(km s$^{-1}$)} & \multicolumn{4}{c}{(mK)} & \\
\hline

C$_{2}$H$_{3}$CN & 77633.8  & 8$_{1,7}$-7$_{1,6}$ & \multicolumn{4}{r}{1.8 (0.4)} & \multicolumn{4}{r}{21.9 (1.2)} & \multicolumn{7}{r}{19.6 (2.9)} & \multicolumn{4}{r}{88.0 (12.0)} & \\
CH$_{3}$CHO & 79099.3 & 4$_{1,3}$-3$_{1,2}$ E  & \multicolumn{4}{r}{4.8 (0.5)} & \multicolumn{4}{r}{20.3 (0.7)} & \multicolumn{7}{r}{20.2 (1.6)} & \multicolumn{4}{r}{222.1 (15.6)} & \\
CH$_{3}$CHO & 79150.2 & 4$_{1,3}$-3$_{1,2}$ A & \multicolumn{4}{r}{4.5 (0.8)} & \multicolumn{4}{r}{17.5 (1.2)} & \multicolumn{7}{r}{19.2 (2.7)} & \multicolumn{4}{r}{220.0$^{a}$(23.9)} & \\
H$^{13}$CCCN & 79350.4 & 9$_K$-8$_K$, K=8-7, 9-8, 10-9 & \multicolumn{4}{r}{2.7 (0.5)} & \multicolumn{4}{r}{16.7 (1.1)} & \multicolumn{7}{r}{19.2 (2.8)} & \multicolumn{4}{r}{131.0$^{a}$(11.7)} & hf\\
NH$_{2}$CN & 79449.7 & 4$_{1,4}$-3$_{1,3}$ & \multicolumn{4}{r}{5.9 (0.4)} & \multicolumn{4}{r}{20.1 (0.4)} & \multicolumn{7}{r}{19.2 (1.0)} & \multicolumn{4}{r}{290.0$^{a}$(11.7)} & \\
C$_{2}$H$_{5}$CN & 79677.5 & 9$_{0,9}$-8$_{0,8}$ & \multicolumn{4}{r}{1.8 (0.3)} & \multicolumn{4}{r}{16.2 (1.2)} & \multicolumn{7}{r}{19.1 (2.9)} & \multicolumn{4}{r}{87.0$^{a}$(9.4)} & \\
HC$_{5}$N & 79876.9 & 30-29 & \multicolumn{4}{r}{2.4 (0.3)} & \multicolumn{4}{r}{18.2 (0.8)} & \multicolumn{7}{r}{19.6 (1.8)} & \multicolumn{4}{r}{115.8 (9.4)} & \\
NH$_{2}$CN & 79963.2 & 4$_{2,3}$-3$_{2,2}$ & \multicolumn{4}{r}{1.3 (0.3)} & \multicolumn{4}{r}{17.6 (1.5)} & \multicolumn{7}{r}{19.0 (3.6)} & \multicolumn{4}{r}{65.0$^{a}$(9.3)} & m\\
NH$_{2}$CN & 79979.5 & 4$_{0,4}$-3$_{0,3}$ & \multicolumn{4}{r}{2.5 (0.4)} & \multicolumn{4}{r}{21.2 (1.0)} & \multicolumn{7}{r}{19.0 (2.4)} & \multicolumn{4}{r}{124.0$^{a}$(13.1)} & \\
CH$_{2}$CO & 80076.7 & 4$_{1,4}$-3$_{1,3}$ & \multicolumn{4}{r}{2.2 (0.2)} & \multicolumn{4}{r}{20.7 (0.5)} & \multicolumn{7}{r}{17.2 (1.1)} & \multicolumn{4}{r}{119.6 (6.2)} & \\
Unidentified & 80282.8 &  & \multicolumn{4}{r}{0.9 (0.2)} & \multicolumn{4}{r}{18.5 (1.0)} & \multicolumn{7}{r}{15.8 (2.3)} & \multicolumn{4}{r}{51.3 (6.2)} & \\
Unidentified & 80373.6 &  & \multicolumn{4}{r}{1.5 (0.4)} & \multicolumn{4}{r}{20.0 (2.5)} & \multicolumn{7}{r}{29.1 (5.5)} & \multicolumn{4}{r}{48.2 (8.3)} & \\
C$_{2}$H$_{5}$CN & 80404.9 & 9$_{2,8}$-8$_{2,7}$ & \multicolumn{4}{r}{1.0 (0.3)} & \multicolumn{4}{r}{17.2 (2.3)} & \multicolumn{7}{r}{24.7$^{a}$(5.5)} & \multicolumn{4}{r}{39.2 (8.1)} & \\
H$_{2}$C$_{2}$N & 80480.9 & 4$_{0,4}$-3$_{0,3}$, J=9/2-7/2 & \multicolumn{4}{r}{2.1 (0.4)} & \multicolumn{4}{r}{17.6 (1.2)} & \multicolumn{7}{r}{19.7 (2.9)} & \multicolumn{4}{r}{100.3 (8.1)} & hf$^{a}$, cd\\
H$_{2}$C$_{2}$N & 80489.9 & 4$_{0,4}$-3$_{0,3}$, J=7/2-5/2 & \multicolumn{4}{r}{2.2 (0.4)} & \multicolumn{4}{r}{20.6 (1.0)} & \multicolumn{7}{r}{18.9 (2.5)} & \multicolumn{4}{r}{105.3 (8.6)} & hf$^{a}$\\
NH$_{2}$CN & 80504.6 & 4$_{1,3}$-3$_{1,2}$ & \multicolumn{4}{r}{5.1 (0.6)} & \multicolumn{4}{r}{21.2 (0.6)} & \multicolumn{7}{r}{18.9 (1.5)} & \multicolumn{4}{r}{260.9 (17.7)} & \\
(CH$_{3}$)$_{2}$O$^{b}$ & 80538.5 & 5$_{2,3}$-5$_{1,4}$ & \multicolumn{4}{r}{1.9 (0.4)} & \multicolumn{4}{r}{20.2 (1.9)} & \multicolumn{7}{r}{28.4 (4.4)} & \multicolumn{4}{r}{61.8 (8.0)} & \\
CH$_{2}$CO & 80832.1 & 4$_{0,4}$-3$_{0,3}$    & \multicolumn{4}{r}{1.3 (0.3)} & \multicolumn{4}{r}{21.2 (1.6)} & \multicolumn{7}{r}{18.8 (3.8)} & \multicolumn{4}{r}{53.0$^{a}$(7.1)} & \\
CH$_{3}$OH & 80993.2 & 7$_{2,6}$-8$_{1,7}$ A- & \multicolumn{4}{r}{$\la$0.3}  & \multicolumn{4}{r}{...}        & \multicolumn{7}{r}{...}        & \multicolumn{4}{r}{$\la$33} & cd\\
HNO & 81477.4 & 1$_{0,1}$-0$_{0,0}$ & \multicolumn{4}{r}{1.2 (0.2)} & \multicolumn{4}{r}{17.2$^{a}$(1.2)} & \multicolumn{7}{r}{20.6 (2.7)} & \multicolumn{4}{r}{53.4 (6.4)} & hf\\
CCS & 81505.1 & 7$_{6}$-6$_{5}$ & \multicolumn{4}{r}{3.4 (0.5)} & \multicolumn{4}{r}{20.8 (1.0)} & \multicolumn{7}{r}{19.9 (2.3)} & \multicolumn{4}{r}{159.2 (13.4)} & \\
HC$^{13}$CCN & 81534.1 & 9$_K$-8$_K$, K=8-7, 9-8, 10-9 & \multicolumn{4}{r}{1.8 (0.2)} & \multicolumn{4}{r}{19.0 (0.6)} & \multicolumn{7}{r}{17.5 (1.5)} & \multicolumn{4}{r}{96.3 (4.8)} & hf\\
HCC$^{13}$CN & 81541.9 & 9$_K$-8$_K$, K=8-7, 9-8, 10-9 & \multicolumn{4}{r}{2.1 (0.2)} & \multicolumn{4}{r}{18.4 (0.5)} & \multicolumn{7}{r}{18.4 (1.2)} & \multicolumn{4}{r}{108.5 (5.0)} & hf\\
CH$_{2}$CO & 81586.2 & 4$_{1,3}$-3$_{1,2}$ & \multicolumn{4}{r}{2.4 (0.2)} & \multicolumn{4}{r}{15.9 (0.4)} & \multicolumn{7}{r}{18.7(0.9)} & \multicolumn{4}{r}{120.0$^{a}$(4.52)} & \\
NH$_{2}$CHO & 81693.5 & 4$_{1,4}$-3$_{1,3}$ & \multicolumn{4}{r}{3.0 (0.3)} & \multicolumn{4}{r}{11.0 (0.7)} & \multicolumn{7}{r}{18.1 (1.7)} & \multicolumn{4}{r}{157.3 (10.0)} & hf\\
HC$_{3}$N & 81881.4 & 9-8 & \multicolumn{4}{r}{50.0 (0.6)} & \multicolumn{4}{r}{17.4 (0.1)} & \multicolumn{7}{r}{18.6 (0.2)} & \multicolumn{4}{r}{2527.7 (19.7)} & \\
c-C$_{3}$H$_{2}$ & 82093.5 & 2$_{0,2}$-1$_{1,1}$ & \multicolumn{4}{r}{4.7 (0.3)} & \multicolumn{4}{r}{19.9 (0.3)} & \multicolumn{7}{r}{17.8 (0.8)} & \multicolumn{4}{r}{246.3 (9.1)} & \\
HNCS, a-type & 82101.8 & 7$_{0,7}$-6$_{0,6}$ & \multicolumn{4}{r}{0.9 (0.3)} & \multicolumn{4}{r}{21.2$^{a}$(2.0)} & \multicolumn{7}{r}{18.5 (4.7)} & \multicolumn{4}{r}{45.0$^{a}$(8.3)} & \\
Unidentified & 82198.8 &  & \multicolumn{4}{r}{0.6 (0.2)} & \multicolumn{4}{r}{17.2 (1.8)} & \multicolumn{7}{r}{18.5 (4.3)} & \multicolumn{4}{r}{30.2 (5.5)} & \\
C$_{2}$H$_{5}$CN & 82458.5 & 9$_{1,8}$-8$_{1,7}$ & \multicolumn{4}{r}{1.4 (0.4)} & \multicolumn{4}{r}{23.9 (1.8)} & \multicolumn{7}{r}{18.5 (4.2)} & \multicolumn{4}{r}{69.0$^{a}$(10.3)} & \\
HCOCH$_{2}$OH & 82470.6 & 8$_{0,8}$-7$_{1,7}$ & \multicolumn{4}{r}{$\la$0.7} & \multicolumn{4}{c}{...} & \multicolumn{7}{c}{...} & \multicolumn{4}{r}{$\la$60)} & \\
HC$_{5}$N & 82539.2 & 31-30 & \multicolumn{4}{r}{1.9 (0.4)} & \multicolumn{4}{r}{18.0 (0.6)} & \multicolumn{7}{r}{21.0 (1.6)} & \multicolumn{4}{r}{76.3 (3.7)} & \\
NH$_{2}$CHO & 82549.5 & 1$_{1,1}$-0$_{0,0}$ & \multicolumn{4}{r}{0.4 (0.1)} & \multicolumn{4}{r}{17.2 (1.4)} & \multicolumn{7}{r}{18.4 (3.4)} & \multicolumn{4}{r}{25.1 (4.0)} & hf, cl\\
c-C$_{3}$H$_{2}$ & 82966.2 & 3$_{1,2}$-3$_{0,3}$ & \multicolumn{4}{r}{2.1 (0.1)} & \multicolumn{4}{r}{18.1 (0.4)} & \multicolumn{7}{r}{18.4 (1.0)} & \multicolumn{4}{r}{108.3 (4.8)} & \\
OC$^{34}$S & 83057.9 & 7-6 & \multicolumn{4}{r}{$\la$0.3} & \multicolumn{4}{c}{...} & \multicolumn{7}{c}{...} & \multicolumn{4}{r}{$\la$39} & \\
HNC$^{18}$O & 83191.5 & 4$_{0,4}$-3$_{0,3}$ & \multicolumn{4}{r}{$\la$0.1} & \multicolumn{4}{c}{...} & \multicolumn{7}{c}{...} & \multicolumn{4}{r}{$\la$30} & hf \\
C$_{2}$H$_{3}$CN & 83207.5 & 9$_{1,9}$-8$_{1,8}$ & \multicolumn{4}{r}{0.7 (0.2)} & \multicolumn{4}{r}{18.2$^{a}$(1.2)} & \multicolumn{7}{r}{17.2 (2.8)} & \multicolumn{4}{r}{37.7 (5.2)} & \\
SO$_{2}$ & 83688.0 & 8$_{1,7}$-8$_{0,8}$ & \multicolumn{4}{r}{1.0 (0.2)} & \multicolumn{4}{r}{18.6 (1.0)} & \multicolumn{7}{r}{17.5 (2.4)} & \multicolumn{4}{r}{55.6 (6.5)} & \\
Unidentified & 83900.3 &  & \multicolumn{4}{r}{2.0 (0.2)} & \multicolumn{4}{r}{14.6 (0.5)} & \multicolumn{7}{r}{16.7 (1.1)} & \multicolumn{4}{r}{111.0 (6.2)} & \\
$^{13}$CCH & 84153.3 & N=1-0, F$_{1}$=1-0, F=3/2-1/2 & \multicolumn{4}{r}{1.0 (0.3)} & \multicolumn{4}{r}{27.9 (2.3)} & \multicolumn{7}{r}{27.1 (5.4)} & \multicolumn{4}{r}{34.1 (5.9)} & hf\\
CH$_{3}$CHO & 84219.7 & 2$_{1,2}$-1$_{0,1}$ A & \multicolumn{4}{r}{0.9 (0.6)} & \multicolumn{4}{r}{22.2$^{a}$(3.4)} & \multicolumn{7}{r}{18.1 (...)} & \multicolumn{4}{r}{47.0$^{a}$(6.1)} & \\
CH$_{3}$OH & 84521.1 & 5$_{-1,5}$-4$_{0,4}$ E & \multicolumn{4}{r}{50.0 (2.0)} & \multicolumn{4}{r}{19.0 (0.2)} & \multicolumn{7}{r}{15.0 (0.5)} & \multicolumn{4}{r}{3140.3 (80.2)} & \\
NH$_{2}$CHO & 84542.4 & 4$_{0,4}$-3$_{0,3}$ & \multicolumn{4}{r}{3.9 (0.3)} & \multicolumn{4}{r}{19.3 (0.4)} & \multicolumn{7}{r}{16.4 (1.0)} & \multicolumn{4}{r}{223.7 (8.2)} & hf\\
$^{30}$SiO & 84745.9 & 2-1 & \multicolumn{4}{r}{1.8 (0.2)} & \multicolumn{4}{r}{19.7 (0.5)} & \multicolumn{7}{r}{17.8 (1.1)} & \multicolumn{4}{r}{95.0 (5.2)} & \\
NH$_{2}$CHO & 84807.9 & 4$_{2,3}$-3$_{2,2}$ & \multicolumn{4}{r}{0.5 (0.1)} & \multicolumn{4}{r}{21.2 (0.6)} & \multicolumn{7}{r}{10.4 (1.3)} & \multicolumn{4}{r}{41.2 (4.7)} & hf\\
O$^{13}$CS & 84865.1 & 7-6 & \multicolumn{4}{r}{$\la$0.3} & \multicolumn{4}{c}{...} & \multicolumn{7}{c}{...} & \multicolumn{4}{r}{$\la$36} & \\
C$_{2}$H$_{3}$CN  & 84946.0 & 9$_{0,9}$-8$_{0,8}$ & \multicolumn{4}{r}{1.0 (0.1)} & \multicolumn{4}{r}{18.8 (0.6)} & \multicolumn{7}{r}{15.4 (1.3)} & \multicolumn{4}{r}{60.3 (4.3)} & \\
$^{13}$CH$_{3}$OH & 84970.2 & 8$_{0,8}$-7$_{1,7}$ A+ & \multicolumn{4}{r}{$\la$0.2}   & \multicolumn{4}{c}{...}        & \multicolumn{7}{c}{...}        & \multicolumn{4}{r}{$\la$24} & \\
NH$_{2}$CHO       & 85093.3 & 4$_{2,2}$-3$_{2,1}$    & \multicolumn{4}{r}{1.1 (0.3)}  & \multicolumn{4}{r}{21.8 (1.5)} & \multicolumn{7}{r}{20.0 (3.5)} & \multicolumn{4}{r}{53.1 (7.6)} & hf\\
OCS & 85139.1 & 7-6 & \multicolumn{4}{r}{10.3 (0.4)} & \multicolumn{4}{r}{17.9 (0.2)} & \multicolumn{7}{r}{18.2 (0.6)} & \multicolumn{4}{r}{534.9 (12.9)} & \\
HC$^{18}$O$^+$ & 85162.1 & 1-0 & \multicolumn{4}{r}{1.2 (0.2)} & \multicolumn{4}{r}{23.4 (1.2)} & \multicolumn{7}{r}{20.9 (2.8)} & \multicolumn{4}{r}{55.6 (6.4)} & \\
HC$_{5}$N & 85201.6 & 32-31 & \multicolumn{4}{r}{2.0 (0.4)} & \multicolumn{4}{r}{22.8 (1.2)} & \multicolumn{7}{r}{17.9 (2.9)} & \multicolumn{4}{r}{104.1 (13.2)} & \\
C$^{13}$CH & 85229.3 & N=1-0, F1=2-1 F=5/2-3/2 & \multicolumn{4}{r}{$\la$0.7} & \multicolumn{4}{c}{...} & \multicolumn{7}{c}{...} & \multicolumn{4}{r}{$\la$81} & hf\\
C$_{2}$H$_{5}$OH & 85265.5 & 6$_{0,6}$-5$_{1,5}$ & \multicolumn{4}{r}{3.3 (0.2)} & \multicolumn{4}{r}{21.2 (0.3)} & \multicolumn{7}{r}{18.7$^{a}$(0.7)} & \multicolumn{4}{r}{163.8 (5.7)} & cl\\
H$^{15}$NCO & 85292.1 & 4$_{0,4}$-3$_{0,3}$ & \multicolumn{4}{r}{$\la$0.3} & \multicolumn{4}{c}{...} & \multicolumn{7}{c}{...} & \multicolumn{4}{r}{$\la$27} & \\
c-C$_{3}$H$_{2}$ & 85338.8 & 2$_{1,2}$-1$_{0,1}$ & \multicolumn{4}{r}{0.8 (0.2)} & \multicolumn{4}{r}{19.2$^{a}$(2.1)} & \multicolumn{7}{r}{18.0 (5.0)} & \multicolumn{4}{r}{40.0$^{a}$(6.0)} & \\
HCS$^+$ & 85347.9 & 2-1 & \multicolumn{4}{r}{2.5 (0.6)} & \multicolumn{4}{r}{21.8 (1.7)} & \multicolumn{7}{r}{17.8 (4.2)} & \multicolumn{4}{r}{133.4 (13.3)} & \\
CH$_{3}^{13}$CCH & 85421.8 & 5$_K$-4$_K$, K=0, 1 & \multicolumn{4}{r}{$\la$0.3} & \multicolumn{4}{c}{...} & \multicolumn{7}{c}{...} & \multicolumn{4}{r}{$\la$39} & m\\
CH$_{3}$CCH & 85442.5 & 5$_{3}$-4$_{3}$ & \multicolumn{4}{r}{0.8 (0.1)} & \multicolumn{4}{r}{21.2$^{a}$(7.4)} & \multicolumn{7}{r}{17.8 (...)} & \multicolumn{4}{r}{42.1 (...)} & \\
CH$_{3}$CCH & 85450.7 & 5$_{2}$-4$_{2}$ & \multicolumn{4}{r}{0.9 (0.2)} & \multicolumn{4}{r}{22.2$^{a}$(1.1)} & \multicolumn{7}{r}{17.8 (2.9)} & \multicolumn{4}{r}{47.8 (5.1)} & bl\\
CH$_{3}$CCH & 85457.2 & 5$_K$-4$_K$, K=0, 1 & \multicolumn{4}{r}{3.7 (0.3)} & \multicolumn{4}{r}{18.5 (0.5)} & \multicolumn{7}{r}{17.8 (1.1)} & \multicolumn{4}{r}{196.7 (9.5)} & m\\
HOCO$^+$ & 85531.5 & 4$_{0,4}$-3$_{0,3}$ & \multicolumn{4}{r}{6.7 (0.2)} & \multicolumn{4}{r}{18.5 (0.2)} & \multicolumn{7}{r}{18.4 (0.5)} & \multicolumn{4}{r}{341.1 (8.3)} & \\
$^{29}$SiO & 85759.0 & 2-1 & \multicolumn{4}{r}{4.1 (0.6)} & \multicolumn{4}{r}{21.8 (1.0)} & \multicolumn{7}{r}{21.8 (2.4)} & \multicolumn{4}{r}{177.0$^{a}$(16.2)} & \\
HC$^{15}$N & 86054.9 & 1-0 & \multicolumn{4}{r}{3.6 (0.4)} & \multicolumn{4}{r}{{[}-0.8 47.2{]}} & \multicolumn{7}{c}{...} & \multicolumn{4}{c}{...} & \\
SO & 86093.9 & 2$_{2}$-1$_{1}$ & \multicolumn{4}{r}{3.3 (0.9)} & \multicolumn{4}{r}{19.9 (1.7)} & \multicolumn{7}{r}{18.9 (4.0)} & \multicolumn{4}{r}{162.0$^{a}$(29.0)} & \\
H$^{13}$CN & 86340.1 & 1-0, F=1-1, 2-1, 0-1 & \multicolumn{4}{r}{27.1 (0.3)} & \multicolumn{4}{r}{{[}-0.8 47.2{]}} & \multicolumn{7}{c}{...} & \multicolumn{4}{c}{...} & hf\\
HCO & 86708.3 & 1$_{0,1}$-0$_{0,0}$, J=3/2-1/2, F=1-0 & \multicolumn{4}{r}{$\la$0.8} & \multicolumn{4}{c}{...} & \multicolumn{7}{c}{...} & \multicolumn{4}{r}{$\la$ 96} & hf$^{a}$\\
H$^{13}$CO$^+$ & 86754.2 & 1-0 & \multicolumn{4}{r}{10.1 (0.3)} & \multicolumn{4}{r}{{[}0.2 36.2{]}} & \multicolumn{7}{c}{...} & \multicolumn{4}{c}{...} & \\
SiO & 86846.9 & 2-1 & \multicolumn{4}{r}{33.2 (0.9)} & \multicolumn{4}{r}{18.2$^{a}$(0.2)} & \multicolumn{7}{r}{20.3 (0.4)} & \multicolumn{4}{r}{1537.8 (27.4)} & \\
HN$^{13}$C & 87090.8 & 1-0 & \multicolumn{4}{r}{14.4 (0.6)} & \multicolumn{4}{r}{16.8 (0.3)} & \multicolumn{7}{r}{19.3 (0.6)} & \multicolumn{4}{r}{697.2 (18.2)} & \\
CCH & 87284.1 & N=1-0, J=3/2-1/2, F=1-1 & \multicolumn{4}{r}{3.9 (0.4)} & \multicolumn{4}{r}{19.8 (0.6)} & \multicolumn{7}{r}{21.3 (1.5)} & \multicolumn{4}{r}{173.8 (11.6)} & hf$^{a}$, cd\\
CCH & 87316.9 & N=1-0, J=3/2-1/2, F=2-1 & \multicolumn{4}{r}{19.3 (0.7)} & \multicolumn{4}{r}{18.5 (0.2)} & \multicolumn{7}{r}{18.6 (0.5)} & \multicolumn{4}{r}{971.7 (22.3)} & hf$^{a}$\\
CCH & 87328.6 & N=1-0, J=3/2-1/2, F=1-0 & \multicolumn{4}{r}{7.0 (0.6)} & \multicolumn{4}{r}{18.3 (0.5)} & \multicolumn{7}{r}{15.7 (1.1)} & \multicolumn{4}{r}{419.0 (24.3)} & hf$^{a}$\\
CCH & 87402.0 & N=1-0, J=1/2-1/2, F=1-1 & \multicolumn{4}{r}{16.4 (0.7)} & \multicolumn{4}{r}{15.6 (0.4)} & \multicolumn{7}{r}{26.1 (0.9)} & \multicolumn{4}{r}{590.0$^{a}$(14.4)} & hf$^{a}$, bl\\
CCH & 87407.1 & N=1-0, J=1/2-1/2, F=0-1 & \multicolumn{4}{r}{4.0 (1.9)} & \multicolumn{4}{r}{20.8 (5.8)} &  \multicolumn{7}{r}{22.5 (7.5)} & \multicolumn{4}{r}{164.8 (58.9)} & hf$^{a}$, bl\\
CCH & 87446.5 & N=1-0, J=1/2-1/2, F=1-0 & \multicolumn{4}{r}{2.9 (0.6)} & \multicolumn{4}{r}{18.1 (1.1)} & \multicolumn{7}{r}{17.5 (2.5)} & \multicolumn{4}{r}{157.4 (19.1)} & hf$^{a}$\\
\hline
\end{tabular}
\end{minipage}
\end{table*}

\begin{table*}
\scriptsize
\begin{minipage}{200mm}
\contcaption{}
\centering
\begin{tabular}{cccccccrrrrrr||r||r||r||r||rr||r||r||rc}
\hline
Molecule  & Frequency & Transition & \multicolumn{4}{c}{Area ($\sigma$)} & \multicolumn{4}{c}{V$_{\rm LSR}$ ($\sigma$)} & \multicolumn{7}{c}{$\Delta_{v_{1/2}}$ ($\sigma$)} & \multicolumn{4}{c}{T$^*_a$ ($\sigma$)} & Notes\\
 & (MHz) &  & \multicolumn{4}{c}{(K km s$^{-1}$)} & \multicolumn{4}{c}{(km s$^{-1}$)} & \multicolumn{7}{c}{(km s$^{-1}$)} & \multicolumn{4}{c}{(mK)} & \\
\hline
HNCO & 87597.3 & 4$_{1,4}$-3$_{1,3}$ & \multicolumn{4}{r}{1.4 (0.5)} & \multicolumn{4}{r}{19.2$^{a}$(2.1)} & \multicolumn{7}{r}{17.4 (5.1)} & \multicolumn{4}{r}{75.0$^{a}$(13.4)} & hf\\
NH$_{2}$CHO & 87848.9 & 4$_{1,3}$-3$_{1,2}$ & \multicolumn{4}{r}{4.1 (0.5)} & \multicolumn{4}{r}{17.9 (0.7)} & \multicolumn{7}{r}{18.4 (1.8)} & \multicolumn{4}{r}{211.3 (17.5)} & hf\\
HC$_{5}$N & 87863.9 & 33-32 & \multicolumn{4}{r}{2.3 (0.4)} & \multicolumn{4}{r}{20.0 (0.9)} & \multicolumn{7}{r}{17.3 (2.2)} & \multicolumn{4}{r}{126.6 (12.5)} & \\
HNCO & 87925.2 & 4$_{0,4}$-3$_{0,3}$ & \multicolumn{4}{r}{62.4 (1.1)} & \multicolumn{4}{r}{17.7 (0.1)} & \multicolumn{7}{r}{17.5 (0.2)} & \multicolumn{4}{r}{3358.4 (39.0)} & hf\\
H$^{13}$CCCN & 88166.8 & 10$_K$-9$_K$, K=10-9, 11-10, 9-8 & \multicolumn{4}{r}{2.1 (0.4)} & \multicolumn{4}{r}{23.6 (1.0)} & \multicolumn{7}{r}{17.3 (2.4)} & \multicolumn{4}{r}{113.3 (11.0)} & hf\\
HCN & 88631.8 & 1-0, F=0-1, 1-1, 2-1 & \multicolumn{4}{r}{123.8 (0.2)} & \multicolumn{4}{r}{{[}-0.8 47.2{]}} & \multicolumn{7}{c}{...} & \multicolumn{4}{c}{...} & hf\\
CH$_{3}$OCHO & 88843.2 & 7$_{1,6}$-6$_{1,5}$ E & \multicolumn{4}{r}{1.2 (0.4)} & \multicolumn{4}{r}{17.7 (2.1)} & \multicolumn{7}{r}{17.1 (5.1)} & \multicolumn{4}{r}{65.0$^{a}$(11.3)} & \\
H$^{15}$NC & 88865.7 & 1-0  & \multicolumn{4}{r}{1.1 (0.2)} & \multicolumn{4}{r}{21.5 (1.3)} & \multicolumn{7}{r}{20.2 (3.2)} & \multicolumn{4}{r}{52.1 (7.8)} & \\
HCO$^+$ & 89188.5 & 1-0 & \multicolumn{4}{r}{74.2 (0.2)} & \multicolumn{4}{r}{{[}0.2 36.2{]}} & \multicolumn{7}{c}{...} & \multicolumn{4}{c}{...} & \\
HCCNC & 89419.3 & 9-8 & \multicolumn{4}{r}{$\la$0.3} & \multicolumn{4}{c}{...} & \multicolumn{7}{c}{...} & \multicolumn{4}{r}{$\la$39} & hf \\
HOC$^+$ & 89487.4 & 1-0 & \multicolumn{4}{r}{$\la$0.3} & \multicolumn{4}{c}{...} & \multicolumn{7}{c}{...} & \multicolumn{4}{r}{$\la$39} & \\
HCOOH & 89579.1 & 4$_{0,4}$-3$_{0,3}$ & \multicolumn{4}{r}{1.2 (0.2)} & \multicolumn{4}{r}{18.6 (0.9)} & \multicolumn{7}{r}{18.4 (2.1)} & \multicolumn{4}{r}{60.6 (6.2)} & cd\\
HCOOH & 89861.4 & 4$_{2,3}$-3$_{2,2}$ & \multicolumn{4}{r}{1.7 (0.3)} & \multicolumn{4}{r}{20.3 (2.3)} & \multicolumn{7}{r}{30.0 (4.1)} & \multicolumn{4}{r}{46.1 (6.7)} & \\
C$_{2}$H$_{5}$OH & 90117.6 & 4$_{1,4}$-3$_{0,3}$ & \multicolumn{4}{r}{2.5 (0.2)} & \multicolumn{4}{r}{18.6 (0.5)} & \multicolumn{7}{r}{16.9 (1.1)} & \multicolumn{4}{r}{138.4 (7.8)} & cd\\
CH$_{3}$COOH & 90203.3 & 8$_{*,8}$-7$_{*,7}$ E & \multicolumn{4}{r}{$\la$0.3} & \multicolumn{4}{c}{...} & \multicolumn{7}{c}{...} & \multicolumn{4}{r}{$\la$36} & \\
HC$_{5}$N & 90526.2 & 34-33 & \multicolumn{4}{r}{2.0 (0.2)} & \multicolumn{4}{r}{19.1 (0.7)} & \multicolumn{7}{r}{18.4 (1.6)} & \multicolumn{4}{r}{102.0 (8.1)} & \\
HC$^{13}$CCN & 90593.0 & 10$_K$-9$_K$, K=9-8, 10-9, 11-10 & \multicolumn{4}{r}{1.9 (0.2)} & \multicolumn{4}{r}{20.5 (0.6)} & \multicolumn{7}{r}{16.8 (1.5)} & \multicolumn{4}{r}{103.3 (7.6)} & hf\\
HCC$^{13}$CN & 90601.7 & 10$_K$-9$_K$, K=9-8, 10-9, 11-10 & \multicolumn{4}{r}{0.8 (0.2)} & \multicolumn{4}{r}{18.7 (1.3)} & \multicolumn{7}{r}{15.1 (3.2)} & \multicolumn{4}{r}{52.2 (8.6)} & hf\\
HNC & 90663.5 & 1-0 & \multicolumn{4}{r}{101.3 (3.6)} & \multicolumn{4}{r}{19.1 (0.3)} & \multicolumn{7}{r}{25.7 (0.7)} & \multicolumn{4}{r}{3709.2 (85.1)} & \\
SiS & 90771.5 & 5-4 & \multicolumn{4}{r}{1.4 (0.3)} & \multicolumn{4}{r}{21.2$^{a}$(1.1)} & \multicolumn{7}{r}{16.8 (2.6)} & \multicolumn{4}{r}{78.5 (11.0)} & \\
(CH$_{3}$)$_{2}$O$^{b}$ & 90938.0 & 6$_{0,6}$-5$_{1,5}$ & \multicolumn{4}{r}{2.5 (0.5)} & \multicolumn{4}{r}{17.8 (2.0)} & \multicolumn{7}{r}{27.7 (4.7)} & \multicolumn{4}{r}{84.0$^{a}$(8.8)} & \\
HC$_{3}$N & 90979.0 & 10-9 & \multicolumn{4}{r}{47.4 (0.8)} & \multicolumn{4}{r}{18.3 (0.1)} & \multicolumn{7}{r}{18.2 (0.2)} & \multicolumn{4}{r}{2441.8 (26.1)} & \\
C$_{2}$H$_{5}$CN & 91549.1 & 10$_{1,9}$-9$_{1,8}$ & \multicolumn{4}{r}{1.1 (0.5)} & \multicolumn{4}{r}{20.3 (2.3)} & \multicolumn{7}{r}{16.6 (6.1)} & \multicolumn{4}{r}{64.0$^{a}$(13.1)} & \\
Unidentified & 91848.0 &  & \multicolumn{4}{r}{3.1 (0.4)} & \multicolumn{4}{r}{23.0 (1.3)} & \multicolumn{7}{r}{24.9 (2.9)} & \multicolumn{4}{r}{115.9 (8.0)} & \\
CH$_{3}$CN, v8=0,1 & 91959.2 & 5$_{4}$-4$_{4}$ & \multicolumn{4}{r}{1.0 (0.3)} & \multicolumn{4}{r}{21.6 (1.3)} & \multicolumn{7}{r}{12.3 (3.2)} & \multicolumn{4}{r}{75.9 (16.6)} & hf\\
CH$_{3}$CN, v8=0,1 & 91971.3 & 5$_{3}$-4$_{3}$ & \multicolumn{4}{r}{4.5 (0.4)} & \multicolumn{4}{r}{20.8 (0.4)} & \multicolumn{7}{r}{15.2 (1.0)} & \multicolumn{4}{r}{277.2 (14.6)} & hf\\
CH$_{3}$CN, v8=0,1 & 91980.0 & 5$_{2}$-4$_{2}$ & \multicolumn{4}{r}{4.6 (0.5)} & \multicolumn{4}{r}{20.2$^{a}$(0.5)} & \multicolumn{7}{r}{16.6 (1.2)} & \multicolumn{4}{r}{259.0$^{a}$(16.4)} & hf\\
CH$_{3}$CN, v8=0,1 & 91987.0 & 5$_K$-4$_K$, K=0, 1 & \multicolumn{4}{r}{17.9 (0.5)} & \multicolumn{4}{r}{17.2$^{a}$(0.2)} & \multicolumn{7}{r}{21.1 (0.5)} & \multicolumn{4}{r}{793.5 (13.0)} & hf, m\\
$^{13}$CS & 92494.3 & 2-1 & \multicolumn{4}{r}{5.7 (0.2)} & \multicolumn{4}{r}{18.7 (0.2)} & \multicolumn{7}{r}{21.3 (0.6)} & \multicolumn{4}{r}{251.8 (5.6)} & \\
Unidentified & 92724.8 &  & \multicolumn{4}{r}{0.8 (0.2)} & \multicolumn{4}{r}{17.2 (1.4)} & \multicolumn{7}{r}{16.4 (3.3)} & \multicolumn{4}{r}{46.5 (7.7)} & \\
NH$_{2}$CHO & 105464.2 & 5$_{0,5}$-4$_{0,4}$ & \multicolumn{4}{r}{2.0 (0.4)} & \multicolumn{4}{r}{22.5 (0.8)} & \multicolumn{7}{r}{14.4 (2.0)} & \multicolumn{4}{r}{127.3 (14.5)} & hf\\
CH$_{2}$NH & 105793.9 & 4$_{0,4}$-3$_{1,3}$ & \multicolumn{4}{r}{1.9 (0.6)} & \multicolumn{4}{r}{17.8 (2.6)} & \multicolumn{7}{r}{20.5 (5.9)} & \multicolumn{4}{r}{90.2 (12.7)} & hf, bl\\
H$^{13}$CCCN & 105799.0 & 12$_K$-11$_K$, K=12-11, 13-12, 11-10 & \multicolumn{4}{r}{1.0 (0.5)} & \multicolumn{4}{r}{13.2$^{a}$(3.3)} & \multicolumn{7}{r}{14.4 (6.7)} & \multicolumn{4}{r}{66.0$^{a}$(18.8)} & hf, bl\\
Unidentified & 106273.2 &  & \multicolumn{4}{r}{6.4 (0.6)} & \multicolumn{4}{r}{17.4 (0.7)} & \multicolumn{7}{r}{24.3 (1.6)} & \multicolumn{4}{r}{247.0 (14.2)} & \\
CCS & 106347.7 & 8$_{9}$-7$_{8}$ & \multicolumn{4}{r}{1.9 (0.5)} & \multicolumn{4}{r}{20.1 (1.6)} & \multicolumn{7}{r}{18.7 (4.1)} & \multicolumn{4}{r}{93.5 (12.5)} & \\
HOCO$^+$ & 106913.5 & 5$_{0,5}$-4$_{0,4}$ & \multicolumn{4}{r}{4.7 (0.4)} & \multicolumn{4}{r}{20.7 (0.4)} & \multicolumn{7}{r}{15.3 (1.0)} & \multicolumn{4}{r}{290.4 (17.5)} & \\
CH$_{3}$OH & 107013.8 & 3$_{1,3}$-4$_{0,4}$ A+ & \multicolumn{4}{r}{-2.7 (...)} & \multicolumn{4}{r}{19.4 (2.3)} & \multicolumn{7}{r}{33.8 (5.0)} & \multicolumn{4}{r}{-76.0 (...)} & al\\
$^{13}$CH$_{3}$CN & 107196.5 & 6$_K$-5$_K$, K=0, 1 & \multicolumn{4}{r}{$\la$0.5} & \multicolumn{4}{r}{...} & \multicolumn{7}{r}{...} & \multicolumn{4}{r}{$\la$69} & hf, m\\
$^{13}$CN & 108636.9 & N=1-0, F1=0, F2=1-0, F=1-1 & \multicolumn{4}{r}{2.6 (0.6)} & \multicolumn{4}{r}{21.0$^{a}$(4.0)} & \multicolumn{7}{r}{35.0 (8.0)} & \multicolumn{4}{r}{69.0$^{a}$(5.8)} & hf, cd\\
$^{13}$CN & 108651.2 & N=1-0, F1=0, F2=1-0, F=2-1 & \multicolumn{4}{r}{1.6 (0.4)} & \multicolumn{4}{r}{15.4$^{a}$(1.7)} & \multicolumn{7}{r}{21.0 (4.0)} & \multicolumn{4}{r}{73.0 (10.0)} & hf\\
HC$^{13}$CCN & 108710.5 & 12$_K$-11$_K$, K=12-11, 13-12, 11-10 & \multicolumn{4}{r}{1.5 (0.3)} & \multicolumn{4}{r}{19.1 (1.5)} & \multicolumn{7}{r}{18.2 (3.5)} & \multicolumn{4}{r}{77.9 (9.4)} & hf\\
HCC$^{13}$CN & 108721.0 & 12$_K$-11$_K$, K=12-11, 13-12, 11-10 & \multicolumn{4}{r}{1.3 (0.3)} & \multicolumn{4}{r}{19.1 (1.2)} & \multicolumn{7}{r}{14.0$^{a}$(2.8)} & \multicolumn{4}{r}{86.0$^{a}$(10.5)} & hf\\
$^{13}$CN & 108780.2 & N=1-0, F1=1, F2=2-1, F=3-2 & \multicolumn{4}{r}{2.1 (0.4)} & \multicolumn{4}{r}{14.7 (1.0)} & \multicolumn{7}{r}{20.9 (2.3)} & \multicolumn{4}{r}{98.5 (11.7)} & hf\\
$^{13}$CN & 108786.9 & N=1-0, F1=1, F2=2-1, F=1-0 & \multicolumn{4}{r}{0.7 (0.3)} & \multicolumn{4}{r}{22.2$^{a}$(0.8)} & \multicolumn{7}{r}{7.0 (2.0)} & \multicolumn{4}{r}{98.0 (20.5)} & hf$^{a}$\\
CH$_{3}$OH & 108893.9 & 0$_{0,0}$-1$_{-1,1}$ E & \multicolumn{4}{r}{10.3 (0.3)} & \multicolumn{4}{r}{17.5 (0.2)} & \multicolumn{7}{r}{20.6 (0.5)} & \multicolumn{4}{r}{471.5 (10.0)} & \\
HC$_{3}$N & 109173.6 & 12-11 & \multicolumn{4}{r}{24.3 (0.5)} & \multicolumn{4}{r}{17.5 (0.1)} & \multicolumn{7}{r}{16.5 (0.3)} & \multicolumn{4}{r}{1389.9 (19.0)} & \\
SO & 109252.2 & 3$_{2}$-2$_{1}$ & \multicolumn{4}{r}{3.7 (1.2)} & \multicolumn{4}{r}{21.2 (2.5)} & \multicolumn{7}{r}{17.7 (4.9)} & \multicolumn{4}{r}{196.0$^{a}$(30.9)} & \\
Unidentified & 109353.8 &  & \multicolumn{4}{r}{1.9 (0.3)} & \multicolumn{4}{r}{20.4 (0.9)} & \multicolumn{7}{r}{20.1 (2.2)} & \multicolumn{4}{r}{83.3 (7.2)} & \\
OCS & 109463.0 & 9-8 & \multicolumn{4}{r}{8.9 (0.4)} & \multicolumn{4}{r}{20.0 (0.3)} & \multicolumn{7}{r}{19.1 (0.7)} & \multicolumn{4}{r}{440.5 (12.9)} & \\
HNCO & 109496.0 & 5$_{1,5}$-4$_{1,4}$ & \multicolumn{4}{r}{0.8 (0.5)} & \multicolumn{4}{r}{14.2$^{a}$(2.7)} & \multicolumn{7}{r}{13.9 (7.4)} & \multicolumn{4}{r}{55.0$^{a}$(13.9)} & hf\\
NH$_{2}$CHO & 109753.5 & 5$_{1,4}$-4$_{1,3}$ & \multicolumn{4}{r}{2.2 (0.6)} & \multicolumn{4}{r}{22.0 (1.2)} & \multicolumn{7}{r}{13.9 (2.8)} & \multicolumn{4}{r}{148.7 (26.1)} & hf\\
C$^{18}$O & 109782.1 & 1-0 & \multicolumn{4}{r}{9.5 (0.8)} & \multicolumn{4}{r}{15.1 (0.6)} & \multicolumn{7}{r}{21.7 (1.4)} & \multicolumn{4}{r}{441.8 (22.7)} & bl\\
 &  &  & \multicolumn{4}{r}{2.5 (0.8)} & \multicolumn{4}{r}{55.2 (2.1)} & \multicolumn{7}{r}{21.7 (5.1)} & \multicolumn{4}{r}{108.0 (22.6)} & \\
HNCO & 109905.7 & 5$_{0,5}$-4$_{0,4}$ & \multicolumn{4}{r}{58.3 (1.1)} & \multicolumn{4}{r}{18.2 (0.1)} & \multicolumn{7}{r}{18.4 (0.3)} & \multicolumn{4}{r}{2980.2 (36.1)} & hf\\
$^{13}$CO & 110201.3 & 1-0 & \multicolumn{4}{r}{83.6 (3.1)} & \multicolumn{4}{r}{16.0 (0.3)} & \multicolumn{7}{r}{21.2 (0.6)} & \multicolumn{4}{r}{3701.9 (87.0)} & \\
 &  &  & \multicolumn{4}{r}{27.2 (3.2)} & \multicolumn{4}{r}{50.2$^{a}$(0.8)} & \multicolumn{7}{r}{20.7 (2.0)} & \multicolumn{4}{r}{1231.5 (82.9)} & \\
CH$_{3}^{13}$CN & 110328.8 & 6$_K$-5$_K$, K=0, 1 & \multicolumn{4}{r}{$\la$0.7} & \multicolumn{4}{r}{...} & \multicolumn{7}{r}{...} & \multicolumn{4}{r}{$\la$87} & hf, m\\
CH$_{3}$CN, v8=0,1 & 110349.7 & 6$_{4}$-5$_{4}$ & \multicolumn{4}{r}{1.0 (0.5)} & \multicolumn{4}{r}{18.7 (2.5)} & \multicolumn{7}{r}{15.6 (5.9)} & \multicolumn{4}{r}{57.6 (21.7)} & hf\\
CH$_{3}$CN, v8=0,1 & 110364.4 & 6$_{3}$-5$_{3}$ & \multicolumn{4}{r}{3.8 (0.6)} & \multicolumn{4}{r}{19.2 (0.8)} & \multicolumn{7}{r}{16.9 (1.9)} & \multicolumn{4}{r}{212.7 (20.87)} & hf\\
CH$_{3}$CN, v8=0,1 & 110374.9 & 6$_{2}$-5$_{2}$ & \multicolumn{4}{r}{3.14 (0.6)} & \multicolumn{4}{r}{19.2$^{a}$(0.9)} & \multicolumn{7}{r}{13.8 (2.2)} & \multicolumn{4}{r}{214.0$^{a}$(21.3)} & hf\\
CH$_{3}$CN, v8=0,1 & 110383.4 & 6$_K$-5$_K$, K=0, 1 & \multicolumn{4}{r}{13.1 (0.7)} & \multicolumn{4}{r}{16.7 (0.4)} & \multicolumn{7}{r}{21.3 (1.0)} & \multicolumn{4}{r}{576.8 (18.3)} & hf, m\\
(CH$_{3}$)$_{2}$O$^{b}$ & 111783.0 & 7$_{0,7}$-6$_{1,6}$ & \multicolumn{4}{r}{3.0 (0.9)} & \multicolumn{4}{r}{19.1 (2.8)} & \multicolumn{7}{r}{27.2 (6.6)} & \multicolumn{4}{r}{103.1 (20.4)} & \\
CH$_{3}$CHO & 112248.7 & 6$_{1,6}$-5$_{1,5}$ A & \multicolumn{4}{r}{1.2 (0.6)} & \multicolumn{4}{r}{17.2$^{a}$(2.4)} & \multicolumn{7}{r}{13.6 (4.8)} & \multicolumn{4}{r}{84.0$^{a}$(28.1)} & bl\\
CH$_{3}$CHO & 112254.5 & 6$_{-1,6}$-5$_{-1,5}$ E & \multicolumn{4}{r}{2.0 (0.6)} & \multicolumn{4}{r}{21.2$^{a}$(1.6)} & \multicolumn{7}{r}{13.6 (3.7)} & \multicolumn{4}{r}{139.0$^{a}$(9.3)} & bl\\
C$^{17}$O & 112359.2 & 1-0 & \multicolumn{4}{r}{3.9 (1.1)} & \multicolumn{4}{r}{15.2$^{a}$(3.7)} & \multicolumn{7}{r}{27.1 (6.3)} & \multicolumn{4}{r}{134.0$^{a}$(20.2)} & bl\\
 &  &  & \multicolumn{4}{r}{2.9 (0.5)} & \multicolumn{4}{r}{42.9 (1.4)} & \multicolumn{7}{r}{20.3 (2.7)} & \multicolumn{4}{r}{135.8 (11.5)} & \\
C$_{2}$H$_{5}$OH & 112807.1 & 2$_{2,1}$-1$_{1,0}$ & \multicolumn{4}{r}{1.6 (0.4)} & \multicolumn{4}{r}{17.5 (1.1)} & \multicolumn{7}{r}{13.5 (2.5)} & \multicolumn{4}{r}{114.3 (14.6)} & \\
C$_{2}$H$_{3}$CN & 112840.6 & 12$_{0,12}$-11$_{0,11}$ & \multicolumn{4}{r}{0.7 (0.3)} & \multicolumn{4}{r}{19.7 (1.0)} & \multicolumn{7}{r}{7.5$^{a}$(2.4)} & \multicolumn{4}{r}{91.0$^{a}$(25.1)} & \\
CN & 113123.3 & N=1-0, J=1/2-1/2, F=1/2-1/2 & \multicolumn{4}{r}{5.5 (0.5)} & \multicolumn{4}{r}{13.9 (0.6)} & \multicolumn{7}{r}{20.6 (1.4)} & \multicolumn{4}{r}{248.8 (15.2)}  & hf$^{a}$, ot, cd\\
CN & 113144.1 & N=1-0, J=1/2-1/2, F=1/2-3/2 & \multicolumn{4}{r}{6.2 (0.5)} & \multicolumn{4}{r}{15.51 (0.6)} & \multicolumn{7}{r}{23.7 (1.5)} & \multicolumn{4}{r}{247.7 (14.6)} & hf$^{a}$ \\
CN & 113170.5 & N=1-0, J=1/2-1/2, F=3/2-1/2 & \multicolumn{4}{r}{11.4 (0.2)} & \multicolumn{4}{r}{{[}-0.2, 32.2{]}} & \multicolumn{7}{c}{...} & \multicolumn{4}{c}{...}   & hf$^{a}$ \\
\hline
\end{tabular}
\end{minipage}
\end{table*}

\begin{table*}
\begin{threeparttable}
\scriptsize
\begin{minipage}{200mm}
\contcaption{}
\centering
\begin{tabular}{cccc||c||c||cr||r||r||rr||r||r||r||r||r||rr||r||r||rc}
\hline
Molecule  & Frequency & Transition & \multicolumn{4}{c}{Area ($\sigma$)} & \multicolumn{4}{c}{V$_{\rm LSR}$ ($\sigma$)} & \multicolumn{7}{c}{$\Delta_{v_{1/2}}$ ($\sigma$)} & \multicolumn{4}{c}{T$^*_a$ ($\sigma$)} & Notes\\
 & (MHz) &  & \multicolumn{4}{c}{(K km s$^{-1}$)} & \multicolumn{4}{c}{(km s$^{-1}$)} & \multicolumn{7}{c}{(km s$^{-1}$)} & \multicolumn{4}{c}{(mK)} & \\
\hline

CN & 113191.3 & N=1-0, J=1/2-1/2, F=3/2-3/2 & \multicolumn{4}{r}{6.5 (0.2)} & \multicolumn{4}{r}{{[}-0.2, 32.2{]}} & \multicolumn{7}{c}{...} & \multicolumn{4}{c}{...}    & hf$^{a}$ \\
CN & 113490.9 & N=1-0, J=3/2-1/2, F=5/2-3/2 & \multicolumn{4}{r}{22.4 (0.4)} & \multicolumn{4}{r}{{[}-0.2, 32.2{]}} & \multicolumn{7}{c}{...} & \multicolumn{4}{c}{...}   & hf       \\
CN & 113508.9 & N=1-0, J=3/2-1/2, F=3/2-3/2 & \multicolumn{4}{r}{7.9 (0.4)} & \multicolumn{4}{r}{{[}-0.2, 32.2{]}} & \multicolumn{7}{c}{...} & \multicolumn{4}{c}{...}    & hf$^{a}$ \\

\hline
\end{tabular}
\begin{tablenotes}
\item Notes: (bl) blended line; (m) multitransition line (frequency refers to the main component of the group); (hf) hyperfine 
structure (frequency refers to the main component of the group); (hf$^{a}$) hyperfine component, it is possible to resolve this hyperfine 
component since its frequency is sufficiently far from the frequencies of the other hyperfine components; (ot) transition less affected by 
opacity; (cl) this line is contaminated by the emission from an unknown molecular species; (al) absorption line; (cd) this transition is used to derive the column 
density (although several transitions of this molecule are detected, there is an insufficient dynamical range in $E_u$ to derive the column density by using a RD).
\item  $^a$ Parameter fixed in the Gaussian fit.
\item  $^b$ Substates EE, AA, EA, AE blended, we show just the most intense transition.
\item  $^c$ Frequency refers to species A.
\end{tablenotes}
\end{minipage}
\end{threeparttable}
\end{table*}


\begin{table*}
\scriptsize
\begin{minipage}{200mm}
\caption{T$_{rot}$, column densities and abundances for both \emph{LOSs}.}\label{table3}
\centering
\begin{tabular}{crrrrcrrccrr}
\hline
Molecule  & \multicolumn{7}{c}{\emph{LOS}+0.693} & \multicolumn{4}{c}{\emph{LOS}$-$0.11}\\
 & \multicolumn{4}{r}{V$_{\rm LSR}$ ($\sigma$)} & T$^a_{rot}$ ($\sigma$) & \multicolumn{1}{c}{N ($\sigma$)} & N/N$_{\rm H_2}$ ($\sigma$) &
 V$_{\rm LSR}$ ($\sigma$) & T$^a_{rot}$ ($\sigma$) & \multicolumn{1}{c}{N ($\sigma$)} & N/N$_{\rm H_2}$ ($\sigma$) \\
 & \multicolumn{4}{c}{(km s$^{-1}$)} & \multicolumn{1}{c}{(K)} & \multicolumn{1}{c}{($\times$10$^{13}$ cm$^{-2}$)} & 
 ($\times$10$^{-9}$) & (km s$^{-1}$) & (K) & \multicolumn{1}{c}{($\times$10$^{13}$ cm$^{-2}$)} & ($\times$10$^{-9}$)\\
\hline
$^{13}$CO      & \multicolumn{4}{c}{...} & ...  & 28100.0 (644.0)$^b$ & 4810.0 (112.0)       & ... & ...  & 11500.0 (953.0)$^b$ & 4790.0 (566.0)\\
               & \multicolumn{4}{c}{...} & ...  & \multicolumn{1}{r}{...} & \multicolumn{1}{r}{...} & ... & ...  &  3010.0 (931.0)$^b$ & 1260.0 (404.0)\\
C$^{18}$O      & \multicolumn{4}{r}{66.8 (0.3)} & ... & 2344.2 (91.2)$^c$ & 400.7 (22.0)   & 15.1 (0.6) & ... & 955.0 (79.4)$^c$ & 400.0 (47.0)\\
               & \multicolumn{4}{c}{...} & ... & \multicolumn{1}{r}{...} & \multicolumn{1}{r}{...} & 55.2 (2.1) & ... & 251.2 (77.6)$^c$   & 105.2 (33.6)\\
C$^{17}$O      & \multicolumn{4}{r}{65.7 (0.7)} & ... & \multicolumn{1}{r}{489.8 (45.7)$^c$} & 83.6 (8.5) & 15.2 (3.7) & ... & 371.5 (102.3)$^c$ & 155.6 (44.7)\\
               & \multicolumn{4}{c}{...} & ... & \multicolumn{1}{r}{...} & \multicolumn{1}{r}{...} & 42.9 (1.4) & ... & 281.8 (44.7)$^c$ & 118.1 (21.1)\\
CN             & \multicolumn{4}{r}{70.5 (0.5)}        & ... & 169.8 (17.4)$^d$ & 29.0 (3.2) & 13.9 (0.6) & ... & 173.8 (19.5)$^d$ & 72.5 (10.1)\\
$^{13}$CN      & \multicolumn{4}{r}{71.3 (1.0)} & ... &   7.4 (1.3)$^d$  &  1.1 (0.2) & 21.0 (4.0) & ... & 7.9 (2.0)$^d$ & 3.3 (0.8)\\
$^{13}$C$^{15}$N$^e$ & \multicolumn{4}{r}{67.9 (1.0)} & ... & 67.6 (5.6)$^c$ & 11.5 (1.1) & ... & ... & \multicolumn{1}{r}{...} & \multicolumn{1}{r}{...}\\
$^{13}$CS      & \multicolumn{4}{r}{[45,80]$^g$} & ... & 2.3 (0.2)$^c$ & 0.4 (0.04) & 18.7 (0.2) & ... & 2.8 (0.1)$^c$ & 1.2 (0.1)\\
SiO            & \multicolumn{4}{c}{...}   & ... & 9.0 (1.8)$^b$    & 1.4 (0.2) & 18.2 (0.2) & ... & 7.2 (0.2)$^c$ & 3.0 (0.3)\\
$^{29}$SiO     & \multicolumn{4}{r}{65.6 (0.6)} & ... & 0.5 (0.1)$^c$ & 0.09 (0.009) & 21.8 (1.0) & ... & 0.9 (0.1)$^c$  & 0.4 (0.1)\\
$^{30}$SiO     & \multicolumn{4}{r}{71.8 (0.6)} & ... & 0.5 (0.1)$^c$ & 0.08 (0.009) & 19.7 (0.5) & ... & 0.4 (0.03)$^c$ & 0.2 (0.02)\\
SiS            & \multicolumn{4}{r}{69.5 (3.7)}   & ...      & 4.5 (1.2)$^c$   & 0.8 (0.2) & 21.2 (1.1) & ... & 1.7 (0.4)$^c$ & 0.7 (0.2)\\
SO             & \multicolumn{4}{r}{$\sim$68$^f$} & 8.1 (2.4) & 61.0 (51.3) & 10.4 (8.8) & $\sim$21$^f$ & 14.2 (8) & 18.6 (3.7) & 7.8 (2.0)\\
CCH            & \multicolumn{4}{r}{65.4 (0.4)} & ... & 316.2 (18.6)$^d$ & 54.1 (3.8) & 19.8 (0.6) & ... & 251.2 (24)$^d$ & 105.0 (13.3)\\
$^{13}$CCH     & \multicolumn{4}{r}{67.6 (2.4)} & ... &  16.6 (3.5)$^d$  &  2.8 (0.6) & 27.9 (2.3) & ... & 11.1 (3.0)$^c$ & 4.8 (1.3)\\
C$^{13}$CH     & \multicolumn{4}{r}{60.6 (1.5)} & ... &  14.8 (2.6)$^c$  &  2.5 (0.5) & ... & ... & $\la$8.7 & $\la$3.6\\
CCS            & \multicolumn{4}{r}{$\sim$67$^f$} & 7.3 (0.5) & 5.6 (1.1) & 1.0 (0.2) & $\sim$21$^f$ & 8.6 (2.3) & 3.6 (2.3) & 1.5 (1.0)\\
HCN            & \multicolumn{4}{c}{...}  & ... & $\ga$240$^b$ & $\ga$36  & ... & ... & $\ga$480$^b$ & $\ga$180\\
               & \multicolumn{4}{c}{...} & ... & $\ga$120$^b$ & $\ga$12  & ... & ... & \multicolumn{1}{r}{...} & \multicolumn{1}{r}{...}\\
H$^{13}$CN     & \multicolumn{4}{c}{...} & ...  & $\ga$11$^b$ & $\ga$2 & ... & ... & $\ga$23$^b$ & $\ga$9\\
               & \multicolumn{4}{c}{...} & ... & $\ga$6$^b$ &  $\ga$1 & ... & ... & \multicolumn{1}{r}{...} & \multicolumn{1}{r}{...}\\
HC$^{15}$N     & \multicolumn{4}{r}{[40, 80]$^g$} & ...  & 0.4 (0.04)$^c$ & 0.06 (0.01) & [-0.8, 47.2]$^g$ & ... & 0.8 (0.1)$^c$ & 0.3 (0.04)\\
               & \multicolumn{4}{r}{[80, 110]$^g$} & ... & 0.2 (0.02)$^c$ & 0.02 (0.01) & ... & ... & \multicolumn{1}{r}{...} & \multicolumn{1}{r}{...}\\
HCO            & \multicolumn{4}{r}{68.5 (1.3)} & ... & 6.9 (1.4)$^d$ & 1.2 (0.3) & ... & ... & $\la$3.3 & $\la$1.4\\
HCS+           & \multicolumn{4}{r}{70.4 (0.7)} & ... & 2.9 (0.3)$^c$ & 0.5 (0.1) & 21.8 (1.7) & ... & 1.4 (0.4)$^c$ & 0.6 (0.2)\\
HNC            & \multicolumn{4}{c}{...}  & ... & $\ga$240$^b$ & $\ga$42 & ... & ... & $\ga$180$^b$ & $\ga$60 \\
HN$^{13}$C     & \multicolumn{4}{c}{...}  & ... & $\ga$11$^b$ & $\ga$2 & ... & ... & $\ga$9$^b$ & $\ga$3\\
H$^{15}$NC     & \multicolumn{4}{r}{[40, 80]$^g$}  & ... & 0.4 (0.05)$^c$ & 0.07 (0.01) & 21.5 (1.3) & ... & 0.3 (0.1)$^c$ & 0.1 (0.01)\\
HNO            & \multicolumn{4}{r}{67.5 (0.8)} & ... & 3.2 (0.4)$^c$ & 0.6 (0.08) & 17.2 (1.2) & ... & 2.5 (0.5)$^c$ & 1.0 (0.2)\\
HOC$^+$        & \multicolumn{4}{r}{68.0 (0.8)} & ... & 0.1 (0.01)$^c$ & 0.02 (0.003) & ... & ... & $\la$0.03 & $\la$0.01\\
HCO$^+$, v=0,1,2 & \multicolumn{4}{c}{...}      & ...  & 57.0 (1.3)$^b$ & 9.6 (2.2)  & ... & ... & 36.0 (0.8)$^b$ & 15.5 (3.0)\\
H$^{13}$CO$^+$ & \multicolumn{4}{c}{...}        & ...  &  2.7 (0.1)$^b$ & 0.5 (0.1)  & ... & ... & 1.7 (0.04)$^b$ & 0.7 (0.1)\\
HC$^{18}$O$^+$ & \multicolumn{4}{r}{74.5 (2.8)} & ... & 0.2 (0.1)$^c$ & 0.04 (0.01)  & 23.4 (1.2) & ... & 0.2 (0.03)$^c$ & 0.06 (0.01)\\
OCS            & \multicolumn{4}{r}{$\sim$65$^f$} & 17.6 (1.1) & 186.2 (4.6)      & 31.8 (1.5) & $\sim$19$^f$ & 15.3 (1.4) & 87.1 (14.1) & 36.4 (6.5)\\
OC$^{34}$S     & \multicolumn{4}{r}{$\sim$66$^f$} & 12.1 (2.9) & 21.4 (3.1)&  3.7 (0.6) & ... & ... & \multicolumn{1}{r}{$\la$4.0} & \multicolumn{1}{r}{$\la$1.7}\\
O$^{13}$CS     & \multicolumn{4}{r}{$\sim$66$^f$} & 11.3 (4.7) & 7.8 (1.2) &  1.3 (0.2) & ... & ... & \multicolumn{1}{r}{$\la$3.5} & \multicolumn{1}{r}{$\la$1.5}\\
O$^{13}$C$^{34}$S & \multicolumn{4}{c}{...}   & ... & $\la$3.0 & $\la$0.5 & ... & ... & \multicolumn{1}{r}{...} & \multicolumn{1}{r}{...}\\
SO$_{2}$       & \multicolumn{4}{r}{65.0 (0.4)} & ... & 83.2 (3.4)$^c$ & 14.2 (1.0) & 18.6 (1.0) & ... & 20.4 (3.7)$^c$ & 8.6 (1.7)\\
HNCO           & \multicolumn{4}{r}{$\sim$66$^f$} & 11.2 (0.2) & 144.5 (5.6) & 24.7 (1.4)& $\sim$18$^f$ & 11.5 (0.5) & 67.6 (4.5) & 28.3 (3.0)\\
H$^{15}$NCO    & \multicolumn{4}{c}{...}   & ...       & $\la$0.4        & $\la$0.1  & ... & ... & $\la$0.3 & $\la$0.1\\
HNC$^{18}$O    & \multicolumn{4}{c}{...}   & ...       & $\la$1.1        & $\la$0.2  & ... & ... & $\la$0.9 & $\la$0.4\\
HNCS, a-type   & \multicolumn{4}{r}{$\sim$66$^f$} & 15.2 (5.6) & 3.1 (2.2) & 0.5 (0.4) & 21.2 (2.0) & ... & 2.0 (0.6)$^c$ & 0.8 (0.3)\\
HOCO$^+$          & \multicolumn{4}{r}{$\sim$66$^f$} & 7.5 (0.4) & 12.0 (1.4) & 2.1 (0.3) & $\sim$20$^f$ & 6.5 (0.8) & 4.8 (1.2) & 2.0 (0.6)\\
CH$_{2}$CO     & \multicolumn{4}{r}{$\sim$66$^f$} & 29.2 (6.8) & 22.9 (6.8) & 3.9 (1.2) & $\sim$19$^f$ & 30.6 (14.8) & 14.8 (12.3) & 6.2 (5.2)\\
CH$_{2}$NH     & \multicolumn{4}{r}{71.5 (1.0)} & ... & 29.5 (4.8)$^c$ & 5.0 (0.8) & 17.8 (2.6) & ... & 24.0 (7.8)$^c$ & 10.1 (3.4)\\
HC$_{3}$N      & \multicolumn{4}{r}{$\sim$70$^f$} & 15.4 (0.6) & 54.9 (3.3)$^b$ & 9.4 (1.0) & $\sim$18$^f$ & 11.2 (0.2) & 26.8 (0.6) & 11.2 (1.0)\\
H$^{13}$CCCN   & \multicolumn{4}{r}{$\sim$70$^f$} & 14.4 (1.6) &  2.6 (0.2) & 0.5 (0.03) & $\sim$18$^f$ & 9.0 (3.0)  & 1.9 (0.3)  & 0.8 (0.2)\\
HC$^{13}$CCN   & \multicolumn{4}{r}{$\sim$66$^f$} & 12.0 (1.9) &  2.0 (0.2) & 0.4 (0.04) & $\sim$20$^f$ & 19.5 (6.1) & 1.3 (0.4)  & 0.5 (0.2)\\
HCC$^{13}$CN   & \multicolumn{4}{r}{$\sim$68$^f$} & 9.4 (1.2)  &  2.7 (0.5) & 0.5 (0.10) & $\sim$19$^f$ & 11.5 (2.2) & 1.4 (0.3)  & 0.6 (0.1)\\
HCCNC              & \multicolumn{4}{r}{70.1 (3.1)} & ... & 0.7 (0.3)$^c$ & 0.1 (0.05) & ... & ... & $\la$0.9 & $\la$0.4\\
HCOOH              & \multicolumn{4}{r}{$\sim$67$^f$} & 18.9 (4.8) & 14.4 (6.2) & 2.5 (1.1) & 18.6 (0.9) & ... & 15.9 (4.2)$^d$ & 6.6 (1.8)\\
H$_{2}$C$_{2}$N    & \multicolumn{4}{r}{70.4 (0.7)} & ...       & 11.0 (1.4)$^d$ & 1.9 (0.3) & 17.6 (1.2) & ...              & 4.8 (1.2)$^d$ & 2.0 (0.5)\\
NH$_{2}$CN         & \multicolumn{4}{r}{$\sim$67$^f$} & 47.6 (3.1) & 11.9 (1.5) & 2.0 (0.3) & $\sim$20$^f$ & 53.6 (10.3)        & 8.3 (3.0) & 3.5 (1.3)\\
c-C$_{3}$H$_{2}$   & \multicolumn{4}{r}{$\sim$69$^f$} & 13.1 (1.9) & 9.4 (2.5) & 1.6 (0.4) & $\sim$20$^f$ & 6.6 (0.4)     & 6.0 (0.7) & 2.5 (0.4)\\
CH$_{3}$CN, v8=0,1 & \multicolumn{4}{r}{69.7 (0.8)}   & 65.2 (6.5)  & 28.0 (3.0)$^b$ & 4.8 (0.6) & $\sim$20$^f$ & 63.5 (8.7) & 7.4 (1.6) & 3.1 (0.7)\\
$^{13}$CH$_{3}$CN  & \multicolumn{4}{r}{$\sim$74$^f$} & 73.3 (32.2) & 1.1 (0.7) & 0.2 (0.1) & ... & 63.5 & $\la$1.5 & $\la$0.6\\
CH$_{3}^{13}$CN    & \multicolumn{4}{r}{69.7 (0.8)} & 73.3 & 1.4 (0.2)$^d$            & 0.2 (0.03) & ... & 63.5 & $\la$1.9 & $\la$0.8\\
CH$_{3}$OH         & \multicolumn{4}{c}{...}          & ...       & 12700.0 (1420.0)$^b$ & 2160.0 (257.0) & ...  & 13.0 & $\la$3981.0 & $\la$1670.0 \\
$^{13}$CH$_{3}$OH  & \multicolumn{4}{r}{70.4 (1.9) }  & 14.0      &   602.6   (67.6)$^c$ & 103.0 (12.2)   & ... & 13.0 & $\la$758.6 & $\la$317.7\\
NH$_{2}$CHO        & \multicolumn{4}{r}{$\sim$68$^f$} & 12.3 (0.6) & 5.9 (0.6) & 1.0 (0.1) & $\sim$19$^f$ & 8.7 (0.8) & 2.5 (0.5) & 1.0 (0.2)\\
C$_{2}$H$_{3}$CN   & \multicolumn{4}{r}{$\sim$70$^f$} & 9.1 (0.9) & 5.1 (1.6) & 0.9 (0.3) & $\sim$20$^f$ & 11.5 (3.2) & 2.0 (1.3) & 0.8 (0.5)\\
CH$_{3}$CCH        & \multicolumn{4}{r}{$\sim$68$^f$} & 61.1 (10.9) & 262.4 (16.9) & 44.9 (3.4) & $\sim$21$^f$ & 54.8 (5.5) & 74.1 (12.3) & 31.0 (5.8)\\
CH$_{3}^{13}$CCH   & \multicolumn{4}{r}{80.9 (2.2)} & 61.1 & 12.6 (3.9)$^d$ & 2.2 (0.7) & ... & 54.8 & $\la$12.8 & $\la$5.4\\
CH$_{3}$CHO-A      & \multicolumn{4}{r}{$\sim$69$^f$} & 5.4 (0.4) & 49.0 (10.2) & 8.4 (1.8) & $\sim$17$^f$ & 4.5 (1.0) & 12.0 (9.1) & 5.0 (3.8)\\
CH$_{3}$CHO-E      & \multicolumn{4}{r}{$\sim$72$^f$} & 6.5 (0.6) & 15.9 (5.0)  & 2.7 (0.9) & $\sim$21$^f$ & 6.2 (1.0) & 20.9 (8.5) & 8.8 (3.6)\\
HC$_{5}$N          & \multicolumn{4}{r}{$\sim$66$^f$} & 32.1 (4.7) & 3.5 (1.2) & 0.6 (0.2) & $\sim$20$^f$ & 44.5 (19.1) & 2.2 (1.8)     & 0.9 (0.8)\\
CH$_{3}$COOH-E     & \multicolumn{4}{c}{...} & ... & $\la$4.0 & $\la$0.7 & ... & ... & $\la$4.3 & $\la$1.8\\
CH$_{3}$OCHO-A     & \multicolumn{4}{r}{$\sim$69$^f$} & 11.5 (1.0) & 34.7 (7.4) & 5.9 (1.3) & ... & ...                & $\la$19.8 & $\la$8.3\\
CH$_{3}$OCHO-E     & \multicolumn{4}{r}{$\sim$70$^f$} & 11.9 (1.1) & 36.3 (7.4) & 6.2 (1.3) & 17.7 (2.1) & ... & 23.4 (10)$^c$ & 9.8 (4.2)\\
\hline
\end{tabular}
\end{minipage}
\end{table*}

\begin{table*}
\begin{threeparttable}
\scriptsize
\begin{minipage}{200mm}
\centering
\begin{tabular}{crrrrcrrccrr}
\hline
Molecule  & \multicolumn{7}{c}{\emph{LOS}+0.693} & \multicolumn{4}{c}{\emph{LOS}$-$0.11}\\
 & \multicolumn{4}{r}{V$_{\rm LSR}$ ($\sigma$)} & T$^a_{rot}$ ($\sigma$) & \multicolumn{1}{c}{N ($\sigma$)} & N/N$_{\rm H_2}$ ($\sigma$) &
 V$_{\rm LSR}$ ($\sigma$) & T$^a_{rot}$ ($\sigma$) & \multicolumn{1}{c}{N ($\sigma$)} & N/N$_{\rm H_2}$ ($\sigma$) \\
 & \multicolumn{4}{c}{(km s$^{-1}$)} & \multicolumn{1}{c}{(K)} & \multicolumn{1}{c}{($\times$10$^{13}$ cm$^{-2}$)} & 
 ($\times$10$^{-9}$) & (km s$^{-1}$) & (K) & \multicolumn{1}{c}{($\times$10$^{13}$ cm$^{-2}$)} & ($\times$10$^{-9}$)\\
\hline
HCOCH$_{2}$OH      & \multicolumn{4}{r}{73.8 (0.9)} & ... & 6.5 (5.6)$^c$ & 1.1 (1.0) & ...     & ... & $\la$4.8 & $\la$2.0\\
C$_{2}$H$_{5}$CN   & \multicolumn{4}{r}{$\sim$67$^f$} & 13.0 (1.0) & 8.1 (1.6) & 1.4 (0.3) & $\sim$18$^f$ & 8.7 (3.6) & 4.8 (3.2) & 2.0 (1.4)\\
(CH$_{3}$)$_{2}$O  & \multicolumn{4}{r}{$\sim$72$^f$} & 20.3 (1.2) & 91.2 (10.0) & 15.5 (1.6) & $\sim$19$^f$ & 15.8 (7.6) & 63.1 (50.1) & 26.4 (21.1)\\
C$_{2}$H$_{5}$OH   & \multicolumn{4}{r}{$\sim$68$^f$} & 5.3 (0.4)  & 44.7  (8.7) &  7.6 (1.5) &   18.6 (0.5) & ...       & 25.1  (2.5)$^d$ & 10.5 (1.4)\\
\hline
\end{tabular}
\begin{tablenotes}
  \item $^a$  T$_{rot}$ derived from RDs or assumed for deriving molecular column densities. The assumed T$_{rot}$ 
  for the $^{13}$C isotopologues of CH$_3$CCH and CH$_3$CN are taken from their other isotopologues. The assumed T$_{rot}$ for the CH$_3$OH and its $^{13}$C 
  isotopologue are taken from \citet{miguel08}. The T$_{rot}$ quoted with uncertainties are determined from RDs. When the T$_{rot}$ is not listed means
  that a T$_{rot}$ of 10 K is assumed, which corresponds to an average value of the low T$_{rot}$ component derived from other molecules by using RDs.
  \item $^b$ We have inferred from the $^{12}$C/$^{13}$C$\la$15, $^{14}$N/$^{15}$N$\la$280 and $^{16}$O/$^{18}$O$\la$186 isotopic ratios given in Table \ref{table6} that the 
  column density of the most abundant isotopologues of this molecule are
  biased by opacity/self-absorption. Thus here we have derived the column density by using either its $^{18}$O, $^{15}$N or $^{13}$C isotopologue for its
  respective velocity component and 
  assuming the $^{16}$O/$^{18}$O=250 or $^{14}$N/$^{15}$N$>$600 ratios (W\&R94) and if necessary our $^{12}$C/$^{13}$C=21 ratio. For \emph{LOS}+0.693 the
  SiO column density is derived from the $^{28}$SiO isotopologue assuming a $^{28}$Si/$^{30}$Si ratio of 18 derived in \emph{LOS}$-$0.11.
  \item $^c$ Only one line of this molecule is detected.
  \item $^d$ Although several transitions of this molecule are detected, there is an insufficient dynamical range in $E_u$ in order to derive the 
  column density from a RD, then we have chosen one transition (see note cd in Tables \ref{table1} and \ref{table2}), usually the less affected by opacity.
  \item $^e$ The observed transition is contaminated by the emission from an unknown molecular species.
  \item $^f$ This velocity is an average of different detected transitions.
  \item $^g$ These velocity ranges are chosen for deriving velocity-integrated intensities used in the molecular column density estimate.
             For \emph{LOS}+0.693 we have used a velocity range for the $^{13}$CS(2-1) line as it is affected by opacity/self-absorption. For the $^{15}$N isotopologues
             of HNC and HCN the velocity ranges are suitable for deriving $^{14}$N/$^{15}$N ratios (see text).
\end{tablenotes}
\end{minipage}
\end{threeparttable}
\end{table*}


\begin{table*}
\begin{threeparttable}
\scriptsize
\begin{minipage}{100mm}
\caption{Isotopic ratios for both GC lines of sight.}\label{table6}
\begin{tabular}{crrrrc}
\hline
Isotope & \multicolumn{1}{c}{molecular}                & \multicolumn{2}{c}{MC+0.693}                   & \multicolumn{1}{c}{MC-0.11} & Bibliographic\\
        & \multicolumn{1}{c}{column density ratio}     & \multicolumn{2}{c}{V$_{\rm LSR}$(km s$^{-1}$)} & \multicolumn{1}{c}{V$_{\rm LSR}$(km s$^{-1}$)} & Data$^{a}$\\
        &                                              & \multicolumn{1}{r}{$\sim$68} & $\sim$85  & \multicolumn{1}{c}{$\sim$20} & \\
\hline
 &     \multicolumn{5}{c}{Ratios/limits unbiased by opacity/self-absorption in both GC sources} \\
\hline
$^{12}$C/$^{13}$C & CN:$^{13}$CN               & 22.9 (4.7) &   ...        & 21.2 (5.7) & $\sim$20\\
                  & CCH:$^{13}$CCH             & 19.1 (4.1) &   ...        & 22.6 (6.5) & \\
                  & CCH:C$^{13}$CH             & 21.4 (3.9) &   ...        & $\ga$29 & \\
                  & OCS:O$^{13}$CS             & 22.8 (3.8) & ... & $\ga$25 & \\
                  & CH$_3$CCH:CH$_3^{13}$CCH   & 20.8 (6.6) & ... & $\ga$6 & \\
$^{14}$N/$^{15}$N & HNCO:H$^{15}$NCO & $\ga$380 & ... & $\ga$214 & $>$600\\
$^{18}$O/$^{17}$O & C$^{18}$O:C$^{17}$O & 4.8 (0.5) & ... & 2.6 (0.9) & 3.2 (0.2)\\
$^{16}$O/$^{18}$O & HNCO:HNC$^{18}$O & $\ga$129 & ... &  $\ga$78 & 250\\
$^{29}$Si/$^{30}$Si & $^{29}$SiO:$^{30}$SiO         & 1.1 (0.2) & ... & 2.2 (0.4) & 1.5\\
\hline
 &     \multicolumn{5}{c}{Ratios/limits biased by opacity/self-absorption/ripples in one/both GC sources}  \\
\hline
$^{12}$C/$^{13}$C & HCN:H$^{13}$CN             &  4.5 (0.7)$^{b}$ & 5.1 (0.7) & 4.4 (0.1)$^{c}$ & $\sim$20\\
                  & HNC:HN$^{13}$C             &  3.0 (0.4)$^{b}$ &  ...  & 5.1 (0.3) & \\
                  & HCO$^+$:H$^{13}$CO$^+$           &  5.0 (0.7)$^{b}$ & 6.8 (1.0) & 7.1 (0.2)$^{c}$ & \\
                  & HC$_3$N:H$^{13}$CCCN       & 11.6 (0.8) &...  & 13.9 (2.4) & \\
                  & HC$_3$N:HC$^{13}$CCN       & 14.8 (1.5) & ... & 21.1 (6.1) & \\
                  & HC$_3$N:HCC$^{13}$CN       & 11.0 (2.1) &...  & 20.0 (4.7) & \\
                  & CH$_3$OH:$^{13}$CH$_3$OH   & 15.1 (3.8) &...  & ... & \\
                  & CH$_3$CN:$^{13}$CH$_3$CN   & 13.6 (9.1) &...  & \multicolumn{1}{r}{$\ga$5} & \\
                  & CH$_3$CN:CH$_3^{13}$CN     & 10.9 (1.6) &...  & $\ga$4 & \\
$^{14}$N/$^{15}$N & $^{12}$C:$^{13}$C$\times$HN$^{13}$C:H$^{15}$NC & 148.7 (23.4)$^{d,e}$ &      ...              & 280.8 (25.2)$^e$ & $>$600\\
                  & $^{12}$C:$^{13}$C$\times$H$^{13}$CN:HC$^{15}$N & 100.5 (14.7)$^{d,e}$ & 163.0 (23.1)$^{d}$ & 141.8 (17.3)$^e$ & \\
$^{16}$O/$^{18}$O & $^{12}$C:$^{13}$C$\times^{13}$CO:C$^{18}$O & 136.3 (7.2)$^{e}$ & ... & 185.5 (17.0)$^{e}$ & 250\\

$^{32}$S/$^{34}$S & OCS:OC$^{34}$S                                 & 8.7 (1.3)$^f$ & ... & $\ga$22 & $\sim$22\\
                  & $^{12}$C:$^{13}$C$\times$O$^{13}$CS:OC$^{34}$S & 7.7 (1.6)$^{e,f}$ &...  & ... & \\
                  & O$^{13}$CS:O$^{13}$C$^{34}$S                   & $\ga$3    &...  & ... & \\
\hline
\end{tabular}
\begin{tablenotes}
\item $^a$ Isotopic ratios from W\&R94.
\item $^b$ The $^{12}$C and $^{13}$C isotopologues are affected by opacity/self-absorption.
\item $^c$ The $^{12}$C isotopologue is affected by opacity/self-absorption.
\item $^d$ The $^{13}$C isotopologue is affected by opacity/self-absorption.
\item $^e$ We have averaged the $^{12}$C/$^{13}$C isotopic ratios unbiased by opacity/self-absorption, then we have obtained the average $^{12}$C/$^{13}$C$\sim$21
ratio over both GC sources. In these ratios we have used the $^{12}$C/$^{13}$C=21 ratio.
\item $^f$ The OC$^{34}$S lines are affected by ripples.
\end{tablenotes}
\end{minipage}
\end{threeparttable}
\end{table*}


\subsection{Isotopic ratios}\label{isotopic_ratios}

We have used the large number of detected isotopologues to derive from their column densities the isotopic ratios of $^{12}$C/$^{13}$C, 
$^{14}$N/$^{15}$N, $^{16}$O/$^{18}$O, $^{18}$O/$^{17}$O, $^{29}$Si/$^{30}$Si and $^{32}$S/$^{34}$S for both GC sources.  Table \ref{table6} summarizes
all the derived isotopic ratios based on optically thin (unbiased group) and optically thick (biased group) emission. When possible, we have 
also used column densities of isotopologues with double isotopic substitution to guarantee optically thin emission. The canonical isotopic ratios 
derived for the GC by W\&R94 are also shown in Table \ref{table6}.
As expected, the isotopic ratios of $\sim$3-15 derived from the biased group are always smaller than the canonical isotopic ratios due to the opacity 
effects on the most abundant isotopic substitution. The unbiased $^{12}$C/$^{13}$C isotopic ratios derived for both GC sources are similar, within 
uncertainties, and its average, $^{12}$C/$^{13}$C=21.3 (1.7), is similar to the canonical value.

We also found a $^{18}$O/$^{17}$O ratio of 3.7 (0.5) and a $^{29}$Si/$^{30}$Si ratio of 1.7 (0.2) averaged over both GC sources, which are also 
within the uncertainties similar to the canonical values. Our lower limits to the $^{16}$O/$^{18}$O and $^{14}$N/$^{15}$N ratios are also consistent 
with the canonical values. The $^{32}$S/$^{34}$S isotopic ratios derived in \emph{LOS}+0.693 are lower than that found in W\&R94, because
the OC$^{34}$S lines are affected by ripples (see Fig.~\ref{fig21}) and likely the OC$^{34}$S column density is overestimated. 
However, in \emph{LOS}$-$0.11 the lower limit of 22 to the OCS/OC$^{34}$S ratio is close to the canonical value.
Our averaged $^{12}$C/$^{13}$C, $^{18}$O/$^{17}$O, $^{29}$Si/$^{30}$Si isotopic ratios are similar to the canonical values (W\&R94) within 1$\sigma$, 
which represents $\sim$10\% in the derived isotopic ratios. This indicates that chemical isotopic fractionation and selective photodissociation are 
negligible, less than $\sim$10\%, for most of the molecules and the physical conditions in the GC, consistent with the results obtained by 
\citet{denise}.

The $^{12}$C isotope is a primary product of nucleosynthesis in stellar cores, while the $^{13}$C isotope is thought to be formed from $^{12}$C
present in stars of later generations \citep{wilson92}. Thus, the $^{12}$C/$^{13}$C ratio can be considered a measure of the degree of gas processing in
the ISM.
Using CCH and their $^{13}$C isotopologues, \citet{sergio2010} derived very large $^{12}$C/$^{13}$C ratios of $>$138 and $>$81 toward the 
starburst galaxies M 82 and NGC 253, respectively. Based on the hyperfine fits to the CCH  line profiles, they ruled out opacity effects in 
the CCH lines. The $^{12}$C/$^{13}$C ratio in the GC is at least a factor of 4 lower than in both starburst galaxies, suggesting that the degree of
gas processing is quite different between them.



\section{Discussion}\label{diss_abundances}

\subsection{Uniform molecular abundances and excitation conditions in the nuclei of galaxies}\label{nucli_galaxies}

\subsubsection{Galactic center}\label{galactic_center}

Early studies of the distribution of the kinetic temperature in the molecular material in the GC using both ammonia and H$_2$ have shown that molecular clouds are warm with 
a mean T$_{kin}$ of $\approx$ 100 K \citep{Hutteme,nemesi01}. More recently, \citet{miguel06} have found 
that the abundance of complex organic molecules is also relatively uniform in the molecular clouds in the GC. In agreement with the findings 
of \citet{miguel06}, we also found that the molecular clouds toward two lines of sight, \emph{LOS}+0.693 and \emph{LOS}$-$0.11, separated by more than $\sim$120 pc  
within the Galactic Center, show a rather uniform chemistry. $\sim$80\% of molecular species detected in both lines of sight show 
similar abundances within a factor of 2, even for the most complex organic molecules like C$_2$H$_5$OH, C$_2$H$_5$CN and (CH$_3$)$_2$O.  

\begin{figure}
\includegraphics[trim=3cm 0 0 0, clip, width=100mm]{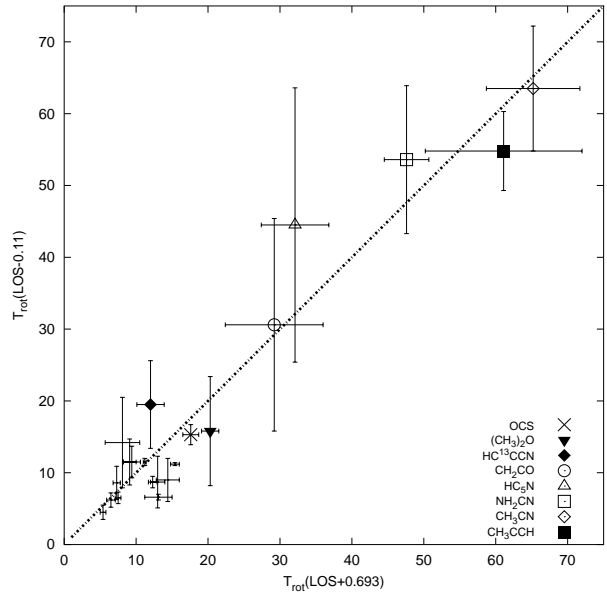}
\caption{Relationship between the T$_{rot}$ of both GC sources. The eight symbols show the highest T$_{rot}$ 
measured in both GC sources. The dashed line shows the line of equal T$_{rot}$.}
\label{fig7}
\end{figure}

Our (CH$_3$)$_2$O, CH$_3$OH, HCOOH, HNCO, $^{13}$CS and C$^{18}$O abundances are also consistent with those derived in previous studies toward both GC 
sources from 2 and 3 mm line data \citep{miguel06,sergio2008}. The ``uniform'' chemistry in the GC was explained for complex molecules as due to 
grain surface chemistry followed by the sputtering of the icy mantles by shocks.
Our database allows now a detailed comparison of the excitation conditions found along the two \emph{LOSs}. Fig.~\ref{fig7} represents the T$_{rot}$ 
derived for \emph{LOS}+0.693 versus the T$_{rot}$ derived for \emph{LOS}$-$0.11. 
It is remarkable that the excitation of the molecular gas is very similar for both GC sources. 
In fact, 86\% of the molecules show the same T$_{rot}$ within 1$\sigma$. 
Another two molecules (c-C$_3$H$_2$ and NH$_2$CHO) show T$_{rot}$ which are consistent within 3$\sigma$.
HC$_3$N is the only molecule that shows a significant difference in its excitation between both GC sources above a 3$\sigma$ level. However, the difference in the T$_{rot}$ 
by a factor of 1.4 corresponds to a change of only a factor of 1.8 in the derived H$_2$ densities of the molecular 
gas (see section \ref{densities}). The similar excitation conditions and molecular complexity found in the molecular gas in the two \emph{LOSs} 
indicate that physical processes operating in the GC are widespread, driving the chemistry and affecting the physical conditions over large scales. 

It is remarkable that the complex organic molecules like CH$_2$CO, HC$_5$N, NH$_2$CN, CH$_3$CCH and CH$_3$CN show the largest T$_{rot}$ in our survey.
We do not believe the high temperatures derived from CH$_2$CO and NH$_2$CN are biased by the range of the energy levels of the observed transitions.
Although the HNCO RDs (see Fig.~\ref{fig5}) cover the same range of energies than that of the NH$_{2}$CN, the derived T$_{rot}$ for both molecules are very 
different. The wide range of derived T$_{rot}$ for the different molecules does not seem to be due to  difference in collisional cross sections, level 
structure and/or dipole moments. For NH$_2$CN and CH$_2$CO, which have virtually identical geometrical cross sections and level structure, but very 
different dipole moments ($\mu_a$=4.32D for NH$_2$CN and $\mu_a$=1.414 D for CH$_2$CO) one would expect to observe a larger T$_{rot}$ for CH$_2$CO. 
In contrast, NH$_2$CN shows a higher T$_{rot}$ than CH$_2$CO.

\begin{figure}
\includegraphics[scale=0.35]{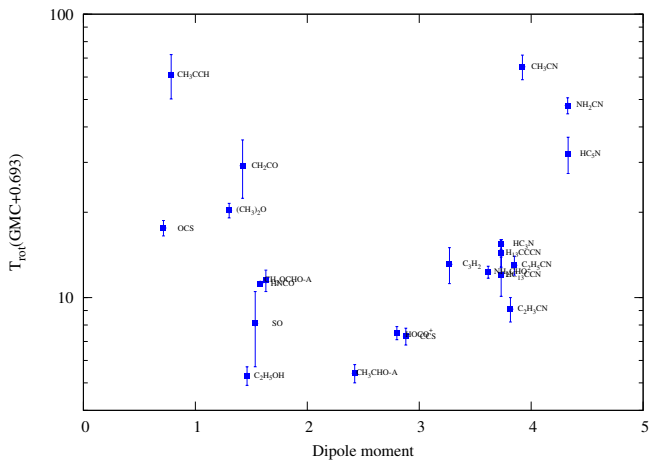}
\caption{The molecular dipole moments represented as a function of the T$_{rot}$ derived for \emph{LOS}+0.693. CH$_3$CN and CH$_3$CCH
are the molecules that show the highest T$_{rot}$ in this survey. When any molecule has more than one dipole moment we have represented
the highest value.}\label{fig8}
\end{figure}

Fig.~\ref{fig8} shows the molecular excitation (T$_{rot}$) derived for \emph{LOS}+0.693 as a function of the dipole moment of the molecule.
We found that there is not any clear dependence of T$_{rot}$ with the dipole moment as expected if the emission from all molecules were arising from the 
same region. 
Although the origin of the difference in T$_{rot}$ is uncertain, we postulate that it may be related to the formation  of the molecular species.
In the shock scenario proposed by \citet{miguel06} and \citet{sergio2008}, the molecular species showing higher T$_{rot}$ will be those produced at early times or just 
ejected from grain and those with lower T$_{rot}$ should be located in the post shocked regions.

\subsubsection{Starburst galaxies}

The bulk of the quiescent molecular clouds in the GC seem to show rather uniform physical conditions and chemical composition. In this section we compare our results obtained for the GC clouds with those derived for nearby extragalactic nuclei with different type of activity. Table \ref{table7} summarizes the T$_{rot}$ and molecular 
abundances of selected species derived for both GC sources and four extragalactic nuclei, NGC 253, M 82, IC 342 and Maffei 2. The molecular species selected in Table \ref{table7} and their corresponding T$_{rot}$ has been successfully used 
by \citet{sergio2009a} and \citet{alad11a} to study the evolution of the nuclei of starburst galaxies. Based on the excitation and the molecular abundances, \citet{sergio2009a} and \citet{alad11a} have classified Maffei 2 as a young starburst galaxy, 
IC 342 and NGC 253 as intermediate-age starburst galaxies and M 82 as an evolved starburst galaxy \citep{alad11a}.

In column 8 of Table \ref{table7}, we show the ratio between the molecular abundance of \emph{LOS}$-$0.11 and those of \emph{LOS}$+$0.63 and the nuclei of nearby galaxies. While the molecular abundances in the two GC \emph{LOSs} clouds are within a factor of 3, those in galactic nuclei are, for most molecules, at least one order of magnitude 
smaller than those measured in the GC. This is expected since the volume sampled in external galaxies is, at least, four orders of magnitude larger than in the GC. Then the ratio between diffuse molecular gas measured by the H$_2$ column densities to dense gas traced by other species is smaller in the GC than in external galaxies. However, some molecular species 
show changes in their abundances by more than one order of magnitude relative to high density tracers like CS. In particular, CH$_3$OH shows the lowest
abundances in galaxies like M 82, IC 342 and Maffei 2 by factors of $\sim$(1-6)$\times$10$^2$ relative to \emph{LOS}$-$0.11. 
NGC 253 is the galaxy that shows the highest methanol abundance between the starburst galaxies in our sample. 
The radiation field in NGC 253 seems to be higher than in M 82, IC 342 and Maffei 2 \citep{Carral, Israel}. However, differences in the evolutionary stage of the 
starbursts and molecular clouds \citep{alad11a}, the distribution of
star forming regions and/or the strength of large-scale shocks could be responsible for the high methanol abundance in NGC 253 as compared to those of the other
galaxies in our sample.
Similar lower methanol abundances, by nearly two orders of magnitude, are also observed toward a sample of molecular clouds in the GC \citep{miguel06} 
that are affected by the UV photodissociation from HII regions. This suggests that this molecule is likely photodissociated by the strong UV radiation field 
of G$_0$$\sim$10$^{2.5-4.0}$ \citep{Carral,Israel} present in these nearby galaxies considered in this comparison.

Another molecule which also shows large abundance differences between our sample of galaxies and \emph{LOS}$-$0.11 is HNCO. It is expected that gas-phase HNCO survives in well-shielded dense molecular clouds, while it is photodissociated in unshielded regions affected 
by UV radiation \citep{sergio2008}. The high HNCO abundances derived in both GC sources compared to those found in the starburst galaxies NGC 253 
and M 82 support the idea that HNCO is a suitable tracer of dense molecular gas ($\ga$10$^6$ cm$^{-3}$) in clouds unaffected by strong UV 
radiation \citep{sergio2008}. This agrees with the finding of \citet{Arancha}, who found that the HNCO abundance in the 20 km s$^{-1}$ cloud 
dominated by shocks is a factor of 20 higher than in the CND surrounding Sgr A$^*$ which is affected by both shocks and photodissociation.

The photodissociation rate of HNCO is a factor of $\sim$1.5 larger than that of CH$_3$OH. Both molecules have photodissociation
rates higher (a factor $\sim$3) and lower (a factor $\sim$4) than those of CCH and HC$_3$N, respectively. The smaller differences
found in the CCH abundances compared to those of CH$_3$OH and HNCO between \emph{LOS}$-$0.11 and the starburst galaxies M 82 
and NGC 253 is likely due to the differences in the photodissociation rates of these molecules. Additionally, CCH could increase
its abundance efficiently in PDRs \citep{Mul}. One expects that HC$_3$N would show larger abundance variations than CH$_3$OH and HNCO 
between \emph{LOS}$-$0.11 and both starburst galaxies due to their differences in the photodissociation
rates. However, this is not observed in Table \ref{table7}. Efficient formation of HC$_3$N through ion-molecule chemistry \citep{Knight} 
could be responsible for the high
HC$_3$N abundance found in both starburst galaxies. The largest differences shown by CH$_3$OH and HNCO in our
comparison is likely due to their photodissociation and that both molecules are not expected to form efficiently in gas phase. CH$_3$OH shows higher
differences in its abundance than HNCO between \emph{LOS}$-$0.11 and the starburst galaxies in our sample. This is probably due to the higher extinction
of the dense clouds, cores where HNCO arises ($\ga$10$^6$ cm$^{-3}$, \citet{sergio2008}) as compared with the more diffuse conditions where CH$_3$OH is 
found ($\sim$5$\times$10$^4$ cm$^{-3}$, \citet{sergio06a}).

Overall, complex molecules in the molecular clouds of this sample of starburst galaxies reveal smaller abundances than those measured in the GC sources 
by nearly 2 orders of magnitude, suggesting that a substantial fraction of the molecular gas in the nearby galaxies in our sample is affected by photodissociation by the UV 
radiation from the starburst.

\begin{table*}
\begin{threeparttable}
\scriptsize
\begin{minipage}{100mm}
\caption{Abundance and excitation of galactic and extragalactic sources.}\label{table7}
\begin{tabular}{cccrrrrrrr}
\hline
Source name and & Type & Molecule & \multicolumn{1}{c}{T$_{rot}$ ($\sigma$)} & N$_{\rm{H_2}}$ ($\sigma$) & \multicolumn{1}{c}{N ($\sigma$)} & \multicolumn{1}{c}{$\;\;\;\;$X=N/N$_{\rm H_{2}}$} & (X$_1$/X)$^A$ \\
Nominal positions & & & \multicolumn{1}{c}{(K)} & (cm$^{-2}$) & \multicolumn{1}{c}{($\times$10$^{13}$ cm$^{-2}$)} & \multicolumn{1}{c}{$\times$10$^{-9}$} & & \\
J2000 ($\alpha,\beta$) & & & & \multicolumn{1}{c}{} & \multicolumn{1}{c}{} & \multicolumn{1}{c}{} & \multicolumn{1}{c}{} \\
\hline
\emph{LOS}+0.693 & Molecular cloud & CS & 10.0 & 5.9 (0.2)$\times$10$^{22}$ & 48.3 (4.2)$^{B}$ & 8.2 (1.0) & 3.0 (0.4) \\
$17^{\rmn{h}}47^{\rmn{m}}22\fs0$, $-28\degr21\arcmin27\farcs0$ & & SO & 8.1 (2.4) & & 61.0 (51.3) & 10.3 (8.7) & 0.8 (0.6) \\
& & SiO & 10.0 & & 9.0 (1.8) & 1.4 (0.2) & 2.1 (0.3)\\
& & OCS & 17.6 (1.1) & & 186.2 (4.6) & 31.6 (1.5) & 1.2 (0.2)\\
& & CCH & 10.0 & & 316.2 (18.6) & 53.6 (3.8) & 2.0 (0.3)\\
& & HCO$^+$ & 10.0 & & 57.0 (1.3) & 9.6 (2.2) & 1.6 (0.5)\\
& & HNCO & 11.2 (0.2) & & 144.5 (5.6) & 24.5 (1.3) & 1.2 (0.1)\\
& & HC$_3$N & 15.4 (0.6) & & 54.9 (3.3) & 9.4 (1.0) & 1.2 (0.1)\\
& & CH$_3$OH & 14.0 (1.3)$^{a}$ & & 12700.0 (1420.3) & 2160.0 (255.0) & 0.5 (0.1)\\
& & CH$_3$CCH & 61.1 (10.9) & & 262.4 (16.9) & 44.5 (3.3) & 0.7 (0.1)\\
\emph{LOS}$-$0.11 & Molecular cloud & CS & 10.0 & 2.4 (0.2)$\times$10$^{22}$ & 58.8 (2.1)$^{B}$ & 24.5 (2.2) & 1\\
$17^{\rmn{h}}45^{\rmn{m}}39\fs0$, $-29\degr04\arcmin05\farcs0$ & & SO & 14.2 & & 18.6 (3.7) & 7.8 (2.0) & 1\\
& & SiO & 10.0 & & 7.2 (0.2) & 3.0 (0.3) & 1 \\
& & OCS & 15.3 (1.4) & & 87.1 (14.1) & 36.3 (6.6) & 1 \\
& & CCH & 10.0 & & 251.2 (24.0) & 104.7 (13.2) & 1 \\
& & HCO$^+$ & 10.0 & & 36.0 (0.8) & 15.5 (3.0) & 1 \\
& & HNCO & 11.5 (0.5) & & 67.6 (4.5) & 28.2 (3.0) & 1 \\
& & HC$_{3}$N & 11.2 (0.2) & & 26.8 (0.6) & 11.2 (1.0) & 1 \\
& & CH$_{3}$OH & 13.0 (1.0)$^{a}$ & & ... & 1100.0 (220)$^{a,C}$ & 1 \\
& & CH$_{3}$CCH & 54.8 (5.5) & & 74.1 (12.3) & 30.9 (5.7) & 1 \\
NGC 253 & Intermediate-age & CS & 12.0 (3)$^{b}$ & 6.2 (0.5)$\times$10$^{22}$$^{b}$&15.0 (6.0)$^{b}$ &2.4 (1.0) & 10.2 (4.2)\\
$00^{\rmn{h}}47^{\rmn{m}}33\fs3$, $-25\degr17\arcmin23\farcs0$ & starburst &SO&40 (24)$^{c}$& &4.5 (3.3)$^{c}$ & 0.73 (0.5) & 10.7 (7.9) \\
& & SiO & 7.4 (0.7)$^{c}$ & & 0.5 (0.1)$^{c}$ & 0.081 (0.02) & 37.0 (8.7) \\
& & OCS & 17 (2)$^{c}$ & & 25.0 (3.0)$^{c}$ & 4.0 (0.6) & 9.1 (2.0) \\
& & CCH & 10.0$^{d}$ & & 73.0 (1.0)$^{d}$ & 11.8 (1.0) & 8.9 (1.3) \\
& & HNCO & 23.0 (6.0)$^{c}$ & & 5.7 (2.7)$^{c}$ & 0.9 (0.4) & 31.3 (15.0) \\
& & HC$_{3}$N & 11.6 (1.8)$^{b}$ & & 8.1 (2.9)$^{b}$ & 1.3 (0.5) & 8.6 (3.2) \\
& & CH$_{3}$OH & 11.6 (0.2)$^{c}$ & & 83 (3)$^{c}$ & 13.4 (1.2) & 82.1 (18.0) \\
& & CH$_{3}$CCH & 44.4 (7.7)$^{b}$ & & 32.0 (10.0)$^{b}$& 5.2 (1.7) & 5.9 (2.2) \\
M 82 & Evolved starburst & CS & 15.1 (0.9)$^{b}$ & 7.9 (0.4)$\times$10$^{22}$$^{b}$ & 3.6 (0.5)$^{b}$ & 0.5 (0.1) & 49.0 (9.3) \\
$09^{\rmn{h}}55^{\rmn{m}}51\fs9$, $69\degr40\arcmin47\farcs1$ & & CCH & 10.0$^{e,D}$ & & 55.0 (1.0)$^{e,D}$ & 6.5 (0.4)$^{D}$ & 16.1 (2.2) \\
& & HNCO & 10.0$^{f,D}$& &$\la$0.7$^{f,D}$ & $\la$0.1$^{D}$ & $\ga$282 \\
& & HC$_{3}$N & 24.7 (3.9)$^{b}$ & &2.5 (1.1)$^{b}$ & 0.3 (0.1) & 37.3 (16.0) \\
& & CH$_{3}$OH & 4.5$^{f}$ & & 15.0$^{f}$ & 1.9 & 579.0 \\
& & CH$_{3}$CCH & 28.1 (1.2)$^{b}$ & & 85.0 (9.0)$^{b}$ & 10.8 (1.3) & 2.9 (0.6) \\
IC 342 & Intermediate-age & CS & 10.6 (0.2)$^{b}$ & 5.8 (0.4)$\times$10$^{22}$$^{b}$ & 6.0 (0.1)$^{b}$ & 1.0 (0.1) & 24.5 (2.6) \\
$03^{\rmn{h}}46^{\rmn{m}}48\fs5$, $68\degr05\arcmin46\farcs0$ & starburst & SO & ... & & ... & $\ga$0.08$^{g,E}$ & $\ga$1.1 \\
& & CCH & ... & & ... & 0.3$^{g,E}$ & 3.5 \\
& & HNCO & 10.2 (0.4)$^{e}$ & & 12.0 (1.0)$^{e}$ & 2.1 (0.2) & 13.4 (2.0) \\
& & HC$_{3}$N & 13.1 (2.3)$^{b}$ & & 2.7 (1.1)$^{b}$ & 0.5 (0.2) & 22.4 (10.0) \\
& & CH$_3$OH & ... & & ... & 5$^{g,E}$ & 220 \\
& & CH$_{3}$CCH & 70.0$^{b}$ & & 45.0$^{b}$ & 7.8 & 4.0 \\
Maffei 2 & Young starburst & CS & 9.7 (8.1)$^{b}$ & 4.4 (0.8)$\times$10$^{22}$$^{b}$ & 6.9 (5.7)$^{b}$ & 1.6 (1.3) & 15.3 (13.3) \\
$02^{\rmn{h}}41^{\rmn{m}}55\fs1$, $59\degr36\arcmin15\farcs0$ & & SiO & ... & & ... & 0.08$^{h,E}$ & 36.2 \\
& & CCH & ... & & ... & 9.6$^{h,E}$ & 10.9 \\
& & HCO$^+$ & ... & & ... & 3.5$^{h,E}$ & 4.4 \\
& & HNCO & 11.5 (0.6)$^{e}$ & & 10 (2)$^{e}$ & 2.3 (0.6) & 12.3 (3.6) \\
& & HC$_{3}$N & 11.6 (0.8)$^{b}$ & & 4.3 (0.6)$^{b}$ & 1.0 (0.2) & 11.2 (2.8) \\
& & CH$_{3}$OH & 10.0$^{f}$ & &33.0$^{f}$ & 7.5 & 147.0 \\
& & CH$_{3}$CCH & $\la$46.9$^{b}$ & &$\la$11$^{b}$ & $\la$2.5 & $\ga$12.4 \\
\hline
\end{tabular}
\begin{tablenotes}
\item $Notes:$ A-Ratio between the molecular abundances for \emph{LOS}$-$0.11 and those for the other sources in this table;
B-The CS abundances for both GC sources are derived from the $^{13}$CS abundances by using our $^{12}$C/$^{13}$C=21 ratio;
C-An uncertainty of 20\% in the methanol abundance is assumed;
D-Offset position of $13\farcs0$, $7\farcs5$ relative to $09^{\rmn{h}}55^{\rmn{m}}51\fs9$,
$69\degr40\arcmin47\farcs1$ (J2000) with N$_{\rm H_2}$=8.5$\times$10$^{22}$ cm$^{-2}$ \citep{sergio06a}.
E-This abundance corresponds to the cloud A in IC 342 or cloud F in Maffei 2, both clouds are located in projection close to the nuclear star clusters.
Uncertainties in the abundances are at least a factor of 3.
\item $^a$ \citet{miguel08}.
\item $^b$ \citet{alad11a}.
\item $^c$ \citet{sergio06b}.
\item $^d$ \citet{sergio2010}.
\item $^e$ \citet{sergio2009a}.
\item $^f$ \citet{sergio06a}.
\item $^g$ \citet{Meier05}.
\item $^h$ \citet{Meier12}. 

\end{tablenotes}
\end{minipage}
\end{threeparttable}
\end{table*}

\begin{figure*}
\subfloat[]{\includegraphics[trim=1cm 0 0 0, clip,width=88mm]{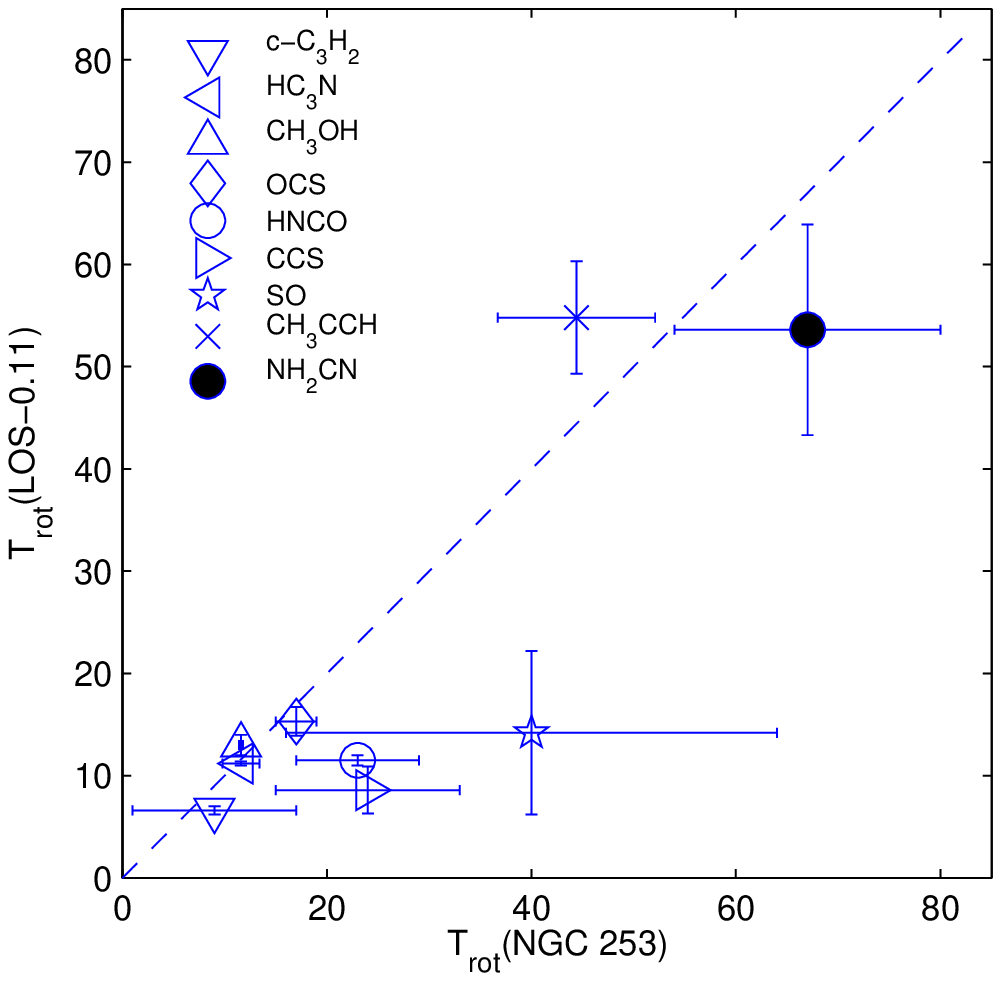}}\hfil
\subfloat[]{\includegraphics[trim=1cm 0 0 0, clip,width=88mm]{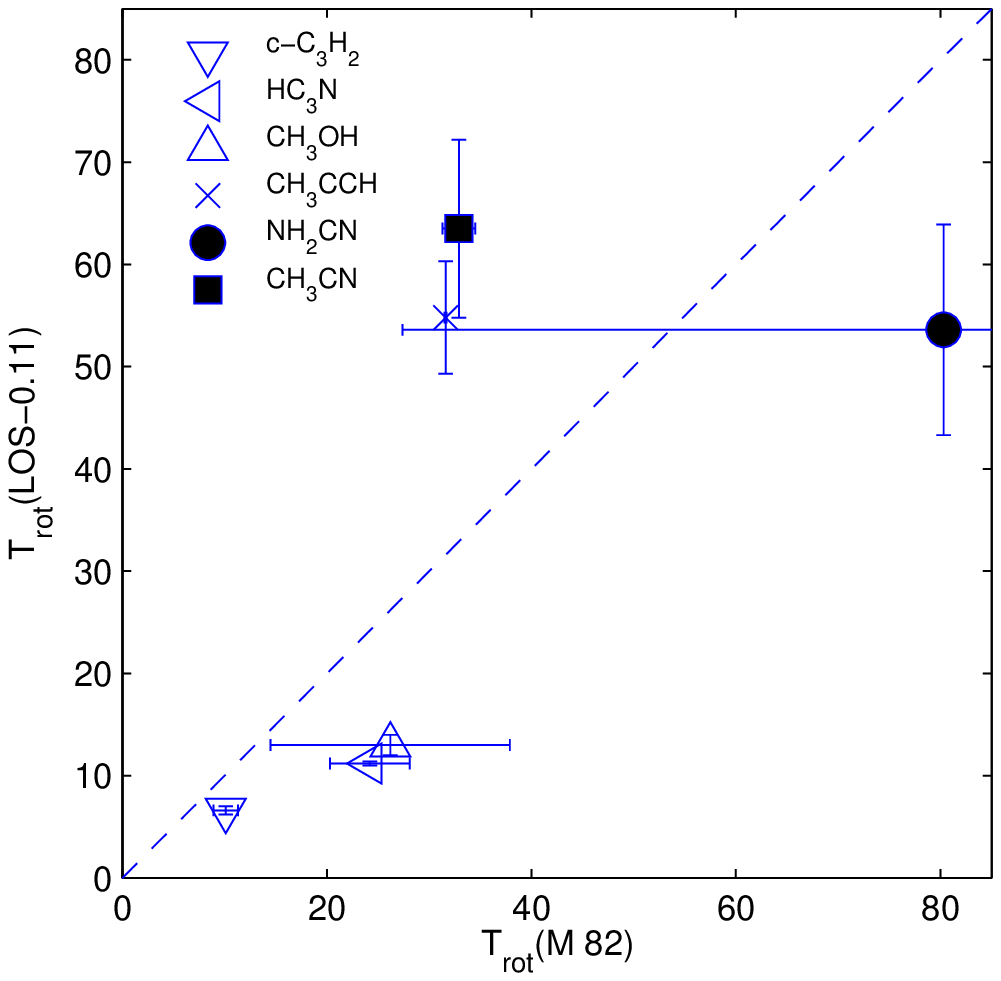}}
\caption{Relationship between the T$_{rot}$ of NGC 253 (left) and M 82 (right) with the T$_{rot}$ of \emph{LOS}$-$0.11. Different symbols correspond to different
molecules. The T$_{rot}$ for NGC 253 are taken from \citet{sergio06b} and \citet{alad11a}, and for M 82 from \citet{alad11b}. For \emph{LOS}$-$0.11, the T$_{rot}$ of
methanol is taken from \citet{miguel08}. The dashed line on each plot is the line of equal T$_{rot}$.}
\label{fig9}
\end{figure*}

Fig.~\ref{fig9} shows the T$_{rot}$ of selected molecules derived in NGC 253 \citep{sergio06b,alad11a} and M 82 \citep{alad11b} versus the T$_{rot}$ derived for \emph{LOS}$-$0.11. In contrast to the large abundance 
difference found between the GC clouds and starburst galaxies, the excitation of complex molecules like c-C$_3$H$_2$, HC$_3$N and CH$_3$OH derived for both starburst 
galaxies, NGC 253 and M 82, is similar within $\sim$3$\sigma$ to the excitation of the molecular cloud found for \emph{LOS}$-$0.11.
Like in the two \emph{LOSs} in the GC, NH$_2$CN and CH$_3$CCH in NGC 253 also show high T$_{rot}$ of 67 K and 44 K, respectively. Interestingly these T$_{rot}$ are within a factor of 1.2 of those of NGC 253. 
In M 82, the T$_{rot}$ derived from CH$_3$CCH and CH$_3$CN show the same trend than in NGC 253 but in this case they are a factor of $\sim$2 lower than those observed in \emph{LOS}$-$0.11.
It is remarkable that the excitation of molecular clouds in the nuclei of galaxies with different activity and evolutionary stage appears relatively uniform, suggesting that the physical conditions, H$_2$ densities 
and kinetic temperatures, must be very similar. If, as suggested in the previous section, the excitation of the molecules is related to the formation mechanism, there must be a common processes driving the 
chemistry in the nuclei of galaxies. However, to explain the underabundance of CS, HC$_3$N, CH$_3$OH, HNCO and other molecules (see Table \ref{table7}) found in our sample of galactic nuclei with respect to the GC, 
the shock scenario proposed by \citet{miguel06} and \citet{sergio2008} for the GC must be combined with the effects of the photodissociation as proposed by \citet{alad11b}.

\begin{figure*}
\subfloat[]{\includegraphics[trim=0 0 0 0,clip, width=88mm,angle=0]{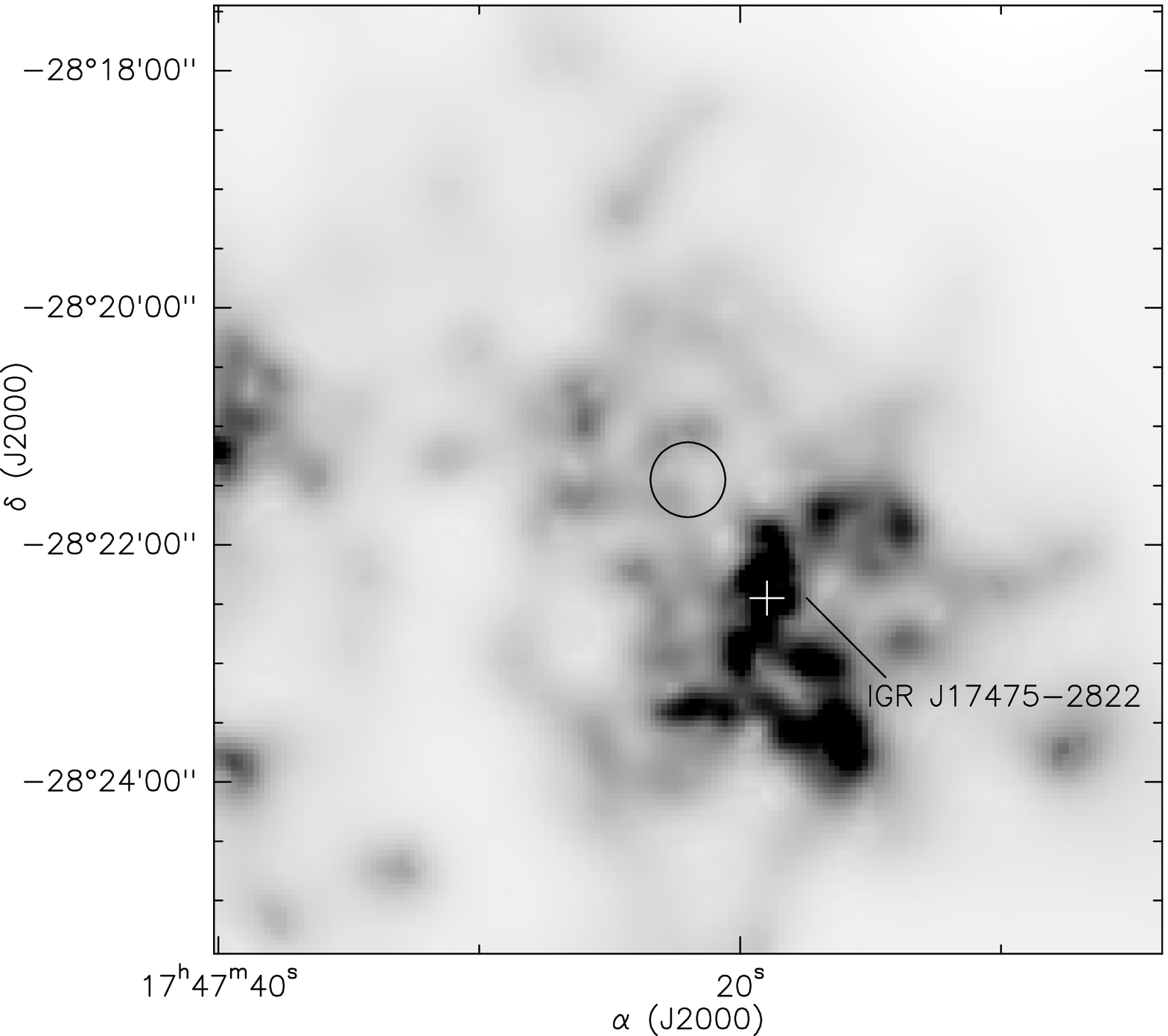}}\hfil
\subfloat[]{\includegraphics[trim=0 0 0 0,clip, width=88mm,angle=0]{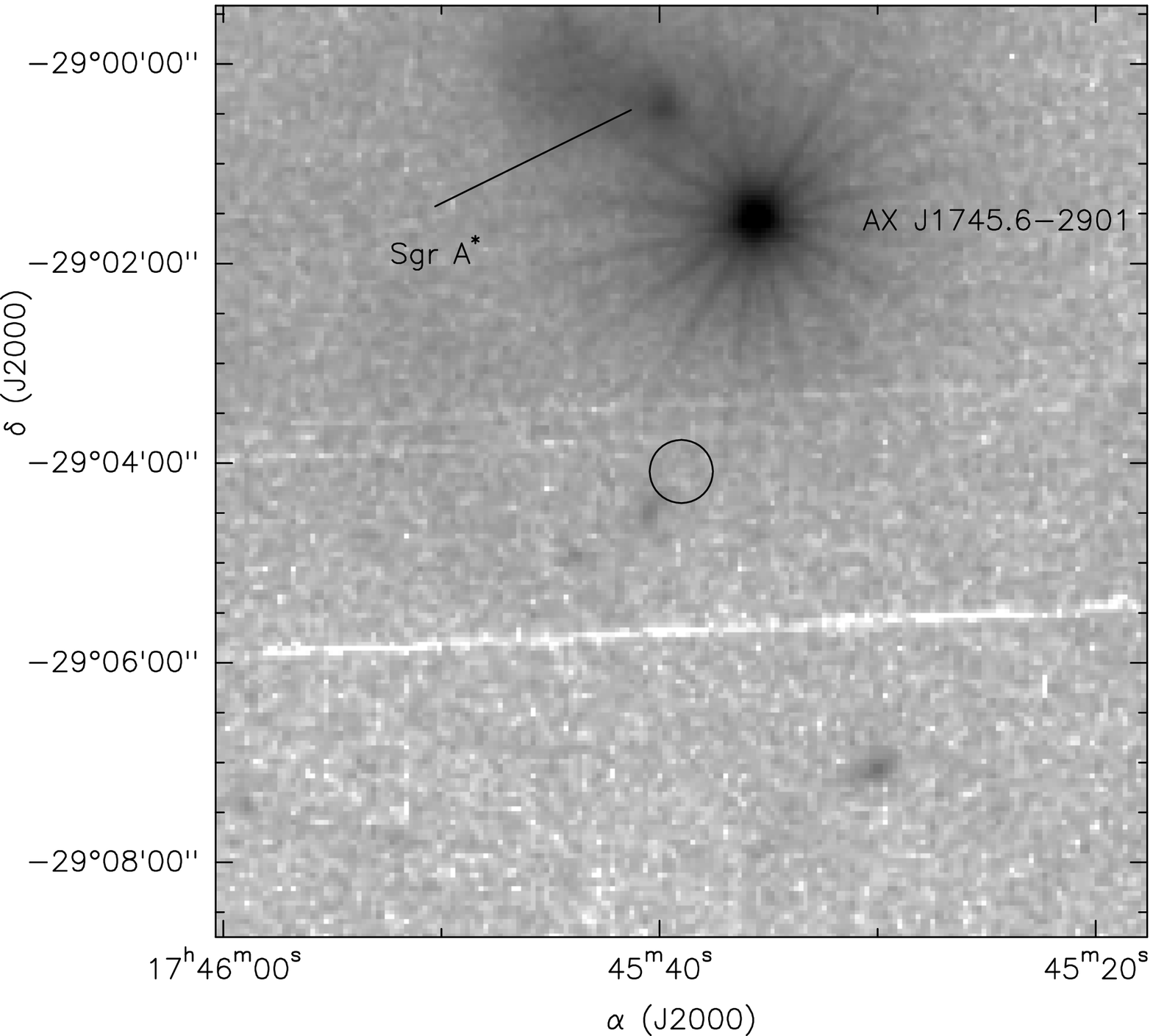}}
\caption{\textbf{(Left panel)} Fe K$\alpha$ line emission observed with the XMM Newton satellite in 2001-2004 toward Sgr B2 \citep{terrier}. The
white cross shows the position of the X-ray source IGR J17475-2822. \textbf{(Right panel)} 0.1-10 keV image observed with the XMM Newton in April 2007
toward Sgr A (image taken from http://xmm.esac.esa.int/xsa/index.shtml). The line shows the position of Sgr A$^*$. AX J1745.6-2901 is the brightest source in the field. The circles with the
beam size of the Mopra telescope (38$^{''}$ at 90 GHz) show \emph{LOS}+0.693 and \emph{LOS}$-$0.11 in the left and right panels, respectively.}
\label{fig10}
\end{figure*}

\subsection{Tracing the UV and X-ray induced chemistry in galactic nuclei }\label{PDR_XDRtracers}

The scenario proposed in the previous section to explain the difference in abundance ratios between these quiescent Galactic center \emph{LOSs} and our sample
of galactic nuclei would indicate that the 
clouds in the two \emph{LOSs} are only marginally affected by the UV radiation. However, in our survey we have detected, for the first time, HCO and 
HOC$^+$ emission toward \emph{LOS}+0.693 in the GC. Emission from these molecules are considered to be tracers of UV and X-ray induced 
chemistry in the associated PDRs \citep{Apponi,Javi09b,sergio2009b} and XDRs \citep{usero,Spaans}. 
It has been proposed that abundance and abundance ratios between key molecules like HCO, HOC$^+$ and HCO$^+$ can be used to 
trace the XDR and PDR chemistries and even to differentiate between them \citep{Meijerink,Spaans}. The high spatial resolution of our data offers an unique 
opportunity to study the PDR/XDR component in nuclei of galaxies using molecular tracers.

\subsubsection{Constrains on the X-ray radiation in the molecular clouds along the GC LOS}

Fig.~\ref{fig10} illustrates the key difference in the X-ray emission observed toward the two \emph{LOSs} (shown as open circles) in the GC. While 
\emph{LOS}$-$0.11 only shows continuum emission in the 0.1-10 keV band, \emph{LOS}+0.693 shows, in addition to the continuum 
emission, one of the strongest Fe K$\alpha$ (6.4 keV) lines observed in our Galaxy \citep{Koyama}. The presence of the strong Fe K$\alpha$ line emission 
is considered as an excellent tracer of XDRs \citep{Jesus00}, showing that X-rays are directly interacting with large column densities of matter. 
As a consequence of this interaction one would expect a chemistry driven by X-rays. In contrast, \emph{LOS}$-$0.11 does not show any 
emission of the Fe K$\alpha$ line \citep{Ponti}, indicating the lack of any relevant XDR.

\subsubsection{Constrains on the UV radiation in the molecular clouds along the GC LOS}\label{PDR_tracer}

As shown in Fig. 1, the two \emph{LOSs} do not show any prominent HII region which would produce associated PDRs. However, \emph{LOS}+0.693 is 
relatively close to the HII regions of Sgr B2N and L, whereas \emph{LOS}$-$0.11 is located near the non-thermal sources Sgr A-E and 
Sgr-F \citep{Lu,Yusef05}. We can use the upper limit to the H$\alpha$ recombination line emission in our survey to constrain the Lyman photons 
in the beam. None of the four H$\alpha$, H39$\alpha$ to H43$\alpha$, recombination lines that fall in the frequency range of our spectral line survey 
are detected toward both \emph{LOSs}. From the 3$\sigma$ upper limit to their intensities and assuming a linewidth of $\sim$35 km s$^{-1}$, we have 
set the upper limits to the thermal continuum fluxes given in Table \ref{table5}. For optically thin emission and the average LTE electron
temperature of 6500 K estimated toward the GC \citep{Goss}, we have derived the upper limits to the Lyman continuum photons \citep{Mezger} given in 
Table \ref{table5}. These upper limits of $\la$10$^{48.4}$ and $\la$10$^{48.7}$ s$^{-1}$ for \emph{LOS}+0.693 and 
\emph{LOS}$-$0.11, respectively, would constrain the spectral type of any ionizing star to be later than O8-O9.

Low angular resolution observations of fine structure lines of CII and OI toward Sgr B2 have shown the presence of an extended FUV radiation field of
G$_0$$\approx$10$^3$-10$^4$, which should produce important PDRs. However, this FUV radiation field would be characterized by a Lyman continuum 
photon flux of $\sim$10$^{50.4}$ s$^{-1}$ \citep{Javi04}. This flux is $\sim$50 times larger than that derived from the recombination lines in 
\emph{LOS}+0.693. Given the large difference in beam size between our observations and those of the fine structure lines, it 
is possible that inhomogeneities in the FUV radiation field could partially explain this discrepancy. 
In any case, the UV radiation field strength of the PDR component in \emph{LOS}+0.693 is rather uncertain. 
The situation for \emph{LOS}$-$0.11 seems to be simpler since there is not nearby 
massive star formation, like Sgr B2N, which could provide a large FUV radiation field. In this \emph{LOS} one expects a negligible PDR component.

\begin{table*}
\begin{threeparttable}
\scriptsize
\begin{minipage}{100mm}
\caption{Physical parameters derived from H$\alpha$ recombination lines.}\label{table5}
\begin{tabular}{crcccccc}
\hline
Recombination &                          & \multicolumn{3}{c}{\emph{LOS}+0.693}    & \multicolumn{3}{c}{\emph{LOS}$-$0.11}\\
line          & $\nu$ & Continuum Flux   & Electron    & Flux of Lyman      & Continuum Flux & Electron  & Flux of Lyman\\
              &                          &                    & Density     & Continuum Photons (N$_{\rm Lyc}$)  &   & Density & Continuum Photons (N$_{\rm Lyc}$)\\
              & (GHz) & (Jy)             & (cm$^{-3}$) & Log(N$_{\rm Lyc}$) (s$^{-1}$)  & (Jy)           & (cm$^{-3}$) & Log(N$_{\rm Lyc}$) (s$^{-1}$)\\
\hline
H43$\alpha$ & 79.912  & ...      & ...         & ...      & $\la$1.2 & $\la$473.2  & $\la$48.6\\
H42$\alpha$ & 85.688  & $\la$1.4 & $\la$513.0 & $\la$48.7 & $\la$2.6 & $\la$715.9  & $\la$49.0\\
H41$\alpha$ & 92.034  & $\la$1.5 & $\la$531.2 & $\la$48.7 & $\la$1.2 & $\la$480.5  & $\la$48.6\\
H39$\alpha$ & 106.737 & $\la$0.8 & $\la$391.2 & $\la$48.4 & $\la$1.3 & $\la$501.5  & $\la$48.7\\
\hline
\end{tabular}
\end{minipage}
\end{threeparttable}
\end{table*}

\subsubsection{The HCO, HOC$^+$, HCO$^+$ and CS abundance as tracers of the PDR and XDR components in galactic nuclei}

Table \ref{table4} shows the HCO$^+$/HOC$^+$, HCO$^+$/HCO and HCO/HOC$^+$ ratios measured for our two \emph{LOSs}, the extragalactic sources (NGC 253, 
M 82 and NGC 1068) and typical galactic PDRs. The HCO$^+$ column densities were calculated from the HC$^{18}$O$^+$ line by 
assuming $^{16}$O/$^{18}$O=250 (W\&R94). We also have included abundances of CS relative to HOC$^+$ and HCO since the CS abundance does not 
seem to change substantially in PDRs and shocked environments in the GC \citep{miguel06,sergio2008}.

\begin{table*}
\begin{threeparttable}
\scriptsize
\begin{minipage}{60mm}
\caption{Ratios of HCO$^+$, CS, HOC$^+$ and HCO.}\label{table4}
\begin{tabular}{crrrrrc}
\hline
               &                & HCO$^+$/HOC$^+$ & HCO$^+$/HCO & HCO/HOC$^+$ & CS/HCO & CS/HOC$^{+}$ \\
               & Velocity       &             &             &        &  \\
Source         & (km s$^{-1}$)  & & & & & \\
\hline
\emph{LOS}+0.693            & $\sim$68  & 546 (175)$^{a}$    & 9 (3)$^{a}$     & 62.9 (30.1) & 6 (1)       &378 (43) \\
\emph{LOS}$-$0.11           & $\sim$20  & $\ga$1134$^{b}$     & $\ga$11$^{b}$    & ...           &$\ga$18  & $\ga$1960\\
NGC 253              & $\sim$180 & 80 (30)$^{c}$      & 5.2 (1.8)$^{c}$ & 15.4 (7.9)  & 2 (1)$^{d}$       & 25 (10)$^{d}$\\
                     & $\sim$280 & 63 (17)$^{c}$      & 5.4 (1.3)$^{c}$ & 11.7 (4.2)  & 1.1 (0.5)$^{d}$ & 13 (5)$^{d}$\\
M 82                 & $\sim$310 & 60 (28)$^{c}$      & 9.6 (2.8)$^{c}$ & 6.3 (3.4)   & 1$^{e}$           & 1$^{f}$\\
NGC 1068             & $\sim$1100& 128 (28)$^{c}$     & 3.2 (1.2)$^{c}$ & 40.0 (17.4) & 0.3 (0.2)$^{g}$ & 6 (4)$^{g}$\\
Horsehead            &   ...     &75-200$^h$ & 1.1$^{i}$            & 68.2-181.8        &   ...         & ...               \\
Orion Bar (PDR peak) & $\sim$9.5 & $<$166$^{j}$ & 2.4$^{k}$ & $<$69.2 & 3$^{l}$ & 100$^{m}$\\
     Molecular peak  & $\sim$10.4& 400$^{j}$ & ... & ... & ... & 4286$^{m}$\\
NGC 7023 (PDR peak)  & $\sim$2.7 & 50-120$^{j}$   & 31$^{k}$ & 1.6-3.9 & 2$^{n}$ & 66$^{o}$\\
     Molecular peak  &   ...     & $>$200$^j$ & ... & ...&... & $\ga$61$^{o}$\\
NGC 2023             & $\sim$10.0&1913$^{p}$ & 12$^{k}$ & 159.4 &... & ... \\
Diffuse clouds       &   ...     &70-120$^q$  & ...     & ... & ... & ... \\
\hline
\end{tabular}
\begin{tablenotes}
\item $^a$ We have derived the N$_{\rm HCO^+}$=5.7 (1.3)$\times$10$^{14}$ cm$^{-2}$ from the N$_{\rm HC^{18}O^+}$ assuming the $^{16}$O/$^{18}$O=250 ratio (W\&R94).
We also have found the N$_{\rm HOC^+}$=1.1 (0.1)$\times$10$^{12}$ cm$^{-2}$ and N$_{\rm HCO}$=6.9 (1.4)$\times$10$^{13}$ cm$^{-2}$ 
in \emph{LOS}+0.693.
\item $^b$ We have derived the N$_{\rm HCO^+}$=3.6 (0.8)$\times$10$^{14}$ cm$^{-2}$ from the N$_{\rm HC^{18}O^+}$ assuming the $^{16}$O/$^{18}$O=250 ratio (W\&R94).
We also have found the N$_{\rm HOC^+}$$\la$0.3$\times$10$^{12}$ cm$^{-2}$ and N$_{\rm HCO}$$\la$3.3$\times$10$^{13}$ cm$^{-2}$ in \emph{LOS}$-$0.11.
\item $^c$ \citet{sergio2009b}
\item $^d$ Estimated from \citet{sergio06b,sergio2009b}.
\item $^e$ Estimated from \citet{Burillo} and \citet{alad11a}.
\item $^f$ Estimated from \citet{Fuen06} and \citet{alad11a}.
\item $^g$ Estimated from \citet{sergio2009a} and \citet{alad13}.
\item $^h$ \citet{Javi09b}.
\item $^i$ \citet{Gerin}.
\item $^j$ \citet{Fuen03}.
\item $^k$ \citet{shilke}.
\item $^l$ Estimated from \citet{Jansen} and \citet{shilke}.
\item $^m$ Estimated from \citet{Jansen} and \citet{Fuen03}.
\item $^n$ Estimated from \citet{Fuen93} and \citet{shilke}.
\item $^o$ Estimated from \citet{Fuen93,Fuen03}.
\item $^p$ \citet{savage}.
\item $^q$ \citet{Liszt}.
\end{tablenotes}
\end{minipage}
\end{threeparttable}
\end{table*}

\subsubsection{The CS/HOC$^+$ and CS/HCO ratios}

Both \emph{LOSs} in the GC show CS/HOC$^+$ ratios which are larger than those measured in Galactic PDRs 
and galaxies by factors of $\ga$4 and $\ga$15, respectively, except for that of the molecular peak of the Orion Bar, which is completely shielded from UV radiation. Although the CS/HCO ratios
also show the same trend as CS/HOC$^+$, the CS/HCO ratios are less conclusive since these ratios are within a factor of $\sim$3 for 
all kinds of objects. This suggests that the PDR/XDR component, traced by HOC$^+$ and to a lesser extent by HCO in both \emph{LOSs} of the GC is smaller than 
those in starburst galaxies and typical unshielded galactic PDRs. The larger CS/HOC$^+$ ratio found in \emph{LOS}$-$0.11 than 
in \emph{LOS}+0.693 suggests a very small (basically negligible) XDR/PDR component along \emph{LOS}-0.11. This is consistent with the measurements of UV and
X-ray emission toward these \emph{LOSs} shown in Fig.~\ref{fig1} and~\ref{fig10}, respectively.
The CS/HOC$^+$ ratios measured in external galaxies are also consistent with this ratio being a good tracer of the PDR/XDR component 
relative to the total gas (see Table \ref{table4}). The most evolved starburst M 82, has the lowest CS/HOC$^+$ ratio (1), consistent with 
its classification as a PDR dominated galaxy while NGC 253 an intermediate-age starburst has a larger CS/HOC$^+$ ratio of 25. The intermediate CS/HOC$^+$ ratio of 6 found 
in NGC 1068, which is considered to be dominated by X-ray chemistry, is much smaller than the ratio found in \emph{LOS}+0.693 which also shows X-rays,
suggesting that the CS/HOC$^+$ ratio could be also a good tracer of strong XDR components since the X-ray luminosity in NGC 1068 \citep{Iwasawa} is nearly four orders
of magnitude higher than in Sgr B2 \citep{Koyama}.

\subsubsection{The HCO$^+$/HOC$^+$ and HCO$^+$/HCO ratios}

The abundance ratios HCO$^+$/HOC$^+$ and HCO$^+$/HCO have also been used to estimate the contribution of the UV radiation to the chemistry of
molecular clouds. HCO$^+$/HOC$^+$ ratios of $<$166 and 75-200 are measured in the Orion Bar \citep{Fuen03} and the Horsehead \citep{Javi09b}, 
respectively, considered prototypical galactic PDRs. We found much larger HCO$^+$/HOC$^+$ ratios of 546 (175) and $\ga$1134 for \emph{LOS}+0.693 
and \emph{LOS}$-$0.11, respectively. These HCO$^+$/HOC$^+$ ratios derived in both GC sources are also higher than those measured in extragalactic sources. 
Like for the CS/HOC$^+$ ratio we find that the HCO$^+$/HOC$^+$ ratio of the molecular peak in the Orion Bar, as well as that of NGC 2023 are close to the values 
observed in the GC. We consider NGC 2023 to represent the conditions in a prototypical giant molecular cloud, as \citet{savage} have claimed that the PDR position observed is 
probably embedded in the molecular cloud, therefore the HCO$^+$/HOC$^+$ ratio would be biased toward the typical cloud unaffected by UV radiation.

The HCO$^+$/HCO ratios show the same trend than the HCO$^+$/HOC$^+$ ratios. The HCO$^+$/HCO ratios are factors of 4-8 (\emph{LOS}+0.693) 
and $\ga$5-10 (\emph{LOS}$-$0.11) higher than the HCO$^+$/HCO ratio of 1-2 in typical Galactic PDRs. 
However, it is not clear why the HCO$^+$/HCO ratios show smaller differences between both GC sources and galaxies.
In our comparison, the PDRs NGC 7023 and NGC 2023 show the highest HCO$^+$/HCO ratios, which are biased because both sources
were observed toward positions where the HCO$^+$ emission arises from the shielded molecular clouds \citep{shilke,savage}.

The difference found in the HCO$^+$/HOC$^+$ and HCO$^+$/HCO ratios between our sample of typical PDRs and \emph{LOS}+0.693 could be due to X-ray 
induced chemistry in the giant XDR observed toward Sgr B2 in the Fe K$\alpha$ (6.4 keV) line. It has been claimed that the HOC$^+$ and HCO 
abundances might increase in regions illuminated by X-rays. \citet{usero} argued that XDR chemistry could provide an explanation for the different 
abundances of HCO$^+$ and HOC$^+$ measured in the Circumnuclear Disk of the AGN NGC 1068, which also shows strong Fe K$\alpha$ (6.4 keV) line like
in \emph{LOS}+0.693. However they found HCO$^+$/HOC$^+$ ratios of $\sim$40-100 and the HCO$^+$/HCO ratio of 3 in the CND of NGC 1068 which are similar to 
those found in our sample of typical PDRs. Therefore XDR chemistry seems unlikely to explain the large HCO$^+$/HOC$^+$ and 
HCO$^+$/HCO ratios observed in the GC \emph{LOS}.

\citet{Meijerink} modeled the chemistry induced by PDRs and XDRs in clouds with different Hydrogen column densities and incident FUV/X-ray 
radiation fields. By using these models they predicted column density ratios for several molecules, including the HCO$^+$/HOC$^+$ and HCO$^+$/HCO 
ratios. For the Hydrogen column density of $\sim$2$\times$10$^{21}$ cm$^{-2}$ and G$_0$ of 10$^3$, the PDR model predicts HCO$^+$/HOC$^+$ ratios 
of $\sim$100, which agree with those derived in NGC 7023 and the Orion Bar \citep{Fuen03}.
For an X-ray flux of 1.6 erg cm$^{-2}$ s$^{-1}$, and a density of 10$^4$ cm$^{-3}$, appropriate to \emph{LOS}+0.693, the XDR models of \citep{Meijerink}
predict an HCO$^+$/HOC$^+$ ratio of $\sim$10 in the cloud interior.
The predicted ratio is at least a factor of 55 lower than those inferred in both GC sources (see Table \ref{table4}), indicating that XDR models cannot reproduce the 
observed HCO$^+$/HOC$^+$ ratios found in both GC sources. This result is consistent with the lack of X-ray emission in \emph{LOS}$-$0.11 (see Fig.~\ref{fig10}).
The derived HCO/HOC$^+$ ratio of $\sim$63 derived in \emph{LOS}+0.693 cannot be used to distinguish between the XDR or PDR scenario 
since this ratio is close to those of both the Horsehead, a typical PDR, and the claimed XDR NGC 1068.
Finally, the HCO$^+$/HCO ratio of $\sim$1 predicted by \citet{Meijerink} for PDRs with G$_0$ of 10$^3$ is consistent with those in the Horsehead and
the Orion Bar \citep{Gerin,shilke}.

In summary, as discussed throughout this section the large HCO$^+$/HOC$^+$, CS/HOC$^+$, HCO$^+$/HCO ratios suggest that the molecular gas 
affected by X-ray/UV radiation fields in the two GC \emph{LOSs} represent a small fraction of the total column density of 
molecular gas. Assuming that the observed HCO$^+$/HCO ratio of 1.1 in the Horsehead represents the actual PDR ratio, we estimated that roughly $\sim$12\%
of the total column density would be affected by the UV radiation in \emph{LOS}+0.693 and $\la$10\% in \emph{LOS}$-$0.11.
These results also support the HNCO/CS diagnostic diagram proposed by \citet{sergio2008,sergio2009a} to establish the dominant 
chemistry and the heating mechanism working in molecular clouds since by using this diagram they found that the chemistry and likely the heating of both GC sources
are mainly dominated by shocks.

\section{Conclusions}\label{conclusions}

We have used the Mopra telescope to carry out a 3 mm spectral line survey in the selected frequency ranges of $\sim$77-93 GHz and $\sim$105-113 GHz 
of two lines of sight, \emph{LOS}+0.693 and \emph{LOS}$-$0.11, toward the Sgr B2 and Sgr A complexes in the Galactic center. The 
main conclusions of our study are the following:

\begin{itemize}
\item We have detected 38 molecular species and 25 isotopologues in \emph{LOS}+0.693 and 34 molecular species and 18 isotopologues in 
\emph{LOS}$-$0.11. We have detected for the first time the PDR/XDR tracers HCO and HOC$^+$ in the quiescent gas in 
\emph{LOS}+0.693. These two species and the complex organic molecules HC$_2$NC and HCOCH$_2$OH have not been detected toward \emph{LOS}$-$0.11.

\item The molecular excitation, T$_{rot}$, and the molecular column densities are derived for all detected molecules by using a LTE 
approximation. The derived T$_{rot}$ varies between $\sim$5 and 73 K for both GC sources, but most molecules show T$_{rot}$$<$20 K, indicating 
subthermal excitation. The symmetric rotors, CH$_3$CN, $^{13}$CH$_3$CN and CH$_3$CCH show the highest T$_{rot}$ of $\sim$55-73 K, consistent 
with a T$_{kin}$ of $\sim$100 K previously derived for the GC clouds.

\item Although \emph{LOS}+0.693 and \emph{LOS}$-$0.11 are separated by more than $\sim$120 pc within the GC, $\sim$80\% of 
molecular species detected in both GC sources reveal similar abundances within a factor of 2 and similar excitation conditions.

\item We have used the large number of detected isotopologues to derive isotopic ratios of $^{12}$C/$^{13}$C, $^{14}$N/$^{15}$N, 
$^{16}$O/$^{18}$O, $^{18}$O/$^{17}$O, $^{29}$Si/$^{30}$Si and $^{32}$S/$^{34}$S for both GC sources. The derived $^{12}$C/$^{13}$C, $^{18}$O/$^{17}$O and
$^{29}$Si/$^{30}$Si ratios averaged over both GC sources agree within uncertainties with the ``canonical'' values for the GC. 
Our results suggest that isotopic fractionation and/or selective photodissociation do not play any role in the determination of isotopic ratios 
from molecular column densities.

\item The comparison of the excitation conditions derived for both \emph{LOSs} in the center of our Galaxy and those found in the starburst galaxies
NGC 253 and M 82 shows that the molecular gas in the nuclei of these galaxies have similar physical conditions.

\item CH$_3$OH is the molecule which shows the highest abundance difference between both GC \emph{LOSs} and starburst galaxies by factors
of $\sim$(1-6)$\times$10$^2$. The large difference is likely due to its photodissociation by UV radiation in starbursts.

\item We have studied the HCO$^+$/HOC$^+$, HCO$^+$/HCO and CS/HOC$^+$ ratios in both GC \emph{LOSs}, typical PDR regions, starburst galaxies and 
the AGN NGC 1068. We find that these abundances ratios cannot be used to distinguish between the effects of the X-ray and UV radiation on the 
molecular clouds.

\item We also propose that the CS/HOC$^+$, HCO$^+$/HCO and HCO$^+$/HOC$^+$ ratios could be used as good tracers of PDR/XDR components in the
molecular clouds in the nuclei of galaxies. These ratios can be used to estimate the fraction of the molecular gas affected by the UV radiation. For example, the large 
HCO$^+$/HCO ratio found in \emph{LOS}+0.693 indicates a PDR component of $\sim$12\% of the total column density.
\end{itemize}

\section*{Acknowledgments}

This work has been partially funded by MICINN grants AYA2010-21697-C05-01 and FIS2012-39162-C06-01, and Astro-Madrid (CAM S2009/ESP-1496). We
also thank the Spanish Ministerio de Ciencia e Innovaci\'on under project ESP2013-47809-C3-1-R. We are very grateful to the 
anonymous referee for suggestions and comments, which have greatly improved the paper.
S.M. acknowledges the co-funding of this work under the Marie Curie Actions of the European Commission (FP7-COFUND).

\end{document}